\RequirePackage{fixltx2e}
\documentclass[british]{article}

\usepackage[T1]{fontenc}
\usepackage[latin9]{inputenc}
\usepackage{geometry}
\usepackage{color}
\usepackage{babel}
\usepackage{enumitem}
\usepackage{amsmath}
\usepackage{amsthm}
\usepackage{amssymb}
\usepackage{mathdots}
\usepackage{setspace}
\usepackage[authoryear]{natbib}
\usepackage[unicode=true,
 bookmarks=true,bookmarksnumbered=true,bookmarksopen=true,bookmarksopenlevel=2,
 breaklinks=true,pdfborder={0 0 1},backref=false,colorlinks=false]
 {hyperref}
\hypersetup{
 pdfauthor={Bernd Funovits}}

\makeatletter
\theoremstyle{plain}
\newtheorem{thm}{\protect\theoremname}[section]
\theoremstyle{remark}
\newtheorem{rem}[thm]{\protect\remarkname}
\theoremstyle{definition}
\newtheorem{example}[thm]{\protect\examplename}

\date{October 2016}

\makeatother

\providecommand{\examplename}{Example}
\providecommand{\remarkname}{Remark}
\providecommand{\theoremname}{Theorem}

\begin{document}
\title{The Dimension of the Set of Causal Solutions of Linear Multivariate
Rational Expectations Models}
\author{Bernd Funovits}
\maketitle

\subsubsection*{Address:}

\begin{singlespace}
University of Helsinki
\end{singlespace}

Department of Economics

\begin{singlespace}
P.O. Box 17 (Arkadiankatu 7)
\end{singlespace}

FIN-00014 Helsinki

Finland

\subsubsection*{E-mail:}

bernd.funovits@helsinki.fi

\subsubsection*{}

\pagebreak{}

\textbf{}
\begin{abstract}
This paper analyses the number of free parameters and solutions of
the structural difference equation obtained from a linear multivariate
rational expectations model. First, it is shown that the number of
free parameters depends on the structure of the zeros at zero of a
certain matrix polynomial of the structural difference equation and
the number of inputs of the rational expectations model. Second, the
implications of requiring that some components of the endogenous variables
be predetermined are analysed. Third, a condition for existence and
uniqueness of a causal stationary solution is given.

JEL Classification: C51, C32, C62

Keywords: Rational expectations, indeterminacy, stochastic singularity,
predetermined variables

\pagebreak{}
\end{abstract}

\section{Introduction}

We investigate structural difference equations (SDE) obtained from
a stochastically singular rational expectations (RE) model whose parameters
are rational functions of deep parameters. Here, stochastically singular
means that the number of endogenous variables is larger than the number
of white noise inputs of the exogenous process driving the economy.
An RE model involves conditional (with respect to the exogenous process
up to a certain point in time) expectations of future endogenous variables. 

The SDE obtained from an RE model distinguishes itself from a usual
SDE in the following way. For every fixed parameter value in the model
class describing the system involving conditional expectations, we
obtain an SDE directly formulated in the endogenous variables (without
conditional expectations) with additional free parameters.

Dynamic stochastic general equilibrium (DSGE) models are a special
kind of RE model and are of particular interest in economics \citep{SmetsWouters03,SmetsWouters07}.
They are obtained from optimizing behaviour of economic agents and
imply cross-equation restrictions on the model parameters \citep{HansenSargent80,Pesaran87}.
(Log)-linearising the (stochastic) first order conditions of the dynamic
stochastic optimization problem around the (non-stochastic) steady
state (see, e.g., \citep{DeJongDave11} Chapter 2) gives a system
of equations involving RE which is linear in the variables and in
general non-linear in the parameters.  

It is of interest whether there exists a unique solution under additional
assumptions on the solution set. Usually, it is required that the
solution $\left(y_{t}\right)_{t\in\mathbb{Z}}$ of an RE model be
non-explosive and causal. Additionally, some model specifications
require that some endogenous variables, so-called predetermined variables,
have trivial forecast errors. Non-explosive behaviour (of at least
some linear combinations) can be justified by the transversality condition
which is a necessary condition for optimality in a dynamic optimization
problem. Causality is often imposed implicitly (see \citep{Sims01}
page 10) by requiring that for a solution $\left(y_{t}\right)_{t\in\mathbb{Z}}$
of an RE model the condition $\mathbb{E}_{t}\left(y_{t}\right)=y_{t}$
holds for an appropriately chosen conditioning set. Assuming causality
can be justified by the fact that the behaviour of economic agents
is fully modelled by a DSGE model, i.e. there are no missing equations
as, e.g., in \citet{LanneSaikkonen13}. In analogy to difference equations
with non-stochastic inputs, the seminal paper by \citep{BlanchardKahn80}
analysed a system where a subset of endogenous variables has a trivial
one-step-ahead forecast error (as a consequence of the timing convention
used in their model). These variables are then called predetermined.
All these restrictions will be imposed transparently in the framework
of the model described in this paper.

The case where multiple solutions satisfy all restrictions imposed
by the modeller attracted some attention as well. It is of great interest
for monetary policy analysis, see, e.g., \citep{ClaridaGaliGertler99}
and \citep{Gali}. \citep{LubikSchorfheide03} analyses the case of
indeterminate equilibria in which it is not possible to get rid of
all additional free parameters in the SDE by imposing non-explosiveness
and causality. However, the authors overlooked that the solution of
this SDE may still be unique, see \citep{FunoDiss} Chapter 6.

Commonly, the problem of stochastic singularity in DSGE models is
put aside by adding measurement noise (see, e.g., \citep{Sargent89,Altug89,Ireland04}),
adding structural shocks (see, e.g., \citep{LeeperSims94,SmetsWouters03}),
or choosing a subset of variables such that the number of structural
shocks and observables coincide. These approaches simplify the problem
of estimation to the usual (non-singular) case. 

However, all these estimation strategies have obvious disadvantages
since they distort the actual estimation problem. An estimation procedure
like adding measurement errors should rather be interpreted as a regularization
strategy\footnote{Adding measurement noise is a technical device unrelated to the underlying
problem of estimating the structural parameters.}. It should be a concern for economists that the model obtained by
adding measurement noise or by adding additional structural shocks
is not necessarily related to economic theory and that there is in
general no reason why the identifiability properties of the distorted
models should be similar to the undistorted, stochastically singular
model. Also, information and estimation efficiency is forfeited by
only using a subset of observable variables. Last and most importantly,
we argue that there is a clean alternative: estimation of the deep
parameters in singular ARMA models. 

In order to understand this connection, we focus on the separation
of the estimation problem of the internal characteristics (e.g. the
parameters in the econometric model) of the SDE obtained by manipulating
the RE model in two separate problems. On the one hand, the problem
of estimating the external characteristics (e.g. second moments or
spectral density of the observations) of the RE model is of a purely
statistical nature. On the other hand, the problem of attaching the
internal characteristics to the external characteristics of the system,
i.e. the identifiability problem, is of an algebraic (and topological)
nature. The essential step is thus the solution of the identifiability
problem. To be more precise, identifiability is defined in \citep{DeistlerSeifert78}
as the existence of an identifying function attaching internal characteristics
of a system (e.g. the parameters in the econometric model) to external
characteristics of a system (e.g. the second moments or spectral density
of the observations). As soon as we know that the external characteristics
are almost everywhere consistently estimable and that the econometric
model is almost everywhere identifiable, then the internal characteristics
are almost everywhere consistently estimable, compare \citep{DeistlerSeifert78}
Theorem 5. Applications of this idea of separating the estimation
problem into these two sub-problems can be found, e.g., in \citep{HannanDeistler12}.
For an example of using this approach in stochastically singular models
see \citep{festschriftbdoanderson2010}, where a (continuous) identifying
function is constructed which attaches the parameters of the a singular
AR model to the second moments of the observations. A similar approach
is conceivable for structural singular ARMA models. 

This paper was inspired by the identifiability analysis for DSGE
models in a state space setting in \citep{KomunjerNg11}. They assume,
however, that the considered state space models have a full rank controllability
and observability matrix, compare Assumption 4-S and Assumption 5-NS
on page 2002 and page 2007 in \citep{KomunjerNg11} respectively.
This assumption is quite strong for structural models, in which the
system matrices (and thus the controllability and observability matrix)
depend on deep parameters and are thus restricted. For example, it
is shown in \citep{AndersonDeistlerFarinaBenvenuti96} that there
are non-pathological examples where the dimension of the state is
minimal among all state space models satisfying the structural restrictions
(in their case non-negativity of the system matrices), but the corresponding
controllability and observability matrices do not have full rank.
Treating the identifiability problem for non-minimal structural state
space models is considerably more difficult\footnote{Compare \citep{GloverWillems74} for identifiability analysis in the
minimal structural state space setting and \citep{Glover73} for examples
as to why identifiability analysis in the non-minimal case is considerably
more difficult.} than the corresponding problem of non-coprime\footnote{Two univariate polynomials are called coprime if they do not have
any non-trivial common factor. Two polynomial matrices are called
coprime if their only common factors are unimodular matrices, i.e.
polynomial matrices with constant non-zero determinant.} (non-singular) structural ARMA models as was done in \citep{DeistlerSchrader79}
for the non-singular ARMA case. The results in this paper provide
a basis for identifiability analysis of structural singular ARMA models
obtained from RE models which is robust with respect to determinacy
and indeterminacy, see also \citep{CastelnuovoFanelli15,QuTkachenko15}.

In this paper, we correct two results in our main reference \citep{BGS95}
and extend on them in three ways. Regarding the corrections, we firstly
correct Property 5 on page 245 in \citep{BGS95} describing the dimension
of the solution set of the RE model. This dimension depends in an
intricate way on the structure of the zeros at zero of the matrix
polynomial of the SDE. Second, Theorem 4 on page 248f in \citep{BGS95}
on the number of ``unstable roots'' of the polynomial matrix (occurring
in the SDE) which is necessary and sufficient for a unique causal
and non-explosive solution is incorrect (and at odds with similar
results in, e.g., \citep{BlanchardKahn80} and \citep{whiteman83}). 

Regarding extensions, we allow for stochastic singularity\footnote{A stochastically singular model has strictly fewer white noise inputs
(of the exogenous process) than endogenous variables.} (as, e.g., in \citep{Qu15Singular}) and derive (Theorem \ref{thm:general-solution-set}
on page \pageref{thm:general-solution-set}) in a transparent way
the number of additional free parameters in the SDE obtained from
an RE model. Secondly, we allow that some components of the endogenous
variables are predetermined which makes our results comparable to
the ones in, e.g., \citep{BlanchardKahn80,KingWatson98,Sims01,LubikSchorfheide03}.
Thirdly, we allow for rational cross-equation restrictions (in contrast
to zero restrictions in \citep{BGS95}). These generalizations allow
for the treatment of the basic New Keynesian monetary model analysed
in \citep{LubikSchorfheide03} and \citep{SmetsWouters07}. 

Additionally, we lay the foundations for the identifiability analysis
of both structural parameters and additional free parameters from
second moment properties of the causal stationary solutions without
imposing minimality or coprimeness in order to generalize the analysis
in \citep{KomunjerNg11} and \citep{deistler75,deistler76,deistler78,DeistlerSchrader79,deistler83,deistlerwang89}.

Some RE models of the form we are treating cannot be transformed
to the normal form used in the papers \citep{BlanchardKahn80} and
\citep{KingWatson98}. All models in these papers can be treated by
the method presented here. Moreover, the solution approach used in
our paper is different. While \citep{BlanchardKahn80} solves the
system involving conditional expectations directly, we firstly transform
the system involving conditional expectations to a difference equation
with additional free parameters and subsequently solve the system
with usual polynomial methods as in \citep{deistler75}.\citep{Sims01}
starts from a model which is already transformed to a first order
difference equation involving a certain number of free parameters.
The goals of \citep{Sims01} are rather of computational nature and
therefore it is not shown how to transform an RE model to Sims' canonical
form and how to obtain certain dependencies among the (in Sims' wording)
``endogenous forecast errors'' which are - as will be shown - of
a complicated nature. \citep{Sims01} gives high-level conditions
for existence and uniqueness which have their counterpart in this
paper, but does not analyse the case of multiple solutions in more
detail. \citep{LubikSchorfheide03} treats the case in which multiple
causal stationary solutions exist. However, \citep{LubikSchorfheide03}
does not analyse the uniqueness condition (which would give a sufficient
condition for the existence of multiple solutions) but only the existence
condition (which gives only a necessary condition for the existence
of multiple solution). In this sense, our analysis of multiple solutions
is more general than the one in \citep{Sims01,LubikSchorfheide03}.
As already mentioned, this paper was inspired by the identifiability
analysis of DSGE models conducted in a state space setting in \citep{KomunjerNg11}.
Their analysis of the problem differs from ours in various ways. First,
they start from Sims' canonical form together with a measurement equation.
Thus, they do not derive dependencies among ``endogenous forecast
errors''. Second, they impose a minimality assumption on the state
space model which is very restrictive in a structural setting, see
\citep{deistler78,DeistlerSchrader79}. Treating non-minimal state
space models is, as already mentioned above, more involved. Our singular
ARMA approach allows for non-coprimeness (which in an ARMA context
is analogous to non-minimality in a state space context) and is thus
more general. 

The paper is structured as follows. In Section \ref{sec:Model},
we introduce the RE model and discuss assumptions relating to it.
In Section \ref{sec:recursive_eq}, we start by deriving an SDE involving
martingale difference sequences (MDS) from the RE model. Subsequently,
it is shown that the set of causal solutions of the RE model and the
SDE involving arbitrary MDS are equivalent under certain conditions
on the MDS. In Section \ref{sec:dimension}, the results regarding
the dimension of the solution set in \citep{BGS95} are corrected
with respect to the number of zeros at zero of a certain polynomial
matrix. In general, there are fewer restrictions on the MDS in order
to render the solution sets of the RE model and the SDEs equivalent.
Furthermore, we generalize the result by allowing for stochastic singularity.
In Section \ref{sec:predetermined}, the RE model is extended to allow
for predetermined components of the endogenous variables. This is
an important generalization of the RE model treated in \citep{BGS95}
because it makes comparison to the methods in, e.g., \citep{BlanchardKahn80,KingWatson98}
possible. Theorems of the same kind as in the case of non-predetermined
endogenous variables are derived for the case allowing for predetermined
variables. In Section \ref{sec:exist_unique}, we show how to obtain
the set of all causal solutions, compare our method to the one outlined
in \citep{Sims01,LubikSchorfheide03}, and illustrate why the result
in \citep{BGS95} is incorrect. 

\pagebreak{}

\section{\label{sec:Model}Model}

We heavily draw on the methods developed in \citep{BS85_fr,BGS85,BGS90_book,BrozeSzafarz91,BGS95}.
In particular, we consider the RE model 
\begin{align}
\sum_{k=0}^{K}\sum_{h=0}^{H}A_{kh}\mathbb{E}_{t-k}\left(y_{t+h-k}\right) & =-u_{t}\label{eq:BGS95_model}\\
\iff\begin{pmatrix}I_{s} & \cdots & I_{s}z^{k} & \cdots & I_{s}z^{K}\end{pmatrix}\begin{pmatrix}A_{00} & \cdots & A_{0h} & \cdots & A_{0H}\\
\vdots & \ddots & \vdots &  & \vdots\\
A_{k0} & \cdots & A_{kh} & \cdots & A_{kH}\\
\vdots &  & \vdots & \ddots & \vdots\\
A_{K0} & \cdots & A_{Kh} & \cdots & A_{KH}
\end{pmatrix}\begin{pmatrix}\mathbb{E}_{t}\left(y_{t}\right)\\
\vdots\\
\mathbb{E}_{t}\left(y_{t+h}\right)\\
\vdots\\
\mathbb{E}_{t}\left(y_{t+H}\right)
\end{pmatrix} & =-u_{t}\nonumber 
\end{align}
where $z$ denotes a complex variable as well as the backward shift
operator, i.e. $z\left(y_{t}\right)_{t\in\mathbb{Z}}=\left(y_{t-1}\right)_{t\in\mathbb{Z}}$,
and thus also $z\left(\mathbb{E}_{t}\left(y_{t+h}\right)\right)_{t\in\mathbb{Z}}=\left(\mathbb{E}_{t-1}\left(y_{t+h-1}\right)\right)_{t\in\mathbb{Z}}$. 

\paragraph{Assumptions on the parameter matrices.}

We assume that there exists an $h\in\left\{ 0,\ldots,H\right\} $
such that $A_{Kh}\neq0$ and a $k\in\left\{ 0,\ldots,K\right\} $
such that $A_{kH}\neq0$. In this way $K$ and $H$ are well defined.
The indices $k$ and $h$ in $A_{kh}$ refer to the $h$-period-ahead
forecast of the endogenous variables, at time $t-k$, i.e. $y_{t-k}$
is forecast $h$ periods ahead with the information\footnote{We will refer to the space $H_{u}(t)$ on which the endogenous variables
are projected as ``the information at time $t$''.} available in period $t-k$.

Furthermore, we assume that there are no redundant equations in the
sense that the determinant of the matrix polynomial 
\begin{equation}
\pi(z)=z^{J_{1}}\left(\sum_{i=J_{0}}^{J_{1}}A_{i}^{*}z^{-i}\right)\label{eq:recursive}
\end{equation}
 where
\begin{gather*}
A_{i}^{*}=\sum_{k=\max\left\{ 0,-i\right\} }^{\min\left\{ K,H-i\right\} }A_{k,k+i},\\
J_{0}=argmin_{i}\left\{ i\ |\ A_{i}^{*}\neq0\right\} ,\,J_{1}=argmax_{i}\left\{ i\ |\ A_{i}^{*}\neq0\right\} ,
\end{gather*}
is not identically zero. The significance of $\pi(z)$ will become
clear in Section \ref{subsec:BGS-recursive-equation}.

\paragraph{Assumptions on the exogenous process.}

We assume that the linearly regular stationary $s$-dimensional exogenous
process $\left(u_{t}\right)_{t\in\mathbb{Z}}$ has a (finite) covariance
matrix $\mathbb{E}\left(u_{t}u_{t}^{T}\right)$, where the superscript
$T$ denotes transposition, of rank $r$ smaller than or equal to
$s$. We denote its Wold representation as $u_{t}=\sum_{j=0}^{\infty}w_{j}\varepsilon_{t-j},\ w_{j}\in\mathbb{R}^{s\times q}$.
Obviously, the rank of the innovation covariance matrix $\mathbb{E}\left(\varepsilon_{t}\varepsilon_{t}^{T}\right)$
is $q$ and the inequalities $q\leq r\leq s$ hold.

\paragraph{The conditional expectations.}

The expression $\mathbb{E}_{t}\left(y_{t+h}\right)$ denotes the projection\footnote{Compare \citet{Doob53} page 155, where the conditional expectation
in the wide sense is defined as the projection on a linear manifold. } of the $s$-dimensional process $y_{t+h}$ on the closure of the
linear\footnote{Note that if all random variables in the conditioning set are Gaussian,
the conditional expectation coincides with the linear projection outlined
here. For more details on conditional expectations see \citet{Billingsley95}
page 445ff.} space spanned by the (present and past) components of $\left\{ u_{t},u_{t-1},\ldots\right\} $
of the exogenous process $\left(u_{t}\right)_{t\in\mathbb{Z}}$, denoted
by $H_{u}(t)=\overline{span\left\{ u_{t-j}^{(i)}\,|\,j\in\mathbb{N},\ i\in\left\{ 1,\ldots,s\right\} \right\} }$,
where the superscript $(i)$ denotes the $i$-th component and $\mathbb{N}=\left\{ 0,1,2,\ldots\right\} $.
To avoid confusion, we will sometimes write more explicitly $\mathbb{E}\left(y_{t+h}|H_{u}(t)\right)$
for the same object\footnote{Some authors, e.g. \citet*{GourierouxLaffontMonfort82} on page 410,
condition on a larger set of variables, containing variables which
are independent with respect to the exogenous process. These variables
are called ``sunspots'' by the authors.}. 

\paragraph{Definition of a solution of an RE model.}

A solution of the RE model (\ref{eq:BGS95_model}) is a stochastic
process $\left(y_{t}\right)_{t\in\mathbb{Z}}$ for which $y_{t}\in H_{u}(t)$,
such that for given exogenous process $\left(u_{t}\right)_{t\in\mathbb{Z}}$
and given parameters $A_{kh},\ k\in\left\{ 0,\ldots,K\right\} ,\,h\in\left\{ 0,\ldots,H\right\} ,$
$\left(y_{t}\right)_{t\in\mathbb{Z}}$ satisfies equation (\ref{eq:BGS95_model})
for all $t\in\mathbb{Z}$. Note that $\left(y_{t}\right)_{t\in\mathbb{Z}}$
is a deterministic function of $\left(u_{t}\right)_{t\in\mathbb{Z}}$,
i.e. there are no additional error terms involved. We restrict ourselves
to solutions which are causal with respect to $\left(u_{t}\right)_{t\in\mathbb{Z}}$.
This is justified by the fact that we have a complete set of equations
describing the behaviour of the economic agents.

\begin{rem}[Consequences of a larger conditioning set]
\label{rem:larger-conditioning-set} Following \citep{GourierouxLaffontMonfort82}
page 411, we consider the univariate model 
\[
y_{t}=a_{01}\mathbb{E}_{t}\left(y_{t+1}\right)+u_{t}.
\]
Here, a process $\left(y_{t}\right)_{t\in\mathbb{Z}}$ for which
the equation above holds for all $t\in\mathbb{Z}$ is not required
to be contained in $H_{u}(t)$, but only in $H_{u,\zeta}(t)$ (the
direct product of the spaces $H_{u}(t)$ and $H_{\zeta}(t)$) where
$\left(\zeta_{t}\right)_{t\in\mathbb{Z}}$ is a $p$-dimensional stochastic
process orthogonal to $\left(u_{t}\right)_{t\in\mathbb{Z}}$. If the
conditional expectation is taken with respect to $H_{u,\zeta}(t)$,
a larger solution set might be obtained. 

The following superposition principle holds: If $\left(y_{t}^{1}\right)_{t\in\mathbb{Z}}$
is a particular solution of 
\[
y_{t}^{1}=a_{01}\mathbb{E}\left(y_{t+1}^{1}|H_{u}(t)\right)+u_{t}
\]
in the sense that it solves the equation above for given $\left(u_{t}\right)_{t\in\mathbb{Z}}$
for every point in time and if $\left(y_{t}^{2}\right)_{t\in\mathbb{Z}}$
is a particular solution of 
\[
y_{t}^{2}=a_{01}\mathbb{E}\left(y_{t+1}^{2}|H_{\zeta}(t)\right)+\zeta_{t}
\]
then (by orthogonality of $\left(u_{t}\right)_{t\in\mathbb{Z}}$ and
$\left(\zeta_{t}\right)_{t\in\mathbb{Z}}$) we obtain that $\left(y_{t}^{1}+y_{t}^{2}\right)_{t\in\mathbb{Z}}$
is a solution of 
\[
\left(y_{t+1}^{1}+y_{t+1}^{2}\right)=a_{01}\mathbb{E}\left[\left(y_{t+1}^{1}+y_{t+1}^{2}\right)|H_{u,\zeta}(t)\right]+u_{t}+\zeta_{t}.
\]
In particular, allowing for a larger conditioning set entails that
the set of solutions of 
\[
y_{t}=a_{01}\mathbb{E}_{t}\left(y_{t+1}|H_{u,\zeta}(t)\right)+u_{t}
\]
 is enlarged (compared to the set of solutions of $y_{t}=a_{01}\mathbb{E}_{t}\left(y_{t+1}|H_{u}(t)\right)+u_{t}$)
by the solutions of the homogeneous equation 
\[
y_{t}=a_{01}\mathbb{E}\left(y_{t+1}|H_{\zeta}(t)\right).
\]
\textcolor{red}{}

\end{rem}

\pagebreak{}

\section{\label{sec:recursive_eq}A Constrained System Equivalent to an RE
Model}

\paragraph{Forecast errors of endogenous variables.}

First, we derive (in analogy to \citep{BGS95}) from the RE model
(\ref{eq:BGS95_model}) an SDE in terms of leads and lags of the endogenous
process. To this end, we write the conditional expectation $\mathbb{E}_{t-k}\left(y_{t-k+h}\right)$
as sum of the endogenous variable $y_{t-k+h}$ and its $h$-step-ahead
forecast error $v_{t-k+h,h}=y_{t-k+h}-\mathbb{E}_{t-k}\left(y_{t-k+h}\right)$.
Obviously, a solution $\left(y_{t}\right)_{t\in\mathbb{Z}}$ of the
RE model (\ref{eq:BGS95_model}) also satisfies the SDE at every point
in time. Subsequently, constraints that are satisfied by the revision
processes $\varepsilon_{t-j}^{j}=\mathbb{E}_{t-j}\left(y_{t}\right)-\mathbb{E}_{t-\left(j+1\right)}\left(y_{t}\right)$
for solutions $\left(y_{t}\right)_{t\in\mathbb{Z}}$ of the RE model
are derived in Section \ref{subsec:bgs-Constraints}. Second, we show
in Section \ref{subsec:bgs-equivalent-solution} that a process $\left(y_{t}\right)_{t\in\mathbb{Z}},\ y_{t}\in H_{u}(t)$
for which the SDE holds and whose revision processes $\varepsilon_{t-j}^{j}=\mathbb{E}_{t-j}\left(y_{t}\right)-\mathbb{E}_{t-\left(j+1\right)}\left(y_{t}\right)$
satisfy the constraints implied by the RE model (\ref{eq:BGS95_model})
is also a solution of the RE model. 

Thus, the problem of finding solutions $\left(y_{t}\right)_{t\in\mathbb{Z}}$
of the RE model (\ref{eq:BGS95_model}) is reduced to the problem
of finding processes $\left(y_{t}\right)_{t\in\mathbb{Z}},\ y_{t}\in H_{u}(t),$
that solve a vector difference equation (for given exogenous process)
involving MDS $\left(\varepsilon_{t}^{j}\right)_{t\in\mathbb{Z}}$
with respect to $H_{u}(t)$ that satisfy certain constraints.

\subsection{\label{subsec:BGS-recursive-equation}Deriving the Structural Difference
Equation from the RE Model}

In the following, we show how to obtain from the RE model (\ref{eq:BGS95_model})
(by substituting for conditional expectations the variables themselves
and the associated prediction errors) the SDE 
\begin{equation}
\underbrace{z^{J_{1}}\left(\sum_{i=J_{0}}^{J_{1}}A_{i}^{*}z^{-i}\right)}_{=\pi(z)}y_{t}=\pi(z)\left(\varepsilon_{t}^{0}+\varepsilon_{t-1}^{1}+\cdots+\varepsilon_{t-H+1}^{H-1}\right)+\zeta_{t-J_{1}}-u_{t-J_{1}}\label{eq:re-poly-1}
\end{equation}
where
\begin{gather*}
A_{i}^{*}=\sum_{k=\max\left\{ 0,-i\right\} }^{\min\left\{ K,H-i\right\} }A_{k,k+i},\\
J_{0}=argmin_{i}\left\{ i\ |\ A_{i}^{*}\neq0\right\} ,\,J_{1}=argmax_{i}\left\{ i\ |\ A_{i}^{*}\neq0\right\} ,
\end{gather*}
and
\begin{align*}
\zeta_{t} & =-\sum_{k=0}^{K}\sum_{j=0}^{H-1}\sum_{h=0}^{j}A_{kh}z^{k+(j-h)}\varepsilon_{t}^{j}\,\textsf{ and }\,\varepsilon_{t}^{j}=\mathbb{E}_{t}\left(y_{t+j}\right)-\mathbb{E}_{t-1}\left(y_{t+j}\right).
\end{align*}
The index $i$ in $A_{i}^{*}$ pertains to the endogenous variables
$y_{t+i}$. All endogenous variables with time index $t+i$, i.e.
occurring in conditional expectations $\mathbb{E}_{t-k}\left(y_{t+h-k}\right)$
for which $h-k=i$ holds, are taken together.

\paragraph{Decomposition of the forecast error of the endogenous variables in
revision processes.}

First, note that 
\begin{align*}
v_{t+h-k,h} & =y_{t+h-k}-\mathbb{E}_{t-k}\left(y_{t+h-k}\right)\\
 & =\left[y_{t+h-k}-\mathbb{E}_{t+(h-k)-1}\left(y_{t+h-k}\right)\right]+\left[\mathbb{E}_{t+(h-k)-1}\left(y_{t+h-k}\right)-\mathbb{E}_{t+(h-k)-2}\left(y_{t+h-k}\right)\right]+\cdots\\
 & \quad\cdots+\left[\mathbb{E}_{t-(h-k)-(h-1)}\left(y_{t+h-k}\right)-\mathbb{E}_{t+(h-k)-h}\left(y_{t+h-k}\right)\right]\\
 & =\varepsilon_{t+h-k}^{0}+\varepsilon_{t+h-k-1}^{1}+\cdots+\varepsilon_{t-k+1}^{h-1}
\end{align*}
and thus equation (\ref{eq:BGS95_model}) is transformed to\footnote{Note that $\mathbb{E}_{t}\left(y_{t}\right)=y_{t}$ because $y_{t}\in H_{u}(t)$.}
\begin{align}
-A_{00}y_{t} & =\sum_{k=0}^{K}\sum_{h=1}^{H}A_{kh}\mathbb{E}_{t-k}\left(y_{t+h-k}\right)+\sum_{k=1}^{K}A_{k0}y_{t-k}+u_{t}\nonumber \\
 & =\sum_{k=0}^{K}\sum_{h=1}^{H}A_{kh}\left(y_{t+h-k}-\sum_{j=0}^{h-1}\varepsilon_{t+h-k-j}^{j}\right)+\sum_{k=1}^{K}A_{k0}y_{t-k}+u_{t}\label{eq:bgs-recursive-equation-intermediate-step}
\end{align}
which is equivalent to 
\begin{align}
z^{J_{1}}\left(\sum_{i=-K}^{H}A_{i}^{*}z^{-i}\right)y_{t} & =z^{J_{1}}\left(\sum_{k=0}^{K}\sum_{h=1}^{H}A_{kh}\sum_{j=0}^{h-1}\varepsilon_{t+h-k-j}^{j}-u_{t}\right)\label{eq:bgs-recursive-equation-intermediate-step2}
\end{align}

where the parameter matrices $A_{i}^{*}$ feature the forecasting
horizon $i$ more prominently, i.e. the matrices $A_{i}^{*},\,i\in\left\{ -K,\ldots,0,\ldots,H\right\} ,$
are obtained by summing over the diagonals of the big matrix in (\ref{eq:BGS95_model})
containing the matrices $A_{kh}$ as elements. 

\paragraph{Changing the order of summation.}

Reordering the sum $z^{J_{1}}\sum_{k=0}^{K}\sum_{h=1}^{H}\sum_{j=0}^{h-1}A_{kh}z^{k+(j-h)}$
appearing on the right hand side of equation (\ref{eq:bgs-recursive-equation-intermediate-step2})
leads to 
\begin{align*}
z^{J_{1}}\sum_{k=0}^{K}\left(\sum_{h=1}^{H}A_{kh}\sum_{j=0}^{h-1}\varepsilon_{t+h-k-j}^{j}\right) & =z^{J_{1}}\sum_{k=0}^{K}\left[\sum_{j=0}^{H-1}\left(\sum_{h=j+1}^{H}A_{kh}\varepsilon_{t+h-k-j}^{j}\right)\right]\\
 & =z^{J_{1}}\sum_{k=0}^{K}\left[\sum_{j=0}^{H-1}\left(\sum_{h=0}^{H}A_{kh}\varepsilon_{t+h-k-j}^{j}-\sum_{h=0}^{j}A_{kh}\varepsilon_{t+h-k-j}^{j}\right)\right]\\
 & =z^{J_{1}}\sum_{k=0}^{K}\sum_{j=0}^{H-1}\sum_{h=0}^{H}A_{kh}\varepsilon_{t+h-k-j}^{j}-z^{J_{1}}\underbrace{\sum_{k=0}^{K}\sum_{j=0}^{H-1}\sum_{h=0}^{j}A_{kh}\varepsilon_{t+h-k-j}^{j}}_{=-\zeta_{t}}\\
 & =\underbrace{\left(z^{J_{1}}\sum_{k=0}^{K}\sum_{h=0}^{H}A_{kh}z^{k-h}\right)}_{=\pi(z)}\sum_{j=0}^{H-1}\varepsilon_{t-j}^{j}+\zeta_{t-J_{1}}
\end{align*}

\begin{rem}[No redundant equations]
\label{rem:redundant-equations} The assumption $A_{00}=-I_{s}$,
imposed in \citep{BGS95}, does not necessarily imply that $\det\left(\pi(z)\right)\not\equiv0$
holds. Consider for example the case where $A_{00}=-I_{s}=-A_{KK}$,
$K=H$ and all other matrices are zero. In this case, $\pi(z)$ is
identically zero. Hence, assuming $A_{00}$ to be non-singular does
not exclude systems with redundant equations. 
\end{rem}

\begin{example}
As an example consider the univariate model with $K=H=2$ and $a_{00}=-1$,
i.e. 
\[
\begin{array}{cccc}
y_{t}= &  & a_{01}\mathbb{E}_{t}\left(y_{t+1}\right) & +a_{02}\mathbb{E}_{t}\left(y_{t+2}\right)\\
 & +a_{10}y_{t-1} & +a_{11}\mathbb{E}_{t-1}\left(y_{t}\right) & +a_{12}\mathbb{E}_{t-1}\left(y_{t+1}\right)\\
 & +a_{20}y_{t-2} & +a_{21}\mathbb{E}_{t-2}\left(y_{t-1}\right) & +a_{22}\mathbb{E}_{t-2}\left(y_{t}\right)+u_{t}.
\end{array}
\]
Replacing the conditional expectations by the variables themselves
and the associated endogenous forecast errors leads to
\[
\begin{array}{cccc}
-u_{t}= & -y_{t} & +a_{01}\left(y_{t+1}-\varepsilon_{t+1}^{0}\right) & +a_{02}\left(y_{t+2}-\varepsilon_{t+2}^{0}-\varepsilon_{t+1}^{1}\right)\\
 & +a_{10}y_{t-1} & +a_{11}\left(y_{t}-\varepsilon_{t}^{0}\right) & +a_{12}\left(y_{t+1}-\varepsilon_{t+1}^{0}-\varepsilon_{t}^{1}\right)\\
 & +a_{20}y_{t-2} & +a_{21}\left(y_{t-1}-\varepsilon_{t-1}^{0}\right) & +a_{22}\left(y_{t}-\varepsilon_{t}^{0}-\varepsilon_{t-1}^{1}\right).
\end{array}
\]
By subtracting and adding forecast errors in a way that in each column
above terms of the same kind appear, we obtain\footnote{The terms which end up in the process $\zeta_{t}$ are in bold face
for the reader's convenience.} {\small{}
\[
\begin{array}{llll}
-u_{t}= & -\left(y_{t}-\varepsilon_{t}^{0}-\varepsilon_{t-1}^{1}\mathbf{+\varepsilon_{t}^{0}+\varepsilon_{t-1}^{1}}\right) & +a_{01}\left(y_{t+1}-\varepsilon_{t+1}^{0}-\varepsilon_{t}^{1}\mathbf{+\varepsilon_{t}^{1}}\right) & +a_{02}\left(y_{t+2}-\varepsilon_{t+2}^{0}-\varepsilon_{t+1}^{1}\right)\\
 & +a_{10}\left(y_{t-1}-\varepsilon_{t-1}^{0}-\varepsilon_{t-2}^{1}\mathbf{+\varepsilon_{t-1}^{0}+\varepsilon_{t-2}^{1}}\right) & +a_{11}\left(y_{t}-\varepsilon_{t}^{0}-\varepsilon_{t-1}^{1}\mathbf{+\varepsilon_{t-1}^{1}}\right) & +a_{12}\left(y_{t+1}-\varepsilon_{t+1}^{0}-\varepsilon_{t}^{1}\right)\\
 & +a_{20}\left(y_{t-2}-\varepsilon_{t-2}^{0}-\varepsilon_{t-3}^{1}\mathbf{+\varepsilon_{t-2}^{0}+\varepsilon_{t-3}^{1}}\right) & +a_{21}\left(y_{t-1}-\varepsilon_{t-1}^{0}-\varepsilon_{t-2}^{1}\mathbf{+\varepsilon_{t-2}^{1}}\right) & +a_{22}\left(y_{t}-\varepsilon_{t}^{0}-\varepsilon_{t-1}^{1}\right).
\end{array}
\]
} This leads to the SDE{\small{}
\[
\begin{array}{l}
\underbrace{a_{2}^{*}}_{=a_{02}}y_{t+2}+\underbrace{a_{1}^{*}}_{=a_{12}+a_{01}}y_{t+1}+\underbrace{a_{0}^{*}}_{=-1+a_{11}+a_{22}}y_{t}+\underbrace{a_{-1}^{*}}_{=a_{10}+a_{21}}y_{t-1}+\underbrace{a_{-2}^{*}}_{=a_{20}}y_{t-2}\\
\quad=\left(a_{2}^{*}z^{-2}+a_{1}^{*}z^{-1}+a_{0}^{*}+a_{-1}^{*}z+a_{-2}^{*}z^{2}\right)\left(\varepsilon_{t}^{0}+\varepsilon_{t-1}^{1}\right)-\cdots\\
\qquad\cdots\underbrace{-\left(a_{00}\varepsilon_{t}^{0}+a_{00}\varepsilon_{t-1}^{1}+a_{01}\varepsilon_{t}^{1}\right)-\left(a_{10}\varepsilon_{t-1}^{0}+a_{10}\varepsilon_{t-2}^{1}+a_{11}\varepsilon_{t-1}^{1}\right)-\left(a_{20}\varepsilon_{t-2}^{0}+a_{20}\varepsilon_{t-3}^{1}+a_{21}\varepsilon_{t-2}^{1}\right)-}_{=\zeta_{t}}\cdots\\
\qquad\cdots-u_{t}
\end{array}
\]
}which in turn is equivalent to 
\[
\left(a_{2}^{*}z^{-2}+a_{1}^{*}z^{-1}+a_{0}^{*}+a_{-1}^{*}z+a_{-2}^{*}z^{2}\right)y_{t}=\left(a_{2}^{*}z^{-2}+a_{1}^{*}z^{-1}+a_{0}^{*}+a_{-1}^{*}z+a_{-2}^{*}z^{2}\right)\left(\varepsilon_{t}^{0}+\varepsilon_{t-1}^{1}\right)+\zeta_{t}-u_{t}.
\]
\end{example}

\subsection{\label{subsec:bgs-Constraints}Constraints on the Revision Process}

In this subsection, the constraints for the revision processes $\varepsilon_{t}^{j}=\mathbb{E}_{t}\left(y_{t+j}\right)-\mathbb{E}_{t-1}\left(y_{t+j}\right)$
for a solution $\left(y_{t}\right)_{t\in\mathbb{Z}}$ of an RE model
are derived. To this end, we take conditional expectations of the
SDE (\ref{eq:re-poly-1}) with respect to different information sets,
and subsequently taking differences. 

We follow \citep{BGS95}, page 244ff. and start from equation (\ref{eq:re-poly-1}),
i.e. 
\[
\underbrace{z^{J_{1}}\left(\sum_{i=J_{0}}^{J_{1}}A_{i}^{*}z^{-i}\right)}_{=\pi(z)}y_{t}=\pi(z)\left(\varepsilon_{t}^{0}+\varepsilon_{t-1}^{1}+\cdots+\varepsilon_{t-H+1}^{H-1}\right)+\zeta_{t-J_{1}}-u_{t-J_{1}}
\]
where $J_{0}=argmin_{i}\left\{ i\,|\,A_{i}^{*}\neq0\right\} ,\,J_{i}=argmax_{i}\left\{ i\,|\,A_{i}^{*}\neq0\right\} ,$
$\zeta_{t}=\sum_{k=0}^{K}\sum_{j=0}^{H-1}\sum_{h=0}^{j}A_{kh}z^{k+(j-h)}\varepsilon_{t}^{j}$,
and $\varepsilon_{t}^{j}=\mathbb{E}_{t}\left(y_{t+j}\right)-\mathbb{E}_{t-1}\left(y_{t+j}\right)$.
We write the Smith canonical form of $\pi(z)$ as
\begin{equation}
\pi(z)=P(z)\alpha(z)\Phi(z)Q(z),\label{eq:smith_form}
\end{equation}
where $P(z)$ and $Q(z)$ are unimodular\footnote{A unimodular matrix is a matrix whose elements are polynomials but
its determinant is a non-zero constant. For further background on
polynomial and rational matrices see \citet{Gant1} Chapter VI, \citet{Kailath}
Chapter 6, \citet{GohbergLancasterRodman06,GohbergLancasterRodman09},
and \citet{HannanDeistler12} Chapter 2. } matrices of dimension $\left(s\times s\right)$, and $\alpha(z)=\begin{pmatrix}\alpha_{1}(z)\\
 & \ddots\\
 &  & \alpha_{s}(z)
\end{pmatrix}$ and $\Phi(z)=\begin{pmatrix}\phi_{1}(z)\\
 & \ddots\\
 &  & \phi_{s}(z)
\end{pmatrix}$ are diagonal polynomial matrices whose $i$-th diagonal element divides
the $(i+1)$-th diagonal element\footnote{If $\phi_{i}(z)$ divides $\phi_{i+1}(z)$, there exists a polynomial
$p(z)$ such that $\phi_{i+1}(z)=p(z)\phi_{i}(z)$.}. Moreover, the entries of $\alpha(z)$ have only zeros at zero. It
follows that the degrees $\left(g_{1},\ldots,g_{s}\right)$, so-called
partial multiplicities\footnote{For more details on partial multiplicities see \citep{GohbergLancasterRodman06}
page 657ff.}, of the diagonal elements $\left(\alpha_{1}(z),\ldots,\alpha_{s}(z)\right)$
of $\alpha(z)$ are non-negative and non-decreasing, i.e. $0\leq g_{1}\leq\cdots\leq g_{s}$.
We denote their sum, the number of zeros at zero of $\det\left(\pi(z)\right)$,
by $G=\sum_{i=1}^{s}g_{i}$. 

We will work with the equation
\begin{equation}
P(z)\alpha(z)\Phi(z)Q(z)y_{t}=P(z)\alpha(z)\Phi(z)Q(z)\left(\varepsilon_{t}^{0}+\varepsilon_{t-1}^{1}+\cdots+\varepsilon_{t-H+1}^{H-1}\right)+\zeta_{t-J_{1}}-u_{t-J_{1}}\label{eq:recursive with smith-1}
\end{equation}

\begin{thm}
\label{thm:multivariate-smith-constraints-1}Assume that $\left(y_{t}\right)_{t\in\mathbb{Z}}\in H_{u}(t)$
is a solution of the RE model (\ref{eq:BGS95_model}). Then, there
is a total number of $H$ revision processes $\left(\varepsilon_{t}^{j}\right)_{t\in\mathbb{Z}},\ j\in\left\{ 0,\ldots,H-1\right\} ,$
of dimension $s$ that satisfy the conditions
\[
\left(\mathbb{E}_{t-i}-\mathbb{E}_{t-(i+1)}\right)\left(\alpha(z)^{-1}P(z)^{-1}\left[\zeta_{t-J_{1}}-u_{t-J_{1}}\right]\right)=0,\quad i\in\left\{ 0,\ldots,H-1\right\} 
\]
or equivalently
\begin{equation}
\left(\mathbb{E}_{t-i}-\mathbb{E}_{t-(i+1)}\right)\left(\alpha(z)^{-1}P(z)^{-1}\zeta_{t-J_{1}}\right)=\left(\mathbb{E}_{t-i}-\mathbb{E}_{t-(i+1)}\right)\left(\alpha(z)^{-1}P(z)^{-1}u_{t-J_{1}}\right),\quad i\in\left\{ 0,\ldots,H-1\right\} .\label{eq:bgs-constraints}
\end{equation}
\end{thm}

\begin{rem}[Perfect foresight solution]
 For arbitrary MDS $\left(\varepsilon_{t}^{j}\right)_{t\in\mathbb{Z}}$,
the solutions $\left(y_{t}\right)_{t\in\mathbb{Z}}$ of (\ref{eq:re-poly-1})
are not necessarily solutions of the RE model (\ref{eq:BGS95_model}).
In particular, the perfect foresight solution for which $\left(\varepsilon_{t}\right)_{t\in\mathbb{Z}}$
is assumed to be identically zero, may not be a solution of the RE
model (\ref{eq:BGS95_model}), compare \citep{BGS85} page 350.
\end{rem}

\begin{rem}
Note that the unimodular matrices $P(z)$ and $Q(z)$ in the Smith-form
of $\pi(z)$ are non-unique. To be more precise, $P(z)$ may be post-multiplied
by any unimodular matrix $r(z)$ and $Q(z)$ may be pre-multiplied
by any unimodular matrix $t(z)$ for which $r(z)\left(\alpha(z)\Phi(z)\right)t(z)=\alpha(z)\Phi(z)$
holds. This non-uniqueness carries over to $\alpha(z)^{-1}P(z)^{-1}$.
Consider for example\footnote{In order to improve readability, only non-zero elements are written. }
$\pi(z)=\begin{pmatrix}z & 1\\
 & z\\
 &  & z & 1\\
 &  &  & z
\end{pmatrix}$ with Smith-forms 
\[
P_{1}(z)\alpha(z)Q_{1}(z)=\begin{pmatrix}1\\
z &  & -1\\
 & 1\\
 & z &  & -1
\end{pmatrix}\begin{pmatrix}1\\
 & 1\\
 &  & z^{2}\\
 &  &  & z^{2}
\end{pmatrix}\begin{pmatrix}z & 1\\
 &  & z & 1\\
1\\
 &  & 1
\end{pmatrix}
\]
and
\[
P_{2}(z)\alpha(z)Q_{2}(z)=\begin{pmatrix}1\\
z & -z^{2} & 1 & -1\\
 & 1\\
 & z & -1
\end{pmatrix}\begin{pmatrix}1\\
 & 1\\
 &  & z^{2}\\
 &  &  & z^{2}
\end{pmatrix}\begin{pmatrix}z & 1\\
 &  & z & 1\\
 &  & 1\\
1 &  & 1-z & -1
\end{pmatrix}.
\]
It follows that
\[
\alpha(z)^{-1}P_{1}(z)^{-1}=\begin{pmatrix}1\\
 &  & 1\\
\frac{1}{z} & -\frac{1}{z^{2}}\\
 &  & \frac{1}{z} & -\frac{1}{z^{2}}
\end{pmatrix}
\]
and
\[
\alpha(z)^{-1}P_{2}(z)^{-1}=\begin{pmatrix}1\\
 &  & 1\\
 &  & \frac{1}{z} & -\frac{1}{z^{2}}\\
\frac{1}{z} & -\frac{1}{z^{2}} & \frac{1-z}{z} & -\frac{1}{z^{2}}
\end{pmatrix}
\]
 
\end{rem}

\subsection{\label{subsec:bgs-equivalent-solution}Constrained Solutions of the
SDE}

In this subsection, we characterize the solutions of the RE model
(\ref{eq:BGS95_model}). They comprise all causal solutions of the
SDE (\ref{eq:re-poly-1}) where the uncorrelated processes $\left(\varepsilon_{t}^{j}\right)_{t\in\mathbb{Z}}$
satisfy the constraints (\ref{eq:bgs-constraints}). We follow \citep{BGS95}
page 244ff. and prove
\begin{thm}
\label{thm:recursive_then_RE}Assume that the process $\left(y_{t}\right)_{t\in\mathbb{Z}}\in H_{u}(t)$
satisfies the equation 
\[
\underbrace{\pi(z)}_{=P(z)\alpha(z)\Phi(z)Q(z)}y_{t}=\pi(z)\left(\varepsilon_{t}^{0}+\varepsilon_{t-1}^{1}+\cdots+\varepsilon_{t-H+1}^{H-1}\right)+\zeta_{t-J_{1}}-u_{t-J_{1}},
\]
where $H$ (arbitrary) $s$-dimensional MDS $\left(\varepsilon_{t}^{j}\right)_{t\in\mathbb{Z}},\ j\in\left\{ 0,\ldots,H-1\right\} ,$
with respect to the information sets $H_{u}(t)$, i.e. $\mathbb{E}_{t}\left(\varepsilon_{t}^{j}\right)=\varepsilon_{t}^{j}$
and $\mathbb{E}_{t-1}\left(\varepsilon_{t}^{j}\right)=0$, satisfy
the conditions
\[
\mathbb{E}_{t-i}\left[\alpha(z)^{-1}P(z)^{-1}\left(\zeta_{t-J_{1}}-u_{t-J_{1}}\right)\right]=\mathbb{E}_{t-(i+1)}\left[\alpha(z)^{-1}P(z)^{-1}\left(\zeta_{t-J_{1}}-u_{t-J_{1}}\right)\right],\quad i\in\left\{ 0,\ldots,H-1\right\} 
\]
or equivalently
\begin{align*}
 & \mathbb{E}_{t-i}\left(\alpha(z)^{-1}P(z)^{-1}\zeta_{t-J_{1}}\right)-\mathbb{E}_{t-(i+1)}\left(\alpha(z)^{-1}P(z)^{-1}\zeta_{t-J_{1}}\right)=\\
 & \quad=-\left[\mathbb{E}_{t-i}\left(\alpha(z)^{-1}P(z)^{-1}u_{t-J_{1}}\right)-\mathbb{E}_{t-(i+1)}\left(\alpha(z)^{-1}P(z)^{-1}u_{t-J_{1}}\right)\right],\quad i\in\left\{ 0,\ldots,H-1\right\} .
\end{align*}
It follows that the process $\left(y_{t}\right)_{t\in\mathbb{Z}}\in H_{u}(t)$
is also a solution of the RE model (\ref{eq:BGS95_model}), i.e. 
\[
\begin{pmatrix}I_{s} & \cdots & I_{s}z^{k} & \cdots & I_{s}z^{K}\end{pmatrix}\begin{pmatrix}A_{00} & \cdots & A_{0h} & \cdots & A_{0H}\\
\vdots & \ddots & \vdots &  & \vdots\\
A_{k0} & \cdots & A_{kh} & \cdots & A_{kH}\\
\vdots &  & \vdots & \ddots & \vdots\\
A_{K0} & \cdots & A_{Kh} & \cdots & A_{KH}
\end{pmatrix}\begin{pmatrix}\mathbb{E}_{t}\left(y_{t}\right)\\
\vdots\\
\mathbb{E}_{t}\left(y_{t+h}\right)\\
\vdots\\
\mathbb{E}_{t}\left(y_{t+H}\right)
\end{pmatrix}=-u_{t}.
\]
\end{thm}

\begin{rem}
Note the similar structure of the proof of Theorem \ref{thm:multivariate-smith-constraints-1}.
While we assumed in Theorem \ref{thm:multivariate-smith-constraints-1}
that the MDS are derived from the solutions of the RE model, we prove
here that for arbitrary MDS (with respect to information sets $H_{u}(t)$)
satisfying the constraints, the causal solutions of the SDE are also
causal solutions of the RE model.
\end{rem}

\pagebreak{}

\section{\label{sec:dimension} The Dimension of the Solution Set of a Stochastically
Singular RE model}

\paragraph{Structure of zeros at zero.}

In this section, we correct Property 5 on page 245 in \citep{BGS95}
describing the dimension of the solution set of the RE model (\ref{eq:BGS95_model})
by rectifying the effect of the structure of zeros at zero of $\det\left(\pi(z)\right)$
on the number of free parameters in the SDE whose causal solution
set is equivalent to the one of a RE model. Note that this changes
the number of restrictions in most cases where $\det\left(\pi(z)\right)$
has more than $J_{1}$ zeros at zero. In the case $H=J_{1}=1$, the
new result is related to the number and sizes of the Jordan blocks
pertaining to the zeros at zero of $\pi(z)$. 

\paragraph{Stochastically singular exogenous process.}

Moreover, we generalize all results in \citep{BGS95} to the case
where the stationary linearly regular exogenous process might have
a singular autocovariance at lag zero and a singular innovation covariance
matrix. Note that their count of ``auxiliary parameters'', i.e.
the number of unrestricted parameters in the SDE, on page 247 below
their formula (4.1) is only correct if the exogenous process has a
spectral density of full rank and if the matrix $\alpha(z)$ describing
the structure of the zeros at zero of $\pi(z)$ is of a particular
form (which is non-generic under the assumptions in \citep{BGS95}).

Lastly, we allow for rational restrictions on the parameter matrices
$A_{kh}$ whereas \citep{BGS95} only allows for zero restrictions.
\begin{thm}
\label{thm:general-solution-set} We consider the RE model (\ref{eq:BGS95_model}),
i.e. 
\[
\begin{pmatrix}I_{s} & \cdots & I_{s}z^{k} & \cdots & I_{s}z^{K}\end{pmatrix}\begin{pmatrix}A_{00} & \cdots & A_{0h} & \cdots & A_{0H}\\
\vdots & \ddots & \vdots &  & \vdots\\
A_{k0} & \cdots & A_{kh} & \cdots & A_{kH}\\
\vdots &  & \vdots & \ddots & \vdots\\
A_{K0} & \cdots & A_{Kh} & \cdots & A_{KH}
\end{pmatrix}\begin{pmatrix}\mathbb{E}_{t}\left(y_{t}\right)\\
\vdots\\
\mathbb{E}_{t}\left(y_{t+h}\right)\\
\vdots\\
\mathbb{E}_{t}\left(y_{t+H}\right)
\end{pmatrix}=-u_{t}
\]
and assume that 

\begin{enumerate}
\item \label{enu:ass1}$\exists h\in\left\{ 0,\ldots,H\right\} $ such that
$A_{Kh}\neq0$ and $\exists k\in\left\{ 0,\ldots,K\right\} $ such
that $A_{kH}\neq0$, that 
\item \label{enu:ass2}$\det\left(\pi(z)\right)\not\equiv0$, where 
\[
\pi(z)=z^{J_{1}}\left(\sum_{i=J_{0}}^{J_{1}}A_{i}^{*}z^{-i}\right)=A_{J_{0}}^{*}z^{J_{1}-J_{0}}+A_{J_{0}+1}^{*}z^{\left(J_{1}-J_{0}\right)-1}+\cdots+A_{0}^{*}z^{J_{1}}+\cdots+A_{J_{1}-1}^{*}z+A_{J_{1}}^{*}
\]
is defined in equation (\ref{eq:recursive}), that 
\item \label{enu:ass3}the entries of the parameter matrices $A_{kh}$ are
of the form $A_{kh}^{ij}=\frac{p_{kh}^{ij}(\theta_{1},\ldots,\theta_{p})}{q_{kh}^{ij}(\theta_{1},\ldots,\theta_{p})}$
where $p_{kh}^{ij}$ and $q_{kh}^{ij}$ are polynomials in $\left(\theta_{1},\ldots,\theta_{p}\right)$
and $q_{kh}^{ij}$ is not identically zero, that 
\item \label{enu:ass4}the matrix $C\in\mathbb{R}^{sH\times sH}$ in the
system\footnote{The matrices $C$ and $D$ are obtained from $P(z)^{-1},\ \alpha(z)^{-1},$
and the matrices $A_{kh}$ appearing in $\zeta_{t}$.} 
\begin{equation}
C\left(\begin{array}{c}
\varepsilon_{t}^{0}\\
\vdots\\
\varepsilon_{t}^{H-1}
\end{array}\right)=D\begin{pmatrix}\left(\mathbb{E}_{t}-\mathbb{E}_{t-1}\right)\left(u_{t}\right)\\
\vdots\\
\left(\mathbb{E}_{t}-\mathbb{E}_{t-1}\right)\left(u_{t+(H-J_{1}+g_{s}-1)}\right)
\end{pmatrix}\label{eq:general_thm_constraints}
\end{equation}
obtained from the constraints in equation (\ref{eq:bgs-constraints})
has rank $w\leq\left(H-J_{1}\right)s+\sum_{i=1}^{s}\min\left(g_{i},J_{1}\right)$,
where $0\leq g_{1}\leq\ldots\leq g_{s}$ denote the partial multiplicities
of $\alpha(z)$ in the Smith-form (\ref{eq:smith_form}) of $\pi(z)$,
and that
\item \label{enu:ass5}the $s$-dimensional stationary exogenous process
$\left(u_{t}\right)_{t\in\mathbb{Z}}$ has Wold decomposition
\[
u_{t}=\sum_{i=0}^{\infty}w_{i}\varepsilon_{t-i}=w(z)\varepsilon_{t},\ w_{i}\in\mathbb{R}^{s\times q},
\]
where $\sum_{i=0}^{\infty}w_{i}w_{i}^{T}<\infty$ (component wise),
$rank\left(\mathbb{E}\left(u_{t}u_{t}^{T}\right)\right)=r$, and $rank\left(\mathbb{E}\left(\varepsilon_{t}\varepsilon_{t}^{T}\right)\right)=q\leq r\leq s$.
\end{enumerate}
For a generic parameter value $\left(\theta_{1},\ldots,\theta_{p}\right)$
satisfying the restrictions above, the SDE 
\begin{equation}
\pi(z)y_{t}=\pi(z)\left(\varepsilon_{t}^{0}+\varepsilon_{t-1}^{1}+\cdots+\varepsilon_{t-H+1}^{H-1}\right)+\zeta_{t-J_{1}}-u_{t-J_{1}}\label{eq:recursive_eq_in_dimension_thm}
\end{equation}
whose set of causal stationary solutions coincides with the ones of
the RE model involves $\left(Hs-w\right)q$ free parameters.

Furthermore, additionally assuming that all non-zero zeros of $\pi(z)$
lie outside the unit circle, it follows that two distinct free parameters
generate distinct causal stationary solutions.
\end{thm}

\begin{rem}[Assumptions on the parameter space]

Firstly, we assume that there exists an $h\in\left\{ 0,\ldots,H\right\} $
such that $A_{Kh}\neq0$ and a $k\in\left\{ 0,\ldots,K\right\} $
such that $A_{kH}\neq0$ in order that $H$ and $K$ be well defined.
Secondly, we assume that
\[
\pi(z)=z^{J_{1}}\left(\sum_{i=J_{0}}^{J_{1}}A_{i}^{*}z^{-i}\right)=A_{J_{0}}^{*}z^{J_{1}-J_{0}}+A_{J_{0}+1}^{*}z^{\left(J_{1}-J_{0}\right)-1}+\cdots+A_{0}^{*}z^{J_{1}}+\cdots+A_{J_{1}-1}^{*}z+A_{J_{1}}^{*}
\]
has determinant not identically zero. Note that the non-singularity
of $A_{00}$ (as assumed in \citep{BGS95}) does not imply that $\det\left(\pi(z)\right)\not\equiv0$,
compare remark \ref{rem:redundant-equations} on page \pageref{rem:redundant-equations}.

Allowing for rational restrictions of the parameter matrices $A_{kh},$
i.e. their entries are of the form $A_{kh}^{ij}=\frac{p_{kh}^{ij}(\theta_{1},\ldots,\theta_{p})}{q_{kh}^{ij}(\theta_{1},\ldots,\theta_{p})}$
where $p_{kh}^{ij}$ and $q_{kh}^{ij}$ are multivariate polynomials
in $\left(\theta_{1},\ldots,\theta_{p}\right)$ and $q_{kh}^{ij}$
is not identically zero, comprise the case of zero restrictions treated
in \citep{BGS95}. These restrictions guarantee that the integer-valued
parameters $J_{1}$, and $\left(g_{1},\ldots,g_{s}\right)$ are well
defined on the parameter space in the sense that they are constant
on the complement of a subset (of the parameter space) of lower dimension. 

Lastly, we impose the high level condition that the matrix $C$ in
equation (\ref{eq:general_thm_constraints}) has rank $w.$ More precise
results as to when the upper bound $\left(H-J_{1}\right)s+\sum_{i=1}^{s}\min\left(g_{i},J_{1}\right)$
is binding will be given later. Note that the rank of the matrix $C$
has only to be checked at one generic point $\left(\theta_{1},\ldots,\theta_{p}\right)$
satisfying the rational parameter restrictions. The matrix $C$ then
has the same rank on an open and dense set in the parameter space
because the determinant of any submatrix of $C$ is a multivariate
rational function of the parameters. 

In \citep{BGS95} it is assumed that $A_{00}=-I_{s}$. The authors
argue on page 255 that this assumption is enough to ensure that the
rank of matrix $C$ in equation (\ref{eq:general_thm_constraints})
attains its upper bound $\left(H-J_{1}\right)s$ in the case $g_{i}=0$.
This is ``likely'' to be true since by only allowing for zero restrictions
the point for which all unrestricted matrices are zero is contained
in the parameter space. For this point, the $\left(H-J_{1}\right)s\times Hs$
dimensional submatrix of $C$ analysed in \citep{BGS95} is indeed
of full rank and thus this property holds for an open and dense set
in the parameter space. 
\end{rem}

\begin{rem}[Exogenous process is linearly regular.]

We search for causal solutions $y_{t}=\sum_{j=0}^{\infty}k_{j}\varepsilon_{t-j},\ k_{j}\in\mathbb{R}^{s\times q}$.The
revision process $\varepsilon_{t}^{j}$ of such a process $y_{t}=\sum_{j=0}^{\infty}k_{j}\varepsilon_{t-j}$
satisfies
\[
\varepsilon_{t-j}^{j}=\mathbb{E}_{t-j}\left(y_{t}\right)-\mathbb{E}_{t-(j+1)}\left(y_{t}\right)=k_{j}\varepsilon_{t-j},\quad j\geq0.
\]
\end{rem}

\paragraph{A different representation of $\zeta_{t}$. }

In order to provide more insights into the structure of the matrix
$C$ in equation (\ref{eq:general_thm_constraints}), we write $\zeta_{t}=-\sum_{k=0}^{K}\sum_{j=0}^{H-1}\sum_{h=0}^{j}A_{kh}z^{k+(j-h)}\varepsilon_{t}^{j}$
as
\[
\zeta_{t}=\sum_{i=0}^{H+K-1}m_{i,\bullet}\varepsilon_{t-i}^{\bullet},
\]
where $\varepsilon_{t}^{\bullet}=\begin{pmatrix}\varepsilon_{t}^{0}\\
\varepsilon_{t}^{1}\\
\vdots\\
\varepsilon_{t}^{H-1}
\end{pmatrix}$ is $sH$-dimensional and $m_{i,\bullet}\in\mathbb{R}^{s\times sH}$.
The matrices $m_{i,\bullet}$ are described in detail in equations
(\ref{eq:zeta_K_kleiner_H}) and (\ref{eq:zeta_K_groesser_H}) on
pages \pageref{eq:zeta_K_kleiner_H} and \pageref{eq:zeta_K_groesser_H}.
\begin{thm}
\label{thm:main_thm_A0_nonsingular} Let all assumptions of Theorem
\ref{thm:general-solution-set} hold. It follows that

\begin{enumerate}
\item for $0\leq g_{1}\leq\cdots\leq g_{j}\leq J_{1}<g_{j+1}\leq\cdots\leq g_{s}$
and if $\begin{pmatrix}m_{0,\bullet}\\
\vdots\\
m_{H-1,\bullet}
\end{pmatrix}\in\mathbb{R}^{sH\times sH}$ has full row rank, the rank of $C$ in equation (\ref{eq:general_thm_constraints})
is bounded from below by $\sum_{k=1}^{j}\left[\left(H-J_{1}\right)+g_{k}\right]+\sum_{k=j+1}^{s}\max\left(H-J_{1}+g_{k},0\right)$,
that 
\item for $g_{i}\leq J_{1}$ and if $\begin{pmatrix}m_{0,\bullet}\\
\vdots\\
m_{H-J_{1}+g_{s}-1,\bullet}
\end{pmatrix}\in\mathbb{R}^{s\left(H-J_{1}+g_{s}\right)\times sH}$ has full row rank, the matrix $C$ generically has row rank $\left(H-J_{1}\right)s+\sum_{i=1}^{s}g_{i}$
, and that 
\item for $g_{i}=0$, the matrices $C$ and $D$ in equation (\ref{eq:general_thm_constraints})
can be transformed such that 
\begin{equation}
\begin{pmatrix}m_{0,\bullet}\\
\vdots\\
m_{H-J_{1}-1,\bullet}
\end{pmatrix}\left(\begin{array}{c}
\varepsilon_{t}^{0}\\
\vdots\\
\varepsilon_{t}^{H-1}
\end{array}\right)=\begin{pmatrix}\left(\mathbb{E}_{t}-\mathbb{E}_{t-1}\right)\left(u_{t}\right)\\
\vdots\\
\left(\mathbb{E}_{t}-\mathbb{E}_{t-1}\right)\left(u_{t+(H-J_{1}-1)}\right)
\end{pmatrix}\label{eq:constraints_no_zeros}
\end{equation}
holds. If $\begin{pmatrix}m_{0,\bullet}\\
\vdots\\
m_{H-J_{1}-1,\bullet}
\end{pmatrix}\in\mathbb{R}^{s\left(H-J_{1}\right)\times sH}$ is of full row rank, the SDE (\ref{eq:recursive_eq_in_dimension_thm})
involves generically $J_{1}sq$ free parameters. 
\end{enumerate}
\end{thm}

\begin{rem}
Assuming non-singularity of $A_{0}^{*}$ would entail that $\det\left(\pi(z)\right)$
has at most $J_{1}s$ zeros at zero. However, it is not sufficient
that $\begin{pmatrix}m_{0,\bullet}\\
\vdots\\
m_{H-1,\bullet}
\end{pmatrix}\in\mathbb{R}^{sH\times sH}$ be of full (row) rank, but only means that there is a non-singular
matrix on the (block) diagonal of $\begin{pmatrix}m_{0,\bullet}\\
\vdots\\
m_{H-1,\bullet}
\end{pmatrix}$. Thus, we have to explicitly assume that this matrix has full (row)
rank. In the case of zero restrictions and assuming that $A_{00}=-I_{s}$
(and of course that there are non-zero matrices such that $K$ and
$H$ are well defined and $\det\left(\pi(z)\right)\not\equiv0)$)
it is easy to show that there is a point in the parameter space such
that $\begin{pmatrix}m_{0,\bullet}\\
\vdots\\
m_{H-1,\bullet}
\end{pmatrix}$ is of full rank by considering equations (\ref{eq:zeta_K_kleiner_H})
and (\ref{eq:zeta_K_groesser_H}). From this it follows that $\begin{pmatrix}m_{0,\bullet}\\
\vdots\\
m_{H-1,\bullet}
\end{pmatrix}$ is of full rank for almost every point in the parameter space because
the determinant is a multivariate rational function of its parameters.
\end{rem}

\begin{example}
We will now give an example (similar to the model in \citep{KingWatson98})
in order to illustrate the causes for the different number of free
parameters here and in \citep{BGS95}. We consider the case $H=J_{1}=1,\ K=0$
such that the RE models takes the form
\begin{align}
A_{00}\mathbb{E}_{t}\left(y_{t}\right)+A_{01}\underbrace{\mathbb{E}_{t}\left(y_{t+1}\right)}_{=y_{t+1}-\varepsilon_{t+1}^{0}} & =-u_{t}\nonumber \\
\iff\underbrace{z\left(A_{00}+A_{01}z^{-1}\right)}_{=\pi(z)}y_{t} & =A_{01}\varepsilon_{t}^{0}-u_{t-1}\nonumber \\
\iff\pi(z)y_{t} & =z\left(A_{00}+A_{01}z^{-1}\right)\varepsilon_{t}^{0}-A_{00}\varepsilon_{t-1}^{0}-u_{t-1}.\label{eq:example_jordan_recursive}
\end{align}
In order that the set of causal solutions of the SDE (\ref{eq:example_jordan_recursive})
coincide with the one of the RE model, the constraint
\begin{equation}
\left[\mathbb{E}_{t}-\mathbb{E}_{t-1}\right]\left(\alpha(z)^{-1}P(z)^{-1}\left(A_{00}\varepsilon_{t-1}^{0}+u_{t-1}\right)\right)=0\label{eq:example_jordan_constraints}
\end{equation}
has to be satisfied. 

If $\det\left(\pi(z)\right)$ is not identically zero, we may represent
this regular matrix pencil in its canonical form according to \citep{Gant2}
(Chapter XII, Section 2) as 
\[
V\left(A_{01}+A_{00}z\right)W^{-1}=\begin{pmatrix}\begin{pmatrix}I_{n(s)}\\
 & I_{n(u)}
\end{pmatrix}\\
 & N
\end{pmatrix}-\begin{pmatrix}\begin{pmatrix}\mathfrak{J}_{s}\\
 & \mathfrak{J}_{u}
\end{pmatrix}\\
 & I
\end{pmatrix}z
\]
where $V$ and $W$ are non-singular matrices, $n_{s}$ and $n_{u}$
are the number of roots of $\det\left(\pi(z)\right)$ outside and
inside the unit circle pertaining to the Jordan blocks in $\mathfrak{J_{s}}$
and $\mathfrak{J}_{u}$ respectively, and $N$ is a quadratic matrix
of dimension $n$ with ones or zeros on the first superdiagonal and
zeros otherwise. For simplicity, we assume that there are no zeros
on the unit circle.

In order to understand the result that there are $\left(s-r\right),\ r\leq G_{1}=\sum_{i=1}^{s}\min\left(g_{i},1\right),$
independent MDS in the SDE (or equivalently that there are $G_{1}$
linear dependencies between the MDS), we need to focus on the Jordan
structure of the matrix $N$ and its relation to the diagonal matrix
$\alpha(z)$ from the Smith-form of $\pi(z)$. To this end, we assume
that there are no non-zero zeros inside or outside the unit circle.
The partial multiplicities $\left(g_{1},\ldots,g_{s}\right)$ of $\alpha(z)$
correspond to the Jordan blocks in the following way: The number of
Jordan blocks of size $k$ corresponds to the number of $g_{i}$s
which are equal to $k$, compare \citep{GohbergLancasterRodman06}
(Appendix A.3, page 656, Proposition A.3.3). It follows, e.g., that
the size of the largest Jordan block corresponds to the highest degree
$g_{s}$. If there are no ones at all on the first superdiagonal,
it follows that $g_{i}=1$ for all $i\in\left\{ 1,\ldots,s\right\} $.
If there are no restrictions on the parameter matrices $A_{00}$ and
$A_{01}$ (other than $\det\left(\pi(z)\right)\not\equiv0$), the
case that $g_{s}$ is equal to the number of zeros at zero of $\det\left(\pi(z)\right)$
and that all other $g_{i}$s are equal to zero is generic. Thus, in
this generic case, there is one linear dependency among the $s$ MDS
according to equation (\ref{eq:example_jordan_constraints}). However,
according to Property 5 on page 245 in \citep{BGS95}, there are no
independent MDS left in the SDE.
\end{example}

\pagebreak{}

\section{\label{sec:predetermined}The Case of Predetermined Variables: Dimension
of the Solution Set}

In the seminal work \citep{BlanchardKahn80}, it is assumed that some
components of the endogenous variables $y_{t}$ are predetermined
in the sense that $\mathbb{E}_{t}\left(x_{t+1}\right)=x_{t+1}$ holds
for the predetermined components $x_{t}$. In \citep{BGS85,BGS95},
there are no predetermined variables, which makes comparison of some
results difficult\footnote{Moreover, the result on the number of ``unstable roots'' of $\pi(z)$
which is necessary and sufficient for a unique causal and non-explosive
solution in \citep{BGS95} is incorrect (unrelated to the error about
restrictions implied by the structure of zeros at zero of $\pi(z)$)
and at odds with the analogous result in \citep{BlanchardKahn80}
as will be discussed in the next section.}.

\paragraph{Predetermined variables. }

We partition the $s$-dimensional endogenous process $\left(y_{t}\right)_{t\in\mathbb{Z}}$
in $\left(H+1\right)$ subprocesses $\left(y_{t}^{s_{i}}\right)_{t\in\mathbb{Z}}$
of respective dimensions $s_{0},\ldots,s_{H}$ and denote the partition
with the multi-index $\gamma=\left(s_{0},s_{1},\ldots,s_{H}\right)$.
Obviously, $\left|\gamma\right|=s_{0}+s_{1}+\cdots+s_{H}=s$ holds.
The $s_{i}$-dimensional subprocess $\left(y_{t}^{s_{i}}\right)_{t\in\mathbb{Z}}$
is predetermined by variables $i$-periods ago, i.e. variables in
$H_{\varepsilon}(t-i)$, such that $y_{t+H}^{s_{i}}=\mathbb{E}_{t+H-i}\left(y_{t+H}^{s_{i}}\right)$
or equivalently $\varepsilon_{t}^{j,s_{i}}=0,\ i>j,$ or in more detail
\begin{align*}
y_{t+H}^{s_{0}} & =\mathbb{E}_{t}\left(y_{t+H}^{s_{0}}\right)+\varepsilon_{t+1}^{H-1,s_{0}}+\varepsilon_{t+2}^{H-2,s_{0}}+\cdots+\varepsilon_{t+H-1}^{1,s_{0}}+\varepsilon_{t+H}^{0,s_{0}}\\
y_{t+H}^{s_{1}} & =\mathbb{E}_{t}\left(y_{t+H}^{s_{1}}\right)+\varepsilon_{t+1}^{H-1,s_{1}}+\varepsilon_{t+2}^{H-2,s_{1}}+\cdots+\varepsilon_{t+H-1}^{1,s_{1}}+\underbrace{\varepsilon_{t+H}^{0,s_{1}}}_{=0}\\
 & \vdots\\
y_{t+H}^{s_{i}} & =\mathbb{E}_{t}\left(y_{t+H}^{s_{i}}\right)+\varepsilon_{t+1}^{H-1,s_{i}}+\varepsilon_{t+2}^{H-2,s_{i}}+\cdots+\varepsilon_{t+H-i}^{i,s_{i}}+\underbrace{\varepsilon_{t+H-(i-1)}^{i-1,s_{i}}+\cdots+\varepsilon_{t+H}^{0,s_{i}}}_{=0}\\
 & \vdots\\
y_{t+H}^{s_{H-1}} & =\mathbb{E}_{t}\left(y_{t+H}^{s_{H-1}}\right)+\varepsilon_{t+1}^{H-1,s_{H-1}}+\underbrace{\varepsilon_{t+2}^{H-2,s_{H-1}}+\cdots+\varepsilon_{t+H}^{0,s_{H-1}}}_{=0}\\
y_{t+H}^{s_{H}} & =\mathbb{E}_{t}\left(y_{t+H}^{s_{H}}\right)+\underbrace{\varepsilon_{t+1}^{H-1,s_{H}}+\cdots+\varepsilon_{t+H}^{0,s_{H}}}_{=0}.
\end{align*}

In \citep{BGS95}, all components of the endogenous variables are
assumed to be non-predetermined, i.e. $s_{0}=s$ and $s_{1}=\cdots=s_{H}=0$
. 

If $y_{t}$ satisfies the restrictions on its revision processes described
above, we write $y_{t}\in H_{\varepsilon}^{Pre}(t;\gamma)$. Restricting
the solution set by requiring some components to be predetermined
is similar to restricting solutions to be causal. For predetermined
components we require that they not depend on $\varepsilon_{t-l+i},\ i>0,$
for an $l\in\left\{ 1,\ldots,H\right\} $; for causal solutions we
require that $y_{t}$ not depend on $\varepsilon_{t+i},\ i>0$. 
\begin{example}
The model in \citep{BlanchardKahn80} has the form 
\[
\begin{pmatrix}\mathbb{E}_{t}\left(y_{t+1}^{s_{0}}\right)\\
\mathbb{E}_{t}\left(y_{t+1}^{s_{1}}\right)
\end{pmatrix}=B\begin{pmatrix}y_{t}^{s_{0}}\\
y_{t}^{s_{1}}
\end{pmatrix}+Cu_{t},\quad t\in\mathbb{Z}.
\]
Thus $s_{0}$ is the number of non-predetermined variables, $s_{1}$
is the number of variables which are predetermined by variables from
one period ago, and $s_{2}=\cdots=s_{H}=0$. The model corresponds
to $K=0,\ H=J_{1}=1,\ A_{00}=-B,\ A_{01}=I_{s}$ in the notation of
\citep{BGS95}.
\end{example}

\paragraph{Equivalence of the solution sets of the RE model and the SDE with
restrictions.}

We will now proceed to prove that the set of all solutions of the
RE model satisfying the predeterminedness conditions coincides with
the set of solutions satisfying the predeterminedness conditions of
the SDE. The SDE involves MDS $\varepsilon_{t-j}^{j}$ which firstly
satisfy the predeterminedness constraints, i.e. $\varepsilon_{t}^{j,s_{i}}=0,\ i>j$,
and secondly are constrained by affine restrictions.
\begin{thm}
\label{thm:predetermined_equivalence_of_solutions}Assume that $y_{t}\in H_{\varepsilon}^{Pre}(t;\gamma)$.
If $\left(y_{t}\right)_{t\in\mathbb{Z}}$ is a solution of (\ref{eq:BGS95_model}),
then it satisfies the SDE
\[
\pi(z)y_{t}=\pi(z)\left(\varepsilon_{t}^{0}+\varepsilon_{t-1}^{1}+\cdots+\varepsilon_{t-H+1}^{H-1}\right)+\zeta_{t-J_{1}}-u_{t-J_{1}}
\]
where $\zeta_{t}=\sum_{k=0}^{K}\sum_{j=0}^{H-1}\sum_{h=0}^{j}A_{kh}z^{k+(j-h)}\varepsilon_{t}^{j}$
and 
\[
\left(\varepsilon_{t}^{0},\ldots,\varepsilon_{t}^{H-1}\right)=\left(\begin{array}{cccccc}
\varepsilon_{t}^{0,s_{0}} & \varepsilon_{t}^{1,s_{0}} & \varepsilon_{t}^{2,s_{0}} & \cdots & \varepsilon_{t}^{H-2,s_{0}} & \varepsilon_{t}^{H-1,s_{0}}\\
0_{s_{1}\times1} & \varepsilon_{t}^{1,s_{1}} & \varepsilon_{t}^{2,s_{1}} &  & \vdots & \vdots\\
\vdots & 0_{s_{2}\times1} & \varepsilon_{t}^{2,s_{2}}\\
 & \vdots & 0_{s_{3}\times1} & \ddots & \vdots & \vdots\\
 &  & \vdots &  & \varepsilon_{t}^{H-2,s_{H-2}} & \varepsilon_{t}^{H-1,s_{H-2}}\\
\vdots & \vdots & \vdots &  & 0_{s_{H-1}\times1} & \varepsilon_{t}^{H-1,s_{H-1}}\\
0_{s_{H}\times1} & 0_{s_{H}\times1} & 0_{s_{H}\times1} & \cdots & 0_{s_{H}\times1} & 0_{s_{H}\times1}
\end{array}\right).
\]
Moreover, the revision processes $\varepsilon_{t}^{j,s_{i}}=\left[\mathbb{E}_{t}-\mathbb{E}_{t-1}\right]\left(y_{t+j}^{s_{i}}\right)$
satisfy the constraints
\begin{equation}
S\left(\mathbb{E}_{t-i}-\mathbb{E}_{t-(i+1)}\right)\left[\alpha(z)^{-1}P(z)^{-1}\left(\zeta_{t-J_{1}}-u_{t-J_{1}}\right)\right]=0,\quad i\in\left\{ 0,\ldots,H-1\right\} \label{eq:predetermined_constraints_zeta}
\end{equation}
where the matrix $S$ of dimension $\underbrace{\left(\sum_{i=0}^{H-1}s_{i}\cdot\left(H-i\right)\right)}_{=P}\times Hs$
has the form
\begin{equation}
S=\left(\begin{array}{cccccc}
\omega_{0,s_{0}}^{\dagger}\\
 & \omega_{0,s_{0}+s_{1}}^{\dagger}\\
 &  & \omega_{0,s_{0}+s_{1}+s_{2}}^{\dagger}\\
 &  &  & \ddots\\
 &  &  &  & \omega_{0,s_{0}+\cdots+s_{H-2}}^{\dagger}\\
 &  &  &  &  & \omega_{0,s_{0}+\cdots+s_{H-1}}^{\dagger}
\end{array}\right),\label{eq:pseudo_inverse_selector}
\end{equation}
and where $\omega_{0,s_{0}+\cdots+s_{i}}^{\dagger}$ denotes the Moore-Penrose
pseudo-inverse of the first $\left(s_{0}+\cdots+s_{i}\right)$ columns
of the first coefficient $\omega_{0}\in\mathbb{R}^{s\times s}$ of
$\Phi(z)Q(z)$ from the Smith-form of $\pi(z)$. 

Likewise, if $\left(y_{t}\right)_{t\in\mathbb{Z}}$ is a solution
of the SDE above, and if the processes $\left(\varepsilon_{t}^{j}\right)_{t\in\mathbb{Z}}$
of the special form above are MDS with respect to $H_{\varepsilon}(t)$
and satisfy the constraints
\[
S\left(\mathbb{E}_{t-i}-\mathbb{E}_{t-(i+1)}\right)\left[\alpha(z)^{-1}P(z)^{-1}\left(\zeta_{t-J_{1}}-u_{t-J_{1}}\right)\right]=0\quad i\in\left\{ 0,\ldots,H-1\right\} ,
\]
then $\left(y_{t}\right)_{t\in\mathbb{Z}}$ is also a solution of
the RE model.
\end{thm}

\paragraph{The dimension of the solution set in the case of predetermined variables.}

Similarly to the case without predetermined variables, we determine
the number of linearly independent MDS and the corresponding number
of free parameters in the SDE by analysing the affine restrictions
among the MDS $\varepsilon_{t}^{s_{i},j}$, i.e. the restrictions
described in equation (\ref{eq:predetermined_constraints_zeta}).
In this way, we characterize the dimension of the set of solutions
$\left(y_{t}\right)_{t\in\mathbb{Z}}$ of the RE model for which $y_{t}\in H_{\varepsilon}^{Pre}(t;\gamma)$. 

The main differences to the non-predetermined case are that the system
of constraints is pre-multiplied by $S$ and that there are fewer
non-zero MDS available.
\begin{thm}
\label{thm:predetermined_dimension} Assume that $y_{t}\in H_{\varepsilon}^{Pre}(t,\gamma)$,
and that assumptions \ref{enu:ass1}, \ref{enu:ass2}, \ref{enu:ass3},
and \ref{enu:ass5} of \ref{thm:general-solution-set} hold. 

\begin{enumerate}
\item Let $0\leq g_{1}\leq\cdots\leq g_{j}\leq J_{1}<g_{j+1}\leq\cdots\leq g_{s}$
and define $\delta_{k}=J_{1}-g_{k}\geq0,\ k\in\left\{ 1,\ldots,j\right\} ,$
and $\gamma_{k}=g_{k}-J_{1}>0,\ k\in\left\{ j+1,\ldots,s\right\} $.
The constraints (\ref{eq:predetermined_constraints_zeta}) take the
form
\begin{equation}
SU^{T}\begin{pmatrix}\mathfrak{P}^{1,\bullet}\\
\vdots\\
\mathfrak{P}^{s,\bullet}
\end{pmatrix}\begin{pmatrix}m_{0,\bullet}\\
\vdots\\
m_{H+\gamma_{s}-1,\bullet}
\end{pmatrix}R^{T}\varepsilon_{t}^{p,\bullet}=SU^{T}\begin{pmatrix}\mathfrak{P}^{1,\bullet}\\
\vdots\\
\mathfrak{P}^{s,\bullet}
\end{pmatrix}\left[\mathbb{E}_{t}-\mathbb{E}_{t-1}\right]\begin{pmatrix}u_{t}\\
\vdots\\
u_{t+(H+\gamma_{s}-1)}
\end{pmatrix}\label{eq:predetermined_constraints_general}
\end{equation}
where $S$ is given in equation (\ref{eq:pseudo_inverse_selector}),
the row selection matrix 
\[
U=\left(\begin{array}{cccc|c|cccc}
1 & 0 & \cdots & 0 & \\
 &  &  &  & \ddots\\
 &  &  &  &  & 1 & 0 & \cdots & 0\\
\hline  &  &  &  & \vdots\\
\hline 0 & \cdots & 0 & 1 & \\
 &  &  &  & \ddots\\
 &  &  &  &  & 0 & \cdots & 0 & 1
\end{array}\right)
\]
selects the $k$-th row of $H$ $s$-dimensional blocks, for $k\in\left\{ 1,\ldots,j\right\} $
\[
\mathfrak{P}^{k,\bullet}=\left(\begin{array}{c|c|c}
0_{\delta_{k}\times s\left(H-\delta_{k}\right)} & 0_{\delta_{k}\times s\delta_{k}} & 0_{\delta_{k}\times s\gamma_{k}}\\
\hline \begin{array}{ccc}
\mathfrak{P}_{k,\bullet|0}\\
\vdots & \ddots\\
\mathfrak{P}_{k,\bullet|H-1-\delta_{k}} & \cdots & \mathfrak{P}_{k,\bullet|0}
\end{array} & 0_{\left(H-\delta_{k}\right)\times s\delta_{k}} & 0_{\left(H-\delta_{k}\right)\times s\gamma_{k}}
\end{array}\right)\in\mathbb{R}^{H\times s\left(H+\gamma_{K}\right)},
\]
for $k\in\left\{ j+1,\ldots,s\right\} $ 
\[
\mathfrak{P}^{k,\bullet}=\left(\begin{array}{ccc|ccc|c}
\mathfrak{P}_{k,\bullet|\gamma_{k}} & \cdots & \mathfrak{P}_{k,\bullet|1} & \mathfrak{P}_{k,\bullet|0} &  & \\
\vdots &  & \vdots & \vdots & \ddots &  & 0_{H\times\left(\gamma_{s}-\gamma_{k}\right)}\\
\mathfrak{P}_{k,\bullet|\gamma_{k}+H-1} & \cdots & \mathfrak{P}_{k,\bullet|H} & \mathfrak{P}_{k,\bullet|H-1} & \cdots & \mathfrak{P}_{k,\bullet|0}
\end{array}\right)\in\mathbb{R}^{H\times s\left(H+\gamma_{K}\right)},
\]
where $\mathfrak{P}_{k,\bullet|m}$ denotes the $k$-the row of the
coefficient pertaining to power $m$ of $z$ in the polynomial matrix
$P^{-1}(z)$ from the Smith-form of $\pi(z)$, and{\tiny{}
\[
R=\left(\begin{array}{cccc}
\left(\begin{array}{cc}
I_{s_{0}} & 0_{s_{0}\times s_{1}+\cdots+s_{H}}\end{array}\right)\\
 & \left(\begin{array}{cc}
I_{s_{0}+s_{1}} & 0_{s_{0}+s_{1}\times s_{2}+\cdots+s_{H}}\end{array}\right)\\
 &  & \ddots\\
 &  &  & \left(\begin{array}{cc}
I_{s_{0}+\cdots+s_{H-1}} & 0_{s_{0}+\cdots+s_{H-1}\times s_{H}}\end{array}\right)
\end{array}\right).
\]
}of dimension $\left(\left(\sum_{i=0}^{H-1}s_{i}\cdot\left(H-i\right)\right)\times Hs\right)$
selects the non-trivial components of $\begin{pmatrix}\varepsilon_{t}^{0}\\
\varepsilon_{t}^{1}\\
\vdots\\
\varepsilon_{t}^{H-1}
\end{pmatrix}$ such that $R$$\begin{pmatrix}\varepsilon_{t}^{0}\\
\varepsilon_{t}^{1}\\
\vdots\\
\varepsilon_{t}^{H-1}
\end{pmatrix}=\varepsilon_{t}^{p,\bullet}$. The number of free parameters corresponds to the dimension of the
right kernel of $\left[SU^{T}\begin{pmatrix}\mathfrak{P}^{1,\bullet}\\
\vdots\\
\mathfrak{P}^{s,\bullet}
\end{pmatrix}\begin{pmatrix}m_{0,\bullet}\\
\vdots\\
m_{H-J_{1}+g_{s}-1,\bullet}
\end{pmatrix}R^{T}\right]$ times $q$.
\item In the case $g_{i}\leq J_{1}$, equation (\ref{eq:predetermined_constraints_general})
simplifies to
\begin{equation}
SU^{T}\begin{pmatrix}\mathfrak{P}^{1,\bullet}\\
\vdots\\
\mathfrak{P}^{s,\bullet}
\end{pmatrix}\begin{pmatrix}m_{0,\bullet}\\
\vdots\\
m_{H-1,\bullet}
\end{pmatrix}R^{T}\varepsilon_{t}^{p,\bullet}=SU^{T}\begin{pmatrix}\mathfrak{P}^{1,\bullet}\\
\vdots\\
\mathfrak{P}^{s,\bullet}
\end{pmatrix}\left(\mathbb{E}_{t}-\mathbb{E}_{t-1}\right)\begin{pmatrix}u_{t}\\
\vdots\\
u_{t+(H-1)}
\end{pmatrix}\label{eq:predetermined_constraints_g_lessthan_J1}
\end{equation}
where 
\[
\mathfrak{P}^{k,\bullet}=\left(\begin{array}{c|c}
0_{\delta_{k}\times s\left(H-\delta_{k}\right)} & 0_{\delta_{k}\times s\delta_{k}}\\
\hline \begin{array}{ccc}
\mathfrak{P}_{k,\bullet|0}\\
\vdots & \ddots\\
\mathfrak{P}_{k,\bullet|H-1-\delta_{k}} & \cdots & \mathfrak{P}_{k,\bullet|0}
\end{array} & 0_{\left(H-\delta_{k}\right)\times s\delta_{k}}
\end{array}\right)\in\mathbb{R}^{H\times sH}.
\]
The number of free parameters corresponds to the dimension of the
right kernel of $\left(SU^{T}\begin{pmatrix}\mathfrak{P}^{1,\bullet}\\
\vdots\\
\mathfrak{P}^{s,\bullet}
\end{pmatrix}\begin{pmatrix}m_{0,\bullet}\\
\vdots\\
m_{H-1,\bullet}
\end{pmatrix}R^{T}\right)$ times $q$.
\item In the case $g_{i}=\bar{g}\leq J_{1}$, equation (\ref{eq:predetermined_constraints_g_lessthan_J1})
further simplifies to
\begin{equation}
\begin{array}{l}
S_{2,\bar{g}}\left(\begin{array}{ccc}
\mathfrak{P}_{\bullet,\bullet|0}\\
\vdots & \ddots\\
\mathfrak{P}_{\bullet,\bullet|H-J_{1}+\bar{g}-1} & \cdots & \mathfrak{P}_{\bullet,\bullet|0}
\end{array}\right)\begin{pmatrix}m_{0,\bullet}\\
\vdots\\
m_{H-J_{1}+\bar{g}-1,\bullet}
\end{pmatrix}R^{T}\varepsilon_{t}^{p,\bullet}=\cdots\\
\qquad=S_{2,\bar{g}}\left(\begin{array}{ccc}
\mathfrak{P}_{\bullet,\bullet|0}\\
\vdots & \ddots\\
\mathfrak{P}_{\bullet,\bullet|H-J_{1}+\bar{g}-1} & \cdots & \mathfrak{P}_{\bullet,\bullet|0}
\end{array}\right)\left[\mathbb{E}_{t}-\mathbb{E}_{t-1}\right]\begin{pmatrix}u_{t}\\
\vdots\\
u_{t+(H-J_{1}+\bar{g}-1)}
\end{pmatrix}
\end{array}\label{eq:predetermined_constraints_gconst}
\end{equation}
where $S_{2,\bar{g}}$ of dimension $\left[\left(H-J_{1}+\bar{g}\right)\sum_{i=0}^{J_{1}-\bar{g}-1}s_{i}+\sum_{i=J_{1}-\bar{g}}^{H-1}s_{i}\left(H-i\right)\right]\times s\left(H-J_{1}+\bar{g}\right)$
is the bottom right submatrix of 
\[
S=\left(\begin{array}{ccc|ccc}
\omega_{0,s_{0}}^{\dagger} &  & \\
 & \ddots & \\
 &  & \omega_{0,s_{0}+\cdots+s_{J_{1}-\bar{g}-1}}^{\dagger}\\
\hline  &  &  & \omega_{0,s_{0}+\cdots+s_{J_{1}-\bar{g}}}^{\dagger}\\
 &  &  &  & \ddots\\
 &  &  &  &  & \omega_{0,s_{0}+\cdots+s_{H-1}}^{\dagger}
\end{array}\right).
\]
If $\left[\begin{pmatrix}m_{0,\bullet}\\
\vdots\\
m_{H-J_{1}+\bar{g}-1,\bullet}
\end{pmatrix}R^{T}\right]$ of dimension $\left[s\left(H-J_{1}+\bar{g}\right)\times\left(\sum_{i=0}^{H-1}s_{i}\cdot\left(H-i\right)\right)\right]$
is of full row rank, there are $\left[\sum_{i=0}^{J_{1}-1}s_{i}\cdot\left(J_{1}-\bar{g}-i\right)\right]q$
free parameters in the SDE. 
\item In the case $g_{i}=0$, equation (\ref{eq:predetermined_constraints_g_lessthan_J1})
further simplifies to
\begin{equation}
\begin{array}{l}
S_{2}\left(\begin{array}{ccc}
\mathfrak{P}_{\bullet,\bullet|0}\\
\vdots & \ddots\\
\mathfrak{P}_{\bullet,\bullet|H-J_{1}-1} & \cdots & \mathfrak{P}_{\bullet,\bullet|0}
\end{array}\right)\begin{pmatrix}m_{0,\bullet}\\
\vdots\\
m_{H-J_{1}-1,\bullet}
\end{pmatrix}R^{T}\varepsilon_{t}^{p,\bullet}=\cdots\\
\qquad=S_{2}\left(\begin{array}{ccc}
\mathfrak{P}_{\bullet,\bullet|0}\\
\vdots & \ddots\\
\mathfrak{P}_{\bullet,\bullet|H-J_{1}-1} & \cdots & \mathfrak{P}_{\bullet,\bullet|0}
\end{array}\right)\left[\mathbb{E}_{t}-\mathbb{E}_{t-1}\right]\begin{pmatrix}u_{t}\\
\vdots\\
u_{t+(H-J_{1}-1)}
\end{pmatrix}
\end{array}\label{eq:predetermined_constraints_gzero}
\end{equation}
where $S_{2}$ of dimension $\left[\left(H-J_{1}\right)\sum_{i=0}^{J_{1}-1}s_{i}+\sum_{i=J_{1}}^{H-1}s_{i}\left(H-i\right)\right]\times s\left(H-J_{1}\right)$
is the bottom right submatrix of 
\[
S=\left(\begin{array}{ccc|ccc}
\omega_{0,s_{0}}^{\dagger} &  & \\
 & \ddots & \\
 &  & \omega_{0,s_{0}+\cdots+s_{J_{1}-1}}^{\dagger}\\
\hline  &  &  & \omega_{0,s_{0}+\cdots+s_{J_{1}}}^{\dagger}\\
 &  &  &  & \ddots\\
 &  &  &  &  & \omega_{0,s_{0}+\cdots+s_{H-1}}^{\dagger}
\end{array}\right).
\]
If $\left[\begin{pmatrix}m_{0,\bullet}\\
\vdots\\
m_{H-J_{1}-1,\bullet}
\end{pmatrix}R^{T}\right]$ of dimension $\left[s\left(H-J_{1}\right)\times\left(\sum_{i=0}^{H-1}s_{i}\cdot\left(H-i\right)\right)\right]$
is of full row rank, there are $\left[\sum_{i=0}^{J_{1}-1}s_{i}\cdot\left(J_{1}-i\right)\right]q$
free parameters in the SDE. 
\item Additionally assuming that all non-zero zeros of $\pi(z)$ lie outside
the unit circle, it follows that two distinct free parameters entail
distinct causal stationary solutions.
\end{enumerate}
\end{thm}

\begin{example}
\label{exa:king_watson_part1}The approach can be illustrated by analysing
the model
\begin{equation}
A\mathbb{E}_{t}\left(\begin{array}{c}
y_{t+1}^{s_{0}}\\
y_{t+1}^{s_{1}}
\end{array}\right)=B\left(\begin{array}{c}
y_{t}^{s_{0}}\\
y_{t}^{s_{1}}
\end{array}\right)+u_{t}\label{eq:example_king_watson}
\end{equation}
 put forward in \citep{KingWatson98}, where $\det\left(A-Bz\right)\not\equiv0$.
After transformation to Kronecker normal form (\citep{Gant2} Chapter
XII), i.e.
\begin{align}
VAW^{-1}W\left(\begin{array}{c}
y_{t}^{s_{0}}-\varepsilon_{t}^{0,s_{0}}\\
y_{t}^{s_{1}}
\end{array}\right) & =VBW^{-1}W\left(\begin{array}{c}
y_{t-1}^{s_{0}}\\
y_{t-1}^{s_{1}}
\end{array}\right)+Vu_{t-1}\nonumber \\
\iff\left[\begin{pmatrix}I_{n(s)}\\
 & I_{n(u)}\\
 &  & N
\end{pmatrix}-\begin{pmatrix}\mathfrak{J}_{s}\\
 & \mathfrak{J}_{u}\\
 &  & I
\end{pmatrix}z\right]W\left(\begin{array}{c}
y_{t}^{s_{0}}\\
y_{t}^{s_{1}}
\end{array}\right) & =\begin{pmatrix}I_{n(s)}\\
 & I_{n(u)}\\
 &  & N
\end{pmatrix}W\left(\begin{array}{c}
\varepsilon_{t}^{0,s_{0}}\\
0
\end{array}\right)+Vu_{t-1},\label{eq:king_watson_number_mds}
\end{align}
where $V$ and $W$ are non-singular matrices, $\mathfrak{J}_{s}$
and $\mathfrak{J}_{u}$ contain Jordan blocks whose diagonal entries
have absolute value smaller and larger than unity\footnote{We assume for simplicity hat there are no Jordan blocks whose diagonal
entries have absolute value equal to one.} respectively, $N$ is a nilpotent matrix which might have ones on
the first super-diagonal and zeros otherwise, we may state the following
facts.

First, the number of non-predetermined variables effectively occurring
in the model is equal to the rank of
\[
\begin{pmatrix}I_{n(s)}\\
 & I_{n(u)}\\
 &  & N
\end{pmatrix}W\left(\begin{array}{c}
I_{s_{0}}\\
0
\end{array}\right)
\]
because that is the number of linearly independent MDS on the right
hand side of equation (\ref{eq:king_watson_number_mds}). Indeed,
in the case where all endogenous variables are non-predetermined,
the number of linearly independent MDS is equal to the dimension of
$A$ minus the geometric multiplicity, i.e. the number of Jordan blocks,
of the eigenvalue zero of $A$. Put differently, the number of Jordan
blocks of the eigenvalue zero of $A$ corresponds to the number of
restrictions among the arbitrary MDS. 

This can be seen as well by considering the Smith-form of $\left(A-Bz\right).$
If we transform equation (\ref{eq:example_king_watson}) to
\[
\begin{array}{l}
A\left(\begin{array}{c}
y_{t}^{s_{0}}-\varepsilon_{t}^{0,s_{0}}\\
y_{t}^{s_{1}}
\end{array}\right)-B\left(\begin{array}{c}
y_{t-1}^{s_{0}}\\
y_{t-1}^{s_{1}}
\end{array}\right)=u_{t-1}\\
\iff\left(A-Bz\right)\left(\begin{array}{c}
y_{t}^{s_{0}}\\
y_{t}^{s_{1}}
\end{array}\right)=\left(A-Bz\right)\left(\begin{array}{c}
I_{s_{0}}\\
0
\end{array}\right)\varepsilon_{t}^{0,s_{0}}+B\left(\begin{array}{c}
I_{s_{0}}\\
0
\end{array}\right)\varepsilon_{t}^{0,s_{0}}+Vu_{t-1}\\
\iff\Phi(z)Q(z)\left(\begin{array}{c}
y_{t}^{s_{0}}\\
y_{t}^{s_{1}}
\end{array}\right)=\Phi(z)Q(z)\left(\begin{array}{c}
I_{s_{0}}\\
0
\end{array}\right)\varepsilon_{t}^{0,s_{0}}+\alpha(z)^{-1}P(z)^{-1}\left[B\left(\begin{array}{c}
I_{s_{0}}\\
0
\end{array}\right)\varepsilon_{t-1}^{0,s_{0}}+Vu_{t-1}\right]
\end{array}
\]
it is easy to see that all partial multiplicities $\left(g_{1},\ldots,g_{s}\right)$
of $\alpha(z)$ that are larger than one do not have any consequences
regarding linear dependencies among MDS in $\varepsilon_{t}^{0,s_{0}}$. 

In order to obtain the exact number of linear dependencies, we need
to take expectations at time $t$ and $t-1$, subtract the latter
from the former, and subsequently pre-multiply the Moore-Penrose pseudo-inverse
of the first $s_{0}$ columns of $\omega_{0}$, the zero lag coefficient
of $\Phi(z)Q(z)$. Thus, the row rank of the matrix obtained from
\[
\omega_{0,s_{0}}^{\dagger}\left(\mathbb{E}_{t}-\mathbb{E}_{t-1}\right)\left[\alpha(z)^{-1}P(z)^{-1}B\left(\begin{array}{c}
I_{s_{0}}\\
0
\end{array}\right)\varepsilon_{t-1}^{0,s_{0}}\right]
\]
is equal to the number of linear dependencies among the MDS in $\varepsilon_{t-1}^{0,s_{0}}$. 
\end{example}

\begin{example}
\label{exa:sims_onatski_1} The following example is taken from \citep{Sims07}.
Even though $\det\left(\pi(z)\right)$ has a zero at zero, there is
no constraint on the MDS. This is a consequence of the fact that one
variable is predetermined. 

We consider the RE Model 
\begin{align*}
\mathbb{E}_{t}\left(y_{t+1}\right) & =\frac{9}{10}y_{t}+v_{t}\\
x_{t} & =\frac{11}{10}x_{t-1}-\frac{1}{100,000}y_{t}+\varepsilon_{t},
\end{align*}
where $\left(v_{t}\right)_{t\in\mathbb{Z}}$ and $\left(\varepsilon_{t}\right)_{t\in\mathbb{Z}}$
are white noise processes. Replacing the conditional expectation by
the endogenous variables and the associated forecast error leads to
\[
y_{t+1}=\frac{9}{10}y_{t}+\eta_{t+1}+v_{t}
\]
where $\eta_{t+1}=y_{t+1}-\mathbb{E}_{t}\left(y_{t+1}\right)$. The
RE system is equivalent to
\begin{align*}
\begin{pmatrix}0 & 0\\
0 & -\frac{11}{10}
\end{pmatrix}\begin{pmatrix}y_{t-1}\\
x_{t-1}
\end{pmatrix}+\begin{pmatrix}-\frac{9}{10} & 0\\
\frac{1}{100,000} & 1
\end{pmatrix}\begin{pmatrix}y_{t}\\
x_{t}
\end{pmatrix}+\begin{pmatrix}1 & 0\\
0 & 0
\end{pmatrix}\begin{pmatrix}y_{t+1}\\
x_{t+1}
\end{pmatrix} & =\begin{pmatrix}1\\
0
\end{pmatrix}\eta_{t+1}+\begin{pmatrix}v_{t}\\
\varepsilon_{t}
\end{pmatrix}\\
\iff\underbrace{z\left[\begin{pmatrix}0 & 0\\
0 & -\frac{11}{10}
\end{pmatrix}z+\begin{pmatrix}-\frac{9}{10} & 0\\
\frac{1}{100,000} & 1
\end{pmatrix}+\begin{pmatrix}1 & 0\\
0 & 0
\end{pmatrix}z^{-1}\right]}_{=\pi(z)}\begin{pmatrix}y_{t}\\
x_{t}
\end{pmatrix} & =\begin{pmatrix}1\\
0
\end{pmatrix}\eta_{t}+\begin{pmatrix}v_{t-1}\\
\varepsilon_{t-1}
\end{pmatrix}
\end{align*}
where the second component is predetermined. The Smith-form of $\pi(z)$
is
\begin{align*}
\pi(z) & =\begin{pmatrix}1-\frac{9}{10}z & 0\\
\frac{z}{100,000} & z\left(1-\frac{11}{10}z\right)
\end{pmatrix}\\
 & =\underbrace{\begin{pmatrix}1-\frac{9}{10}z & -89,000\\
\frac{z}{100,000} & \frac{99}{100}
\end{pmatrix}}_{=P(z)}\underbrace{\begin{pmatrix}1 & 0\\
0 & z
\end{pmatrix}}_{=\alpha(z)}\underbrace{\begin{pmatrix}1 & 0\\
0 & \frac{1}{99}\left(9z-10\right)\left(11z-10\right)
\end{pmatrix}}_{=\Phi(z)}\underbrace{\begin{pmatrix}1 & z\left(90,000-99,000z\right)\\
0 & 1
\end{pmatrix}.}_{=Q(z)}
\end{align*}
In order to obtain constraints on the MDS $\eta_{t}$ in a way that
the set of causal solutions of the SDE coincides with the one of the
RE model, we proceed in four steps. First, we transform the recursive
such that $\pi(z)$ appears on both sides, i.e. 
\begin{align*}
\pi(z)\begin{pmatrix}y_{t}\\
x_{t}
\end{pmatrix} & =\left[\begin{pmatrix}0 & 0\\
0 & -\frac{11}{10}
\end{pmatrix}z^{2}+\begin{pmatrix}-\frac{9}{10} & 0\\
\frac{1}{100,000} & 1
\end{pmatrix}z+\begin{pmatrix}1 & 0\\
0 & 0
\end{pmatrix}\right]\begin{pmatrix}1\\
0
\end{pmatrix}\eta_{t}+\cdots\\
 & \quad\cdots+\underbrace{\left[-\begin{pmatrix}0 & 0\\
0 & -\frac{11}{10}
\end{pmatrix}\begin{pmatrix}1\\
0
\end{pmatrix}\eta_{t-2}-\begin{pmatrix}-\frac{9}{10} & 0\\
\frac{1}{100,000} & 1
\end{pmatrix}\begin{pmatrix}1\\
0
\end{pmatrix}\eta_{t-1}\right]}_{=\zeta_{t-1}}+\begin{pmatrix}v_{t-1}\\
\varepsilon_{t-1}
\end{pmatrix}.
\end{align*}
Second, we left-multiply 
\begin{align*}
\alpha(z)^{-1}P(z)^{-1} & =\begin{pmatrix}1 & 0\\
0 & \frac{1}{z}
\end{pmatrix}\begin{pmatrix}1 & 90,000\\
-\frac{z}{99,000} & \frac{100}{99}-\frac{10}{11}z
\end{pmatrix}\\
 & =\begin{pmatrix}1 & 90,000\\
-\frac{1}{99,000} & \frac{100}{99}\frac{1}{z}-\frac{10}{11}
\end{pmatrix}\\
 & =\begin{pmatrix}1 & 90,000\\
-\frac{1}{99,000} & -\frac{10}{11}
\end{pmatrix}+\begin{pmatrix}0 & 0\\
0 & \frac{100}{99}
\end{pmatrix}z^{-1},
\end{align*}
i.e.
\[
\left[\Phi(z)Q(z)\right]y_{t}=\left[\Phi(z)Q(z)\right]\begin{pmatrix}1\\
0
\end{pmatrix}\eta_{t}+\alpha(z)^{-1}P(z)^{-1}\left[\zeta_{t-1}+\begin{pmatrix}v_{t-1}\\
\varepsilon_{t-1}
\end{pmatrix}\right].
\]
Third, we take the difference of the the conditional expectations
at $t$ and $\left(t-1\right)$, i.e. we apply $\left(\mathbb{E}_{t}-\mathbb{E}_{t-1}\right)$
such that we obtain  
\begin{align*}
\left(\mathbb{E}_{t}-\mathbb{E}_{t-1}\right)\left\{ \alpha(z)^{-1}P(z)^{-1}\left[\zeta_{t-1}+\begin{pmatrix}v_{t-1}\\
\varepsilon_{t-1}
\end{pmatrix}\right]\right\}  & =0\\
\iff\begin{pmatrix}0\\
-\frac{100}{99}\frac{1}{^{100,000}}
\end{pmatrix}\eta_{t}+\begin{pmatrix}0\\
\frac{100}{99}
\end{pmatrix}\varepsilon_{t} & =0.
\end{align*}
Last, we left-multiply $\omega_{0,s_{0}}^{\dagger}=\begin{pmatrix}1 & 0\end{pmatrix}$,
the Moore-Penrose pseudo-inverse of the columns of the zero-lag coefficient
of
\[
\Phi(z)Q(z)=\begin{pmatrix}1 & 1,000\times9\times z\times\left(10-11z\right)\\
0 & \frac{1}{99}\left(9z-10\right)\left(11z-10\right)
\end{pmatrix}=\begin{pmatrix}1 & 0\\
0 & \frac{100}{99}
\end{pmatrix}+\begin{pmatrix}0 & 90,000\\
0 & -\frac{200}{99}
\end{pmatrix}z+\begin{pmatrix}0 & -99,000\\
0 & 1
\end{pmatrix}z^{2}
\]
pertaining to non-predetermined variables and eventually obtain that
there are no constraints in this example since
\begin{align*}
\omega_{0,s_{0}}^{\dagger}\left(\mathbb{E}_{t}-\mathbb{E}_{t-1}\right)\left\{ \alpha(z)^{-1}P(z)^{-1}\left[\zeta_{t-1}+\begin{pmatrix}v_{t-1}\\
\varepsilon_{t-1}
\end{pmatrix}\right]\right\}  & =0\\
\iff\begin{pmatrix}1 & 0\end{pmatrix}\left[\begin{pmatrix}0\\
-\frac{100}{99}\frac{1}{^{100,000}}
\end{pmatrix}\eta_{t}+\begin{pmatrix}0\\
\frac{100}{99}
\end{pmatrix}\varepsilon_{t}\right] & =0.
\end{align*}
is trivially satisfied.
\end{example}

\pagebreak{}

\section{\label{sec:exist_unique}Existence and Uniqueness of Non-Explosive
Solutions}

Imposing non-explosiveness conditions as general as in \citep{Sims01}
for a process $y_{t}\in H_{u}(t)$ for which the RE equation (\ref{eq:BGS95_model})
holds for every $t\in\mathbb{Z}$ is straightforward in our framework.

First, since we are only interested in causal solutions of the SDE
(\ref{eq:re-poly-1}) (in which redundant MDS have been replaced),
we consider only solutions for which the determinant of $\pi(z)$
is developed in terms of non-negative powers of the backward shift,
i.e.
\[
y_{t}=\det\left(\pi(z)\right)^{-1}adj(\pi(z))g(\varepsilon),
\]
where $g(\varepsilon)$ denotes a polynomial matrix depending on present
and past values of the innovations of the exogenous process. Note
that some zeros at zero (the ones pertaining to $g_{i}>J_{1}$) relate
to finite non-causality whereas others (those with $g_{i}\leq J_{1}$)
do not. Second, the non-explosiveness conditions, which are given
in the form of an $\left(r\times s\right)$-dimensional, $r\leq s$,
matrix $G$ of full (row) rank, are taken into account by requiring
that $Gy_{t}$ does not explode faster than a given rate\footnote{In this framework proposed by \citet{Sims01}, systems with unit roots
satisfy the non-explosiveness condition. A treatment of RE models
involving unit roots can be found, e.g., in \citet{AlSadoon16}. } of growth $\xi>1$. If it is possible to cancel roots $\lambda$
of $\det\left(\pi(z)\right)$ (by adjusting free parameters) for which
$\left|\lambda\right|^{-1}>\xi$ a causal, non-explosive solution
exists. This solution is unique if there are no remaining free parameters.
Note that \citep{BP97,BP95} give sufficient conditions for a ``unique
stable solution'' of a blocked version of (\ref{eq:BGS95_model})
(although without considering predetermined variables).

\paragraph{Factorization of $\pi(z)$ in a stable and an unstable part.}

To be more precise, after taking the restrictions on $\left(k_{0},\ldots,k_{H-1}\right)$
in 
\[
\pi(z)y_{t}=\pi(z)\left(k_{0}\varepsilon_{t}+k_{1}\varepsilon_{t-1}+\cdots+k_{H-1}\varepsilon_{t-H+1}\right)+\zeta_{t-J_{1}}-w(z)\varepsilon_{t-J_{1}}
\]
into account and noting that it is always possible (if there are no
zeros of $\pi(z)$ on the unit circle) to factorize $\pi(z)$ as $\pi(z)=\pi_{u}(z)\pi_{s}(z)$,
where the determinant of $\pi_{s}(z)$ has only roots outside the
unit circle and the determinant of $\pi_{u}(z)$ has only roots inside
the unit circle, we obtain that 
\[
y_{t}=\pi_{s}(z)^{-1}\pi_{u}(z)^{-1}\left(R_{\theta}(z)-w(z)z^{J_{1}}\right)\varepsilon_{t},
\]
where $R_{\theta}(z)$ is a polynomial matrix depending on the deep
parameters $\left(\theta_{1},\ldots,\theta_{p}\right)$. If we can
tune the deep parameters in $R_{\theta}(z)$ in a way that $R_{\theta}(z)=w(z)z^{J_{1}}+\pi_{u}(z)A_{\theta}(z)$,
where $A_{\theta}(z)$ is a polynomial matrix, then we obtain 
\[
y_{t}=\pi_{s}(z)^{-1}A_{\theta}(z)\varepsilon_{t}.
\]
If there is only one parameter value $\left(\theta_{1},\ldots,\theta_{p}\right)$
such that the equation above holds, then the solution is unique. 

\paragraph{A comment on the incorrect number of free parameters given in \citep{BGS95}.}

\citep{BGS95} claim in their Theorem 4 on page 249 and 250 that,
under the assumption that the exogenous process admits a stationary
(finite or infinite) moving-average representation, there exists a
unique solution to the RE model if the number of unstable roots equals
the number of free parameters in the SDE (\ref{eq:re-poly-1}). This
is incorrect because one may only cancel at most as many unstable
roots as there are MDS appearing in different equations (as is the
case in \citep{BlanchardKahn80} under their full rank assumption).
 Indeed, they argue that their result holds by claiming that one
can cancel an unstable root of a certain polynomial matrix by fixing
one of the free parameters. However, this is in general not correct
since the roots to be cancelled out have to lie in the same space,
taking account of which requires additional free parameters. \citep{whiteman83}
proves (starting on page 91) a similar result (using similar methods
as above) to \citep{BGS95} under more restrictive assumptions\footnote{Whiteman does not allow for zeros at infinity and requires that all
zeros of $\pi(z)$ are distinct and that $H=J_{1}$.}. Whiteman's result is correct and therefore consistent with the result
in \citep{BlanchardKahn80} (compare Propositions 1, 2, and 3 on page
1308 in \citep{BlanchardKahn80}) but not consistent with the one
in Theorem 4 on page 249 and 250 in \citep{BGS95} mentioned above.
\begin{example}
In order to illustrate the process of cancelling unstable roots (and
the number of free parameters required for it), we analyse the Blanchard
and Kahn model 
\[
\begin{pmatrix}\mathbb{E}_{t}\left(y_{t+1}^{s_{0}}\right)\\
\mathbb{E}_{t}\left(y_{t+1}^{s_{1}}\right)
\end{pmatrix}=B\begin{pmatrix}y_{t}^{s_{0}}\\
y_{t}^{s_{1}}
\end{pmatrix}+C\varepsilon_{t},\quad t\in\mathbb{N},
\]
where $\mathbb{E}_{t}\left(y_{t+1}^{s_{1}}\right)=y_{t+1},\ \mathbb{E}_{t}\left(y_{t+1}^{s_{0}}\right)=y_{t+1}-\varepsilon_{t+1}^{0},$
and $\left(\varepsilon_{t}\right)_{t\in\mathbb{N}}$ is an $s_{0}$-dimensional
white noise process with uncorrelated components with unit variance.
Assuming that the number $s_{0}$ of non-predetermined variables $y_{t+1}^{s_{0}}$
coincides with the number $n(u)$ of unstable roots of the $\left(s\times s\right)$-dimensional
matrix $B$, and that a full rank assumption holds, we show that in
general $\left(s_{0}\right)^{2}$ free parameters are needed in order
to cancel $s_{0}$ unstable roots. We want to find a matrix $K$ of
free parameters in equation (\ref{eq:BK-as-BGS-cancelling-roots})
below such that $\left(T_{u,s_{0}}K+T_{u\bullet}Cz\right)$ can be
factored as $\left(I_{n(u)}-\mathfrak{J}_{u}z\right)A(z),$ where
$A(z)$ is a polynomial matrix of appropriate dimensions.

Writing the system above as
\[
\begin{pmatrix}y_{t+1}^{s_{0}}-\varepsilon_{t+1}^{0}\\
y_{t+1}^{s_{1}}
\end{pmatrix}=B\begin{pmatrix}y_{t}^{s_{0}}\\
y_{t}^{s_{1}}
\end{pmatrix}+C\varepsilon_{t},\quad t\in\mathbb{N},
\]
and pre-multiplying this equation by a left basis $T$ of the invariant
subspace of $B=T^{-1}\begin{pmatrix}\mathfrak{J}_{s}\\
 & \mathfrak{J}_{u}
\end{pmatrix}T$, where $\mathfrak{J}_{u}$ contains the Jordan blocks pertaining
to unstable roots, we obtain after rearranging
\begin{equation}
\begin{pmatrix}\left(I_{n(s)}-\mathfrak{J}_{s}z\right)s_{t}\\
\left(I_{n(u)}-\mathfrak{J}_{u}z\right)u_{t}
\end{pmatrix}=T_{\bullet,s_{0}}\varepsilon_{t}^{0}+\begin{pmatrix}T_{s\bullet}\\
T_{u\bullet}
\end{pmatrix}C\varepsilon_{t-1},\label{eq:BK-as-BGS}
\end{equation}
where $u_{t}=T_{u,\bullet}\begin{pmatrix}y_{t}^{s_{0}}\\
y_{t}^{s_{1}}
\end{pmatrix}$ and $s_{t}=T_{s,\bullet}\begin{pmatrix}y_{t}^{s_{0}}\\
y_{t}^{s_{1}}
\end{pmatrix}$ in obvious notation.

First, note that the endogenous forecast error $\varepsilon_{t}^{0}=y_{t}^{s_{0}}-\mathbb{E}_{t-1}\left(y_{t}^{s_{0}}\right)$
is a linear function of the white noise process $\left(\varepsilon_{t}\right)_{t\in\mathbb{N}}$,
i.e. we may write $\varepsilon_{t}^{0}=K\varepsilon_{t}$, where $K$
is of dimension $\left(s_{0}\times s_{0}\right)$. 

Consider the unstable part of the system (\ref{eq:BK-as-BGS}), i.e.
\begin{align}
\left(I_{n(u)}-\mathfrak{J}_{u}z\right)u_{t} & =\left(T_{u,s_{0}}K+T_{u\bullet}Cz\right)\varepsilon_{t}\label{eq:BK-as-BGS-cancelling-roots}\\
 & =\left(T_{u,s_{0}}K+\mathfrak{J}_{u}\left(\mathfrak{J}_{u}\right)^{-1}T_{u\bullet}Cz\right)\varepsilon_{t},\nonumber 
\end{align}
and note that (in accordance to Sims' existence condition, see equation
(40) on page 11 in \citep{Sims01}) the matrix polynomial can be factorized
in the desired way if $span\left(\left(\mathfrak{J}_{u}\right)^{-1}T_{u\bullet}C\right)\subseteq span\left(T_{u,s_{0}}\right)$.
In order to fix ideas, we assume $T_{u,s_{0}}$ to be invertible,
take
\[
K=-\left(T_{u,s_{0}}\right)^{-1}\left(\mathfrak{J}_{u}\right)^{-1}T_{u\bullet}C,
\]
obtain
\begin{align*}
\left(I_{n(u)}-\mathfrak{J}_{u}z\right)u_{t} & =\left(-T_{u,s_{0}}\left(T_{u,s_{0}}\right)^{-1}\left(\mathfrak{J}_{u}\right)^{-1}T_{u\bullet}C+\mathfrak{J}_{u}\left(\mathfrak{J}_{u}\right)^{-1}T_{u\bullet}Cz\right)\varepsilon_{t}\\
 & =\left(I_{n(u)}-\mathfrak{J}_{u}z\right)\left[-\mathfrak{J}_{u}^{-1}T_{u\bullet}C\right]\varepsilon_{t},
\end{align*}
and thus it follows that 
\begin{align*}
u_{t} & =\frac{adj\left(I_{n(u)}-\mathfrak{J}_{u}z\right)\left(I_{n(u)}-\mathfrak{J}_{u}z\right)}{\det\left(I_{n(u)}-\mathfrak{J}_{u}z\right)}\left[-\left(\mathfrak{J}_{u}\right)^{-1}T_{u\bullet}C\right]\varepsilon_{t}\\
 & =\frac{\det\left(I_{n(u)}-\mathfrak{J}_{u}z\right)I_{n(u)}}{\det\left(I_{n(u)}-\mathfrak{J}_{u}z\right)}\left[-\left(\mathfrak{J}_{u}\right)^{-1}T_{u\bullet}C\right]\varepsilon_{t}=-\left(\mathfrak{J}_{u}\right)^{-1}T_{u\bullet}C\varepsilon_{t}.
\end{align*}
\end{example}

\begin{example}
\label{exa:king_watson_part2} We now complete the treatment of the
model in \citep{KingWatson98} that we started in Example \ref{exa:king_watson_part1}.
It is interesting that King and Watson did not formulate a theorem
analogous to Proposition 1 in \citep{BlanchardKahn80} which states
that (under a certain rank condition) there exists a unique solution
if the number of unstable eigenvalues coincides with number of non-predetermined
variables. By further analysing the structure of the zeros at zero
of $\pi(z)$ (or in the terminology of \citep{KingWatson98} the structure
of the zeros at infinity), one obtains the number of non-predetermined
variables effectively occurring in the model in \citep{KingWatson98}.
This number can subsequently be compared to the number of unstable
roots in the same way as is done in \citep{BlanchardKahn80}.

Of course, the largest power of $\left(\frac{1}{z}\right)$ occurring
in the inverse of $\left(N-Iz\right)$ depends on the structure of
the nilpotent matrix $N$. If this degree is larger than one, (finite)
non-causalities are introduced into the solution. Since we are searching
for causal solutions, these roots have to be cancelled using the free
parameters describing the MDS. E.g., if $N_{1}$ is a $\left(3\times3\right)$-dimensional
matrix and has ones on the first superdiagonal and zeros otherwise,
then 
\[
\left(N_{1}-zI\right)^{-1}=\begin{pmatrix}-\frac{1}{z} & -\frac{1}{z^{2}} & -\frac{1}{z^{3}}\\
 & -\frac{1}{z} & -\frac{1}{z^{2}}\\
 &  & -\frac{1}{z}
\end{pmatrix}.
\]
 However, if $N_{2}=0$, it follows that 
\[
\left(N_{2}-zI\right)^{-1}=\begin{pmatrix}-\frac{1}{z}\\
 & -\frac{1}{z}\\
 &  & -\frac{1}{z}
\end{pmatrix}.
\]
The same fact is reflected in the Smith-forms of $\left(N_{1}-zI\right)$
and $\left(N_{2}-zI\right)$ whose diagonal matrices are $\begin{pmatrix}1\\
 & 1\\
 &  & z^{3}
\end{pmatrix}$ and $\begin{pmatrix}z\\
 & z\\
 &  & z
\end{pmatrix}$ respectively. 

When we factor the matrix polynomial $\pi(z)=\pi_{u}(z)\pi_{s}(z)$
into an unstable and stable part, the zeros at zero of $\pi(z)$ pertain
to the first zero at zero of any Jordan block pertaining to the eigenvalue
zero of $A$ (or equivalently the number of partial indices $\left(g_{1},\ldots,g_{s}\right)$
of $\alpha(z)$ which are larger or equal to one) belong to the stable
part because they do not introduce finite non-causalities. The other
$\sum_{i:g_{i}>1}\left(g_{i}-1\right)$ zeros at zero of $\pi(z)$
belong to $\pi_{u}(z)$ and have to be cancelled by the free parameters
in the SDE because they would otherwise introduce finite non-causalities.

Thus, under a certain rank condition, there exists a unique solution
of the model in \citep{KingWatson98} if the number of non-predetermined
variables effectively occurring in the model (which is equal to the
rank of $\begin{pmatrix}I_{n(s)}\\
 & I_{n(u)}\\
 &  & N
\end{pmatrix}W\left(\begin{array}{c}
I_{s_{0}}\\
0
\end{array}\right)$) coincides with the number of ``unstable'' roots, i.e. the dimension
of $\mathfrak{J}_{u}$ plus the $\sum_{i:g_{i}>1}\left(g_{i}-1\right)$
zeros at zero of $\pi(z)$ which introduce finite non-causalities.
\end{example}

\begin{example}
As another example we want to illustrate the method presented here
and compare it to the method in \citep{Sims01,LubikSchorfheide03}.
We continue with Example \ref{exa:sims_onatski_1}. In order to obtain
a difference equation without unstable roots, we proceed in three
steps. First, we represent the MDS $\eta_{t}=\begin{pmatrix}k_{v} & k_{\varepsilon}\end{pmatrix}\begin{pmatrix}v_{t}\\
\varepsilon_{t}
\end{pmatrix}$ as function of the innovations of the exogenous process and obtain
\begin{align*}
\pi(z)\begin{pmatrix}y_{t}\\
x_{t}
\end{pmatrix} & =\begin{pmatrix}1\\
0
\end{pmatrix}\eta_{t}+\begin{pmatrix}v_{t-1}\\
\varepsilon_{t-1}
\end{pmatrix}\\
\iff\pi(z)\begin{pmatrix}y_{t}\\
x_{t}
\end{pmatrix} & =\begin{pmatrix}k_{v}+z & k_{\varepsilon}\\
0 & z
\end{pmatrix}\begin{pmatrix}v_{t}\\
\varepsilon_{t}
\end{pmatrix}.
\end{align*}
Second, we factor $\pi(z)=\pi_{u}(z)\pi_{s}(z)$ in order to separate
unstable from stable roots and define 
\begin{align*}
\pi_{u}(z) & =P(z)\begin{pmatrix}1 & 0\\
0 & (11z-10)
\end{pmatrix}\\
 & =\begin{pmatrix}\left(1-\frac{9}{10}z\right) & -99\times90\times10\times\left(11z-10\right)\\
\frac{z}{100,000} & \frac{99}{100}\left(11z-10\right)
\end{pmatrix}
\end{align*}
and
\begin{align*}
\pi_{s}(z) & =\begin{pmatrix}1 & 0\\
0 & z
\end{pmatrix}\begin{pmatrix}1 & 0\\
0 & \frac{1}{99}\left(9z-10\right)
\end{pmatrix}Q(z)\\
 & =\begin{pmatrix}1 & 9\times1000\times z\times\left(10-11z\right)\\
0 & \frac{1}{99}\times z\times\left(9z-10\right)
\end{pmatrix}.
\end{align*}
Note that the zero at $z=0$ is a stable zero because it does not
induce (finite) non-causality since $g_{i}\leq J_{1}$ holds. Third,
we factor $B_{\theta}(z)=\left(R_{\theta}(z)-zI\right)=\begin{pmatrix}k_{v}+z & k_{\varepsilon}\\
0 & z
\end{pmatrix}$ such that $B_{\theta}(z)=\pi_{u}(z)A_{\theta}(z)$ where $A_{\theta}(z)$
is a polynomial matrix. To this end, we consider 
\begin{align*}
\pi_{u}(z)^{-1}B_{\theta}(z) & =\left[\begin{pmatrix}1 & 0\\
0 & \frac{1}{11z-10}
\end{pmatrix}P(z)^{-1}\right]\begin{pmatrix}k_{v}+z & k_{\varepsilon}\\
0 & z
\end{pmatrix}\\
 & =\begin{pmatrix}1 & 90,000\\
\frac{-z}{90,000\left(11z-10\right)} & \frac{100}{99\left(11z-10\right)}-\frac{10z}{11\left(11z-10\right)}
\end{pmatrix}\begin{pmatrix}k_{v}+z & k_{\varepsilon}\\
0 & z
\end{pmatrix}\\
 & =\begin{pmatrix}1 & 90,000\\
\frac{-z}{90,000\left(11z-10\right)} & \frac{100-90z}{99\left(11z-10\right)}
\end{pmatrix}\begin{pmatrix}k_{v}+z & k_{\varepsilon}\\
0 & z
\end{pmatrix}.
\end{align*}
In order to obtain a polynomial matrix $A_{\theta}(z)$, we try to
tune the free parameters in a way such that the term $\left(11z-10\right)$
in the denominators of elements $\left(2,1\right)$ and $(2,2)$ of
$\pi_{u}(z)^{-1}$ cancels out. Element $(2,1)$ is equal to $\frac{-\left(k_{v}+z\right)z}{99,000(11z-10)}$.
Thus, for $k_{v}=-\frac{10}{11}$ we obtain that element $\left(2,1\right)$
is equal to $\frac{-z}{1000\times9\times11^{2}}=-\frac{z}{1,089,000}$.
Element $(2,2)$ is equal to $\frac{100-90z}{99\left(11z-10\right)}$.
Thus, for $k_{\varepsilon}=\frac{200,000}{11}$ we obtain that element
$\left(2,2\right)$ is equal to $-\frac{10z}{121}$. Eventually, the
matrix $A_{\theta}(z)$ is indeed polynomial and equal to 
\[
A_{\theta}(z)=\begin{pmatrix}z-\frac{10}{11} & \frac{200,000}{11}+\frac{90,000\times11}{11}z\\
\frac{-z}{121\times1,000\times9} & \frac{-10z}{121}
\end{pmatrix}.
\]

The causal solutions are now obtained by inverting $\pi_{s}(z)$ in
\begin{align*}
\pi_{s}(z)y_{t} & =A_{\theta}(z)\begin{pmatrix}v_{t}\\
\varepsilon_{t}
\end{pmatrix}\\
\iff\begin{pmatrix}1 & 9\times1000\times z\times\left(10-11z\right)\\
0 & \frac{1}{99}\times z\times\left(9z-10\right)
\end{pmatrix}y_{t} & =\begin{pmatrix}z-\frac{10}{11} & \frac{200,000}{11}+\frac{90,000\times11}{11}z\\
\frac{-z}{121\times1,000\times9} & \frac{-10z}{121}
\end{pmatrix}\begin{pmatrix}v_{t}\\
\varepsilon_{t}
\end{pmatrix}
\end{align*}
such that we obtain 
\[
y_{t}=\frac{1}{\left(1-\frac{9}{10}z\right)}\frac{1}{11}\begin{pmatrix}-10\left(1-\frac{11}{10}z\right) & 20,000\\
\frac{1}{10,000} & 9
\end{pmatrix}\begin{pmatrix}v_{t}\\
\varepsilon_{t}
\end{pmatrix}.
\]
 Note that $\pi_{s}(z)$ as well as $A_{\theta}(z)$ have a zero at
zero in their second row.
\end{example}

\begin{example}
In \citep{Sims01}, one starts with a system where all linear dependencies
among the MDS are already taken into account. Still, as a matter of
comparison we want to show for the same example as above how the method
in \citep{Sims01} can be used.

We define a new variable $\xi_{t}=\mathbb{E}_{t}\left(y_{t+1}\right)$
such that $\eta_{t+1}=y_{t+1}-\xi_{t}$ and consider the system
\[
\begin{pmatrix}x_{t}\\
\xi_{t}\\
y_{t}
\end{pmatrix}=\underbrace{\begin{pmatrix}\frac{11}{10} & -\frac{1}{100,000} & 0\\
0 & \frac{9}{10} & 0\\
0 & 1 & 0
\end{pmatrix}}_{=\Gamma_{1}}\begin{pmatrix}x_{t-1}\\
\xi_{t-1}\\
y_{t-1}
\end{pmatrix}+\underbrace{\begin{pmatrix}1 & 0\\
0 & 1\\
0 & 0
\end{pmatrix}}_{=\Psi}\begin{pmatrix}\varepsilon_{t}\\
v_{t}
\end{pmatrix}+\underbrace{\begin{pmatrix}-\frac{1}{100,000}\\
\frac{9}{10}\\
1
\end{pmatrix}}_{=\Pi}\eta_{t}.
\]
In order to obtain a difference equation which has only causal stationary
solutions we proceed in five steps. First, we calculate the eigendecomposition
of $\Gamma_{1}=VSV^{-1}$ and pre-multiply $V^{-1}=\begin{pmatrix}1 & -\frac{1}{200,000} & 0\\
0 & \frac{9}{10} & 0\\
0 & 1 & 1
\end{pmatrix}$ in order to decouple the system in an unstable and stable part, i.e.
\[
\begin{pmatrix}u_{t}\\
s_{t}^{1}\\
s_{t}^{2}
\end{pmatrix}=\begin{pmatrix}\frac{11}{10} & 0 & 0\\
0 & \frac{9}{10} & 0\\
0 & 0 & 0
\end{pmatrix}\begin{pmatrix}u_{t-1}\\
s_{t-1}^{1}\\
s_{t-1}^{2}
\end{pmatrix}+\begin{pmatrix}1 & -\frac{1}{200,000}\\
0 & \frac{10}{9}\\
0 & -\frac{10}{9}
\end{pmatrix}\begin{pmatrix}\varepsilon_{t}\\
v_{t}
\end{pmatrix}+\begin{pmatrix}-\frac{11}{200,000}\\
1\\
0
\end{pmatrix}\eta_{t}
\]
where $\begin{pmatrix}u_{t}\\
s_{t}^{1}\\
s_{t}^{2}
\end{pmatrix}=V^{-1}\begin{pmatrix}x_{t}\\
\xi_{t}\\
y_{t}
\end{pmatrix}$. Second, we choose $\eta_{t}$ as a linear function of $\varepsilon_{t}$
and $v_{t}$ such that there is no exogenous influence on the unstable
part of the system and $u_{t}\equiv0$ can be chosen as a solution.
It follows that for
\begin{align*}
\frac{-11}{200,000}\eta_{t} & =-\varepsilon_{t}+\frac{1}{200,000}v_{t}\\
\iff\eta_{t} & =\frac{200,000}{11}\varepsilon_{t}-\frac{10}{11}v_{t}
\end{align*}
Sims' existence condition is satisfied. Moreover, there are no further
free parameters that might influence the stable part of the equation.
Thus, the uniqueness condition is satisfied. 

Third, we set $u_{t}\equiv0$ by setting the $\left(1,1\right)$ element
of $S$ equal to zero and obtain
\[
\begin{pmatrix}u_{t}\\
s_{t}^{1}\\
s_{t}^{2}
\end{pmatrix}=\begin{pmatrix}0 & 0 & 0\\
0 & \frac{9}{10} & 0\\
0 & 0 & 0
\end{pmatrix}\begin{pmatrix}u_{t-1}\\
s_{t-1}^{1}\\
s_{t-1}^{2}
\end{pmatrix}+\begin{pmatrix}0 & 0\\
\frac{200,000}{11} & \frac{10}{9}\\
0 & -\frac{10}{9}
\end{pmatrix}\begin{pmatrix}\varepsilon_{t}\\
v_{t}
\end{pmatrix}.
\]
Fourth, we pre-multiply $V$ in order to obtain a system in the original
variables, i.e.
\[
\begin{pmatrix}x_{t}\\
\xi_{t}\\
y_{t}
\end{pmatrix}=\begin{pmatrix}0 & \frac{9}{200,000} & 0\\
0 & \frac{9}{10} & 0\\
0 & 1 & 1
\end{pmatrix}\begin{pmatrix}x_{t-1}\\
\xi_{t-1}\\
y_{t-1}
\end{pmatrix}+\begin{pmatrix}\frac{9}{11} & \frac{1}{110,000}\\
\frac{180,000}{11} & \frac{2}{11}\\
\frac{200,000}{11} & -\frac{10}{11}
\end{pmatrix}\begin{pmatrix}\varepsilon_{t}\\
v_{t}
\end{pmatrix}.
\]
Last, we get rid of $\xi_{t}$ by noting that the first two rows of
both matrices on the right hand side of the equation above are linearly
dependent, i.e. we pre-multiply $\begin{pmatrix}20,000 & -1 & 0\end{pmatrix}$
and obtain
\begin{align*}
20,000x_{t}-\xi_{t} & =\begin{pmatrix}20,000 & -1 & 0\end{pmatrix}\left[\begin{pmatrix}0 & \frac{9}{200,000} & 0\\
0 & \frac{9}{10} & 0\\
0 & 1 & 1
\end{pmatrix}\begin{pmatrix}x_{t-1}\\
\xi_{t-1}\\
y_{t-1}
\end{pmatrix}+\begin{pmatrix}\frac{9}{11} & \frac{1}{110,000}\\
\frac{180,000}{11} & \frac{2}{11}\\
\frac{200,000}{11} & -\frac{10}{11}
\end{pmatrix}\begin{pmatrix}\varepsilon_{t}\\
v_{t}
\end{pmatrix}\right]\\
\iff\xi_{t} & =20,000x_{t}
\end{align*}
 such that finally
\[
\begin{pmatrix}y_{t}\\
x_{t}
\end{pmatrix}=\begin{pmatrix}0 & 20,000\\
0 & \frac{9}{10}
\end{pmatrix}\begin{pmatrix}y_{t-1}\\
x_{t-1}
\end{pmatrix}+\begin{pmatrix}-\frac{10}{11} & \frac{200,000}{11}\\
\frac{1}{110,000} & \frac{9}{11}
\end{pmatrix}\begin{pmatrix}v_{t}\\
\varepsilon_{t}
\end{pmatrix}.
\]
The (unique) causal solution is given by
\begin{align*}
\begin{pmatrix}y_{t}\\
x_{t}
\end{pmatrix} & =\begin{pmatrix}1 & -20,000z\\
0 & 1-\frac{9}{10}z
\end{pmatrix}^{-1}\begin{pmatrix}-\frac{10}{11} & \frac{200,000}{11}\\
\frac{1}{110,000} & \frac{9}{11}
\end{pmatrix}\begin{pmatrix}v_{t}\\
\varepsilon_{t}
\end{pmatrix}\\
 & =\frac{1}{1-\frac{9}{10}z}\begin{pmatrix}\left(1-\frac{9}{10}z\right) & 20,000z\\
0 & 1
\end{pmatrix}\begin{pmatrix}-\frac{10}{11} & \frac{200,000}{11}\\
\frac{1}{11\times10,000} & \frac{9}{11}
\end{pmatrix}\begin{pmatrix}v_{t}\\
\varepsilon_{t}
\end{pmatrix}.
\end{align*}
\end{example}

\pagebreak{}

\section{Summary and Conclusion}

In this paper, we corrected results in \citep{BGS95} regarding firstly
the number of free parameters in the SDE whose set of causal solution
is equivalent to the one of the associated RE model and secondly regarding
the number of free parameters necessary and (under conditions) sufficient
for the existence of a unique solution. The first correction is related
to the structure of the zeros at zero in the Smith-form of the polynomial
matrix in the SDE. Understanding the structure of the zeros at zero
of the polynomial matrix appearing in the SDE also connects the results
in \citep{KingWatson98} and \citep{BlanchardKahn80} since up to
now it was not clear what the equivalent of Blanchard and Kahn's ``number
of non-predetermined'' variables in the framework of \citep{KingWatson98}
is. 

Furthermore, we extended these results firstly by allowing for predetermined
components in the vector of endogenous variables, secondly by allowing
for stochastic singularity, i.e. the number of exogenous variables
and the dimension of their innovations may be strictly smaller than
the number of endogenous variables, and thirdly by allowing for general
parameter restrictions. These extensions make the results comparable
to the other ones in the literature on econometric treatment of RE
models.

While the approach in \citep{Sims01} features a procedure as to how
to solve an RE model that is already in the form of an SDE of order
one (after blocking some variables) and where all linear dependencies
among MDS are taken into account, our paper is mainly concerned with
the preceding step about how to obtain an SDE containing arbitrary
MDS whose set of causal solutions is equivalent to the one of the
associated RE model.

Last, our contribution is the basis for identifiability analysis of
causal solutions of RE models which is robust with respect to determinacy
and indeterminacy. Identifiability analysis of DSGE models without
imposing minimality (as was done in \citep{KomunjerNg11}) will be
part of a future research project.

\section{Acknowledgements}

Research for this paper was supported by the Jubiläumsfond of the
OeNB under the contract number 16546 and the Austrian Science Fund
(FWF). A previous version of this paper is part of my Ph.D. thesis
at the University of Vienna. I am hughly endebted to my thesis adviser
Manfred Deistler for continuing guidance and support as well as helpful
comments on this paper. All errors remain mine.

\pagebreak{}


\pagebreak{}

\appendix

\section{Proof of Theorem \ref{thm:multivariate-smith-constraints-1}}

The proof is divided into several steps.

\paragraph{Step 1: Left-multiply $\left(P(z)\alpha(z)\right)^{-1}$ on the SDE
(\ref{eq:recursive with smith-1}) (which was derived from (\ref{eq:BGS95_model})).}

The equation we will work with is
\begin{equation}
\underbrace{\Phi(z)Q(z)}_{=\Omega(z)}y_{t}=\underbrace{\Phi(z)Q(z)}_{=\Omega(z)}\left(\varepsilon_{t}^{0}+\varepsilon_{t-1}^{1}+\cdots+\varepsilon_{t-H+1}^{H-1}\right)+\alpha(z)^{-1}P(z)^{-1}\left(\zeta_{t-J_{1}}-u_{t-J_{1}}\right).\label{eq:smith-proof-1}
\end{equation}

\paragraph{Step 2: Take conditional expectations of $\Omega(z)y_{t}$ with respect
to the information at time $\left(t-i\right),\ i\in\left\{ 0,\ldots,H\right\} ,$
and subtract equation $\left(i+1\right)$ from equation $i$ for $i\in\left\{ 0,\ldots,H-1\right\} .$}

Note that lags of $y_{t}$ appearing in $\Omega(z)y_{t}=\omega_{0}y_{t}+\omega_{1}y_{t-1}+\cdots+\omega_{i}y_{t-i}+\omega_{i+1}y_{t-(i+1)}+\cdots+\omega_{\deg\left(\Omega(z)\right)}y_{t-\deg\left(\Omega(z)\right)}$
which are larger than $i$, have the same conditional expectation
with respect to information sets up to time $(t-i)$ and up to time
$\left(t-(i+1)\right)$ because we assumed that $y_{t}\in H_{u}(t)$
. Thus, we obtain for the left hand side of equation (\ref{eq:smith-proof-1})
\begin{align}
\left(\mathbb{E}_{t-i}-\mathbb{E}_{t-(i+1)}\right)\left(\Omega(z)y_{t}\right) & =\left(\mathbb{E}_{t-i}-\mathbb{E}_{t-(i+1)}\right)\left(\omega_{0}y_{t}+\omega_{1}y_{t-1}+\cdots+\omega_{i}y_{t-i}\right)\nonumber \\
 & =\omega_{0}\left(\mathbb{E}_{t-i}-\mathbb{E}_{t-(i+1)}\right)\left(y_{t}\right)+\cdots+\omega_{i}\left(\mathbb{E}_{t-i}-\mathbb{E}_{t-(i+1)}\right)\left(y_{t-i}\right)\nonumber \\
 & =\omega_{0}\varepsilon_{t-i}^{i}+\omega_{1}\varepsilon_{t-i}^{i-1}+\dots+\omega_{i}\varepsilon_{t-i}^{0}.\label{eq:bgs-constraints-proof-intermediate}
\end{align}
Note that we need the causality of $\left(y_{t}\right)_{t\in\mathbb{Z}}$
here. Otherwise, the expression 
\[
\omega_{i+1}\left(\mathbb{E}_{t-i}-\mathbb{E}_{t-(i+1)}\right)\left(y_{t-(i+1)}\right)
\]
 would not be zero. 

\paragraph{Step 3: Take the conditional expectation of $\Omega(z)\left(\varepsilon_{t}^{0}+\varepsilon_{t-1}^{1}+\cdots+\varepsilon_{t-H+1}^{H-1}\right)$
with respect to information at time $\left(t-i\right),\ i\in\left\{ 0,\ldots,H\right\} ,$
and subtract equation $\left(i+1\right)$ from equation $i$ for $i\in\left\{ 0,\ldots,H-1\right\} .$}

Considering the term $\Omega(z)\varepsilon_{t-j}^{j}$, we note that
lags larger than $\left(i-j\right)$ are contained in both information
sets which contain information up to time $(t-i)$ and up to time
$\left(t-(i+1)\right)$. Thus, we obtain for $i\geq j$
\[
\begin{array}{l}
\left(\mathbb{E}_{t-i}-\mathbb{E}_{t-(i+1)}\right)\left(\Omega(z)\varepsilon_{t-j}^{j}\right)=\cdots\\
\qquad=\left(\mathbb{E}_{t-i}-\mathbb{E}_{t-(i+1)}\right)\left(\omega_{0}\varepsilon_{t-j}^{j}+\omega_{1}\varepsilon_{t-j-1}^{j}+\cdots+\omega_{i-j-1}\varepsilon_{\underbrace{t-j-(i-j-1)}_{=t-i+1}}^{j}+\omega_{i-j}\varepsilon_{\underbrace{t-j-(i-j)}_{=t-i}}^{j}\right)\\
\qquad=\omega_{i-j}\varepsilon_{t-i}^{j}
\end{array}
\]
such that
\[
\left(\mathbb{E}_{t-i}-\mathbb{E}_{t-(i+1)}\right)\left(\Omega(z)\left(\varepsilon_{t}^{0}+\varepsilon_{t-1}^{1}+\cdots+\varepsilon_{t-H+1}^{H-1}\right)\right)=\omega_{i}\varepsilon_{t-i}^{0}+\omega_{i-1}\varepsilon_{t-i}^{1}+\cdots+\omega_{1}\varepsilon_{t-i}^{i-1}+\omega_{0}\varepsilon_{t-i}^{i}
\]
which is equal to (\ref{eq:bgs-constraints-proof-intermediate}),
i.e. $\left(\mathbb{E}_{t-i}-\mathbb{E}_{t-(i+1)}\right)\left(\Omega(z)y_{t}\right)$,
from above. 

\paragraph{Step 4: Conclude.}

On the right hand side of equation (\ref{eq:smith-proof-1}) remains
thus
\[
\mathbb{E}_{t-i}\left[\alpha(z)^{-1}P(z)^{-1}\left(\zeta_{t-J_{1}}-u_{t-J_{1}}\right)\right]=\mathbb{E}_{t-(i+1)}\left[\alpha(z)^{-1}P(z)^{-1}\left(\zeta_{t-J_{1}}-u_{t-J_{1}}\right)\right]
\]
from which the theorem follows.

\pagebreak{}

\section{Proof of Theorem \ref{thm:recursive_then_RE}}

We start by pre-multiplying $\alpha^{-1}(z)P^{-1}(z)$ on $\pi(z)y_{t}=\pi(z)\left(\varepsilon_{t}^{0}+\varepsilon_{t-1}^{1}+\cdots+\varepsilon_{t-H+1}^{H-1}\right)+\zeta_{t-J_{1}}-u_{t-J_{1}}$
in order to obtain 
\[
\underbrace{\Phi(z)Q(z)}_{=\Omega(z)}y_{t}=\Omega(z)\left(\varepsilon_{t}^{0}+\varepsilon_{t-1}^{1}+\cdots+\varepsilon_{t-H+1}^{H-1}\right)+\alpha^{-1}(z)P^{-1}(z)\zeta_{t-J_{1}}-\alpha^{-1}(z)P^{-1}(z)u_{t-J_{1}}.
\]

\paragraph{Step 1: Take conditional expectations of $\Omega(z)y_{t}=\Omega(z)\left(\varepsilon_{t}^{0}+\varepsilon_{t-1}^{1}+\cdots+\varepsilon_{t-H+1}^{H-1}\right)+\alpha^{-1}(z)P^{-1}(z)\zeta_{t-J_{1}}-\alpha^{-1}(z)P^{-1}(z)u_{t-J_{1}}$
with respect to $H_{u}(t-i),\ i\in\left\{ 0,\ldots,H\right\} ,$ and
subtract each projection from the preceding.}

The left hand side of the equation is evidently $\left(\mathbb{E}_{t-i}-\mathbb{E}_{t-(i+1)}\right)\left(\Omega(z)y_{t}\right)$,
where only lags of $y_{t}$ up to time $\left(t-i\right)$ have to
be considered because $y_{t-j},\ j>i$ is contained in both $H_{u}(t-i)$
and $H_{u}\left(t-(i+1)\right)$ and thus cancels out. For this to
be true, the causality of $\left(y_{t}\right)_{t\in\mathbb{Z}}\in H_{u}(t)$
is required.

For the right hand side, we consider the term $\Omega(z)\varepsilon_{t-j}^{j}$
and note that lags larger than $\left(i-j\right)$ are contained in
both information sets which contain information up to time $(t-i)$
and up to time $\left(t-(i+1)\right)$. Thus, we obtain for $i\geq j$
\[
\begin{array}{l}
\left(\mathbb{E}_{t-i}-\mathbb{E}_{t-(i+1)}\right)\left(\Omega(z)\varepsilon_{t-j}^{j}\right)=\cdots\\
\qquad=\left(\mathbb{E}_{t-i}-\mathbb{E}_{t-(i+1)}\right)\left(\omega_{0}\varepsilon_{t-j}^{j}+\omega_{1}\varepsilon_{t-j-1}^{j}+\cdots+\omega_{i-j-1}\varepsilon_{\underbrace{t-j-(i-j-1)}_{=t-i+1}}^{j}+\omega_{i-j}\varepsilon_{\underbrace{t-j-(i-j)}_{=t-i}}^{j}\right)\\
\qquad=\omega_{i-j}\varepsilon_{t-i}^{j}
\end{array}
\]
such that
\[
\left(\mathbb{E}_{t-i}-\mathbb{E}_{t-(i+1)}\right)\left[\Omega(z)\left(\varepsilon_{t}^{0}+\varepsilon_{t-1}^{1}+\cdots+\varepsilon_{t-H+1}^{H-1}\right)\right]=\omega_{i}\varepsilon_{t-i}^{0}+\omega_{i-1}\varepsilon_{t-i}^{1}+\cdots+\omega_{1}\varepsilon_{t-i}^{i-1}+\omega_{0}\varepsilon_{t-i}^{i}.
\]
Likewise, applying conditional expectations with respect to information
up to time $\left(t-i\right)$ and up to time $\left(t-(i+1)\right)$
on $\alpha^{-1}(z)P^{-1}(z)\zeta_{t-J_{1}}-\alpha^{-1}(z)P^{-1}(z)u_{t-J_{1}}$
gives $\left(\mathbb{E}_{t-i}-\mathbb{E}_{t-(i+1)}\right)\left[\alpha(z)^{-1}P(z)^{-1}\left(\zeta_{t-J_{1}}-u_{t-J_{1}}\right)\right]=0,\quad i\in\left\{ 0,\ldots,H-1\right\} .$

\paragraph{Step 2: Use the constraints and obtain a system of equations relating
the MDS to some conditional expectations.}

Using the constraints and denoting the matrix-valued coefficients
of $\Omega(z)$ by $\omega_{i}$, we thus obtain 
\[
\left(\mathbb{E}_{t-i}-\mathbb{E}_{t-(i+1)}\right)\left[\left(\omega_{0}+\omega_{1}z+\cdots+\omega_{i}z^{i}\right)y_{t}\right]=\omega_{i}\varepsilon_{t-i}^{0}+\omega_{i-1}\varepsilon_{t-i}^{1}+\cdots+\omega_{1}\varepsilon_{t-i}^{i-1}+\omega_{0}\varepsilon_{t-i}^{i},\quad i\in\left\{ 0,\ldots,H-1\right\} ,
\]
or equivalently the following system of equations
\begin{align*}
\omega_{0}y_{t}-\mathbb{E}_{t-1}\left(\omega_{0}y_{t}\right) & =\omega_{0}\varepsilon_{t}^{0}\\
\left(\mathbb{E}_{t-1}-\mathbb{E}_{t-2}\right)\left(\omega_{0}y_{t}+\omega_{1}y_{t-1}\right) & =\omega_{0}\varepsilon_{t-1}^{1}+\omega_{1}\varepsilon_{t-1}^{0}\\
\left(\mathbb{E}_{t-2}-\mathbb{E}_{t-3}\right)\left(\omega_{0}y_{t}+\omega_{1}y_{t-1}+\omega_{2}y_{t-2}\right) & =\omega_{0}\varepsilon_{t-2}^{2}+\cdots+\omega_{2}\varepsilon_{t-2}^{0}\\
 & \vdots\\
\left(\mathbb{E}_{t-i}-\mathbb{E}_{t-(i+1)}\right)\left(\omega_{0}y_{t}+\cdots+\omega_{i}y_{t-i}\right) & =\omega_{0}\varepsilon_{t-i}^{i}+\cdots+\omega_{i}\varepsilon_{t-i}^{0}\\
 & \vdots\\
\left(\mathbb{E}_{t-(H-1)}-\mathbb{E}_{t-H}\right)\left(\omega_{0}y_{t}+\cdots+\omega_{H-1}y_{t-(H-1)}\right) & =\omega_{0}\varepsilon_{t-(H-1)}^{H-1}+\cdots+\omega_{H-1}\varepsilon_{t-(H-1)}^{0}
\end{align*}

\paragraph{Step 3: Reorder the system of equations in order to conclude.}

If the $i$-th equation is shifted $H-i$ periods backwards, this
system can be written as 
\begin{align*}
z^{H-1}\omega_{0}\left(y_{t}-\mathbb{E}_{t-1}\left(y_{t}\right)\right) & =\omega_{0}\varepsilon_{t-H+1}^{0}\\
z^{H-2}\left[\omega_{0}\left(\mathbb{E}_{t-1}-\mathbb{E}_{t-2}\right)\left(y_{t}\right)+\omega_{1}\left(\mathbb{E}_{t-1}-\mathbb{E}_{t-2}\right)\left(y_{t-1}\right)\right] & =\omega_{0}\varepsilon_{t-H+1}^{1}+\omega_{1}\varepsilon_{t-H+1}^{0}\\
z^{H-3}\left[\omega_{0}\left(\mathbb{E}_{t-2}-\mathbb{E}_{t-3}\right)\left(y_{t}\right)+\cdots+\omega_{2}\left(\mathbb{E}_{t-2}-\mathbb{E}_{t-3}\right)\left(y_{t-2}\right)\right] & =\omega_{0}\varepsilon_{t-H+1}^{2}+\cdots+\omega_{2}\varepsilon_{t-H+1}^{0}\\
 & \vdots\\
z^{H-1-(i+1)}\left[\omega_{0}\left(\mathbb{E}_{t-i}-\mathbb{E}_{t-(i+1)}\right)\left(y_{t}\right)+\cdots+\omega_{i}\left(\mathbb{E}_{t-i}-\mathbb{E}_{t-(i+1)}\right)\left(y_{t-i}\right)\right] & =\omega_{0}\varepsilon_{t-H+1}^{i}+\cdots+\omega_{i}\varepsilon_{t-H+1}^{0}\\
 & \vdots\\
\omega_{0}\left(\mathbb{E}_{t-(H-1)}-\mathbb{E}_{t-H}\right)\left(y_{t}\right)+\cdots+\omega_{H-1}\left(\mathbb{E}_{t-(H-1)}-\mathbb{E}_{t-H}\right)\left(y_{t-(H-1)}\right) & =\omega_{0}\varepsilon_{t-H+1}^{H-1}+\cdots+\omega_{H-1}\varepsilon_{t-(H-1)}^{0}
\end{align*}
or equivalently
\[
\begin{pmatrix}\omega_{0} & 0 & \cdots & 0\\
\omega_{1} & \omega_{0} & \ddots & \vdots\\
\vdots &  & \ddots & 0\\
\omega_{H-1} & \cdots & \omega_{1} & \omega_{0}
\end{pmatrix}\left(\mathbb{E}_{t-H+1}-\mathbb{E}_{t-H}\right)\begin{pmatrix}y_{t-H+1}\\
\vdots\\
y_{t-H-i+2}\\
\vdots\\
y_{t}
\end{pmatrix}=\begin{pmatrix}\omega_{0} & 0 & \cdots & 0\\
\omega_{1} & \omega_{0} & \ddots & \vdots\\
\vdots &  & \ddots & 0\\
\omega_{H-1} & \cdots & \omega_{1} & \omega_{0}
\end{pmatrix}\begin{pmatrix}\varepsilon_{t-(H-1)}^{0}\\
\vdots\\
\varepsilon_{t-(H-1)}^{i}\\
\vdots\\
\varepsilon_{t-(H-1)}^{H-1}
\end{pmatrix}.
\]
Since $\omega_{0}$ is a non-singular matrix, we obtain that $\mathbb{E}_{t-i}\left(y_{t}\right)-\mathbb{E}_{t-(i+1)}\left(y_{t}\right)=\varepsilon_{t-i}^{i}$
and can reconstruct the RE model.

\pagebreak{}

\section{\label{sec:Proof-of-Theorem General}Proof of Theorem \ref{thm:general-solution-set}}

\paragraph{Step 1: \label{par:main_proof_step1_permute}Rewrite the constraints
in equation (\ref{eq:general_thm_constraints}).}

We consider the constraints derived from the RE model in Theorem \ref{thm:multivariate-smith-constraints-1},
i.e. 
\[
\begin{pmatrix}\left(\mathbb{E}_{t}-\mathbb{E}_{t-1}\right)\left[\begin{pmatrix}z^{-g_{1}}\\
 & \ddots\\
 &  & z^{-g_{s}}
\end{pmatrix}P(z)^{-1}\left(\zeta_{t-J_{1}}-u_{t-J_{1}}\right)\right]\\
\vdots\\
\left(\mathbb{E}_{t-(H-1)}-\mathbb{E}_{t-H}\right)\left[\begin{pmatrix}z^{-g_{1}}\\
 & \ddots\\
 &  & z^{-g_{s}}
\end{pmatrix}P(z)^{-1}\left(\zeta_{t-J_{1}}-u_{t-J_{1}}\right)\right]
\end{pmatrix}=0.
\]
We define $\mathfrak{P}(z)=P(z)^{-1}$, denote its $j$-th row by
$\mathfrak{P}_{j,\bullet}(z)$ and the coefficient pertaining to $z^{k}$
as $\mathfrak{P}_{j,\bullet|k}$. Shifting the $i$-th block in the
equation system above $\left(i-1\right)$ periods ahead, we obtain
\[
\left[\mathbb{E}_{t}-\mathbb{E}_{t-1}\right]\left\{ \begin{pmatrix}1\\
z^{-1}\\
\vdots\\
z^{-H+1}
\end{pmatrix}\otimes\left[\begin{pmatrix}z^{-g_{1}}\\
 & \ddots\\
 &  & z^{-g_{s}}
\end{pmatrix}\begin{pmatrix}\cdots & \mathfrak{P}_{1,\bullet}(z) & \cdots\\
 & \vdots\\
\cdots & \mathfrak{P}_{s,\bullet}(z) & \cdots
\end{pmatrix}\left(\zeta_{t-J_{1}}-u_{t-J_{1}}\right)\right]\right\} =0.
\]
Next, we permutate the rows in this equation in order to treat the
rows in $\begin{pmatrix}\cdots & \mathfrak{P}_{1,\bullet}(z) & \cdots\\
 & \vdots\\
\cdots & \mathfrak{P}_{s,\bullet}(z) & \cdots
\end{pmatrix}$ separately. To this end, we pre-multiply
\begin{equation}
U=\left(\begin{array}{cccc|c|cccc}
1 & 0 & \cdots & 0 & \\
 &  &  &  & \ddots\\
 &  &  &  &  & 1 & 0 & \cdots & 0\\
\hline  &  &  &  & \vdots\\
\hline 0 & \cdots & 0 & 1 & \\
 &  &  &  & \ddots\\
 &  &  &  &  & 0 & \cdots & 0 & 1
\end{array}\right)\in\mathbb{R}^{sH\times sH}\label{eq:permutation_matrix}
\end{equation}
 in order to extract the $j$-th component in each of the $H$ blocks
with $s$ rows. Eventually, we will analyse each $H\times sH$ dimensional
block in
\begin{equation}
\left(\mathbb{E}_{t}-\mathbb{E}_{t-1}\right)\left\{ \begin{pmatrix}I_{H}\otimes\left(z^{-g_{1}}\mathfrak{P}_{1,\bullet}(z)\right)\\
\vdots\\
I_{H}\otimes\left(z^{-g_{s}}\mathfrak{P}_{s,\bullet}(z)\right)
\end{pmatrix}\left[\begin{pmatrix}1\\
z^{-1}\\
\vdots\\
z^{-H+1}
\end{pmatrix}\otimes\left(\zeta_{t-J_{1}}-u_{t-J_{1}}\right)\right]\right\} =0\label{eq:main_proof_permutated_restrictions}
\end{equation}
and derive explicit constraints (in the guise of the matrices $C$
and $D$ in equation (\ref{eq:general_thm_constraints})) on the MDS
$\left(\varepsilon_{t}^{i}\right)_{t\in\mathbb{Z}},\,i\in\left\{ 0,\ldots,H-1\right\} ,$
by writing $\zeta_{t}=-\sum_{k=0}^{K}\sum_{j=0}^{H-1}\sum_{h=0}^{j}A_{kh}z^{k+(j-h)}\varepsilon_{t}^{j}$
in an intelligent way. 

\paragraph{Step 2: Rewrite $\zeta_{t}=-\sum_{k=0}^{K}\sum_{j=0}^{H-1}\sum_{h=0}^{j}A_{kh}z^{k+(j-h)}\varepsilon_{t}^{j}$
as $\zeta_{t}=\sum_{i=0}^{H+K-1}m_{i,\bullet}\varepsilon_{t-i}^{\bullet}$.}

We denote the $sH$-dimensional vector $\begin{pmatrix}\varepsilon_{t}^{0}\\
\varepsilon_{t}^{1}\\
\vdots\\
\varepsilon_{t}^{H-1}
\end{pmatrix}$ by $\varepsilon_{t}^{\bullet}$ and distinguish the cases $K\leq H-1$
and $K>H-1$. In the case $K\leq H-1$, we obtain (where $\Delta_{1}=\left(H-1\right)-K${\footnotesize{}
\begin{align}
\zeta_{t} & =-\sum_{k=0}^{K}\sum_{j=0}^{H-1}\sum_{h=0}^{j}A_{kh}z^{k+(j-h)}\varepsilon_{t}^{j}\label{eq:zeta_K_kleiner_H}\\
\hline  & =-\left(A_{00},A_{01},\ldots,A_{0,H-1}\right)\varepsilon_{t}^{\bullet}\nonumber \\
 & \quad-\begin{pmatrix}I_{s} & I_{s}\end{pmatrix}\begin{pmatrix}0 & A_{00} & A_{01} & \cdots & A_{0,H-2}\\
A_{10} & A_{11} & \cdots & A_{1,H-2} & A_{1,H-1}
\end{pmatrix}\varepsilon_{t-1}^{\bullet}\nonumber \\
 & \quad-\begin{pmatrix}I_{s} & I_{s} & I_{s}\end{pmatrix}\begin{pmatrix}0 & 0 & A_{00} & A_{01} & \cdots & A_{0,H-3}\\
0 & A_{10} & A_{11} & \cdots & A_{1,H-3} & A_{1,H-2}\\
A_{20} & A_{21} & \cdots &  & A_{2,H-2} & A_{2,H-1}
\end{pmatrix}\varepsilon_{t-2}^{\bullet}\nonumber \\
 & \vdots\nonumber \\
 & \quad-\underbrace{\begin{pmatrix}I_{s} & \cdots & I_{s}\end{pmatrix}}_{K\text{ elements}}\left(\begin{array}{ccc|ccc}
0 &  & A_{00} & A_{01} & \cdots & A_{0,\Delta_{1}+1}\\
 & \iddots & \vdots & \vdots &  & \vdots\\
A_{K-1,0} & \cdots & A_{K-1,K-1} & A_{K-1,K} & \cdots & A_{K-1,H-1}
\end{array}\right)\varepsilon_{t-(K-1)}^{\bullet}\nonumber \\
\hline  & \quad-\underbrace{\begin{pmatrix}I_{s} & \cdots & I_{s}\end{pmatrix}}_{(K+1)\text{ elements}}\left(\begin{array}{ccc|ccc}
0 &  & A_{00} & A_{01} & \cdots & A_{0,\Delta_{1}}\\
 & \iddots & \vdots & \vdots &  & \vdots\\
A_{K,0} & \cdots & A_{K,K} & A_{K,K+1} & \cdots & A_{K,H-1}
\end{array}\right)\varepsilon_{t-K}^{\bullet}\nonumber \\
 & \vdots\nonumber \\
 & \quad-\underbrace{\begin{pmatrix}I_{s} & \cdots & I_{s}\end{pmatrix}}_{(K+1)\text{ elements}}\left(\begin{array}{ccc|ccc|ccc}
0 & \cdots & 0 & 0 & 0 & A_{00} & A_{01} & \cdots & A_{0,\Delta_{1}-r}\\
\vdots &  & \vdots & \iddots & \iddots & \vdots & \vdots &  & \vdots\\
0 & \cdots & 0 & A_{K,0} & \cdots & A_{K,K} & A_{K,K+1} & \cdots & A_{K,K+\Delta_{1}-r}
\end{array}\right)\varepsilon_{t-(K+r)}^{\bullet}\nonumber \\
 & \vdots\nonumber \\
 & \quad-\underbrace{\begin{pmatrix}I_{s} & \cdots & I_{s}\end{pmatrix}}_{(K+1)\text{ elements}}\left(\begin{array}{ccc|ccc}
0 & \cdots & 0 & 0 &  & A_{00}\\
\vdots &  & \vdots &  & \iddots & \vdots\\
0 & \cdots & 0 & A_{K,0} & \cdots & A_{K,K}
\end{array}\right)\varepsilon_{t-(K+\Delta_{1})}^{\bullet}\nonumber \\
\hline  & \quad-\underbrace{\begin{pmatrix}I_{s} & \cdots & I_{s}\end{pmatrix}}_{K\text{ elements}}\left(\begin{array}{c|ccc|ccc}
0 & 0 & \cdots & 0 & 0 &  & A_{10}\\
\vdots & \vdots &  & \vdots &  & \iddots & \vdots\\
0 & 0 & \cdots & 0 & A_{K,0} & \cdots & A_{K,K-1}
\end{array}\right)\varepsilon_{t-(H-1)-1}^{\bullet}\nonumber \\
 & \vdots\nonumber \\
 & \quad-\underbrace{\begin{pmatrix}I_{s} & \cdots & I_{s}\end{pmatrix}}_{(K+1-r)\text{ elements}}\left(\begin{array}{ccc|ccc|ccc}
0 & \cdots & 0 & 0 & \cdots & 0 & 0 &  & A_{r0}\\
\vdots &  & \vdots & \vdots &  & \vdots &  & \iddots & \vdots\\
0 & \cdots & 0 & 0 & \cdots & 0 & A_{K,0} & \cdots & A_{K,K-r}
\end{array}\right)\varepsilon_{t-(H-1)-r}^{\bullet}\nonumber \\
 & \vdots\nonumber \\
 & \quad-\left(\begin{array}{ccc|ccc|c}
0 & \cdots & 0 & 0 & \cdots & 0 & A_{K,0}\end{array}\right)\varepsilon_{t-(H-1)-K}^{\bullet}\nonumber 
\end{align}
}In the case $K>H-1$, we obtain (where $\Delta_{2}=K-\left(H-1\right)$){\footnotesize{}
\begin{align}
\zeta_{t} & =-\sum_{k=0}^{K}\sum_{j=0}^{H-1}\sum_{h=0}^{j}A_{kh}z^{k+(j-h)}\varepsilon_{t}^{j}\label{eq:zeta_K_groesser_H}\\
\hline  & =-\left(A_{00},A_{01},\ldots,A_{0,H-1}\right)\varepsilon_{t}^{\bullet}\nonumber \\
 & \quad-\begin{pmatrix}I_{s} & I_{s}\end{pmatrix}\begin{pmatrix}0 & A_{00} & A_{01} & \cdots & A_{0,H-2}\\
A_{10} & A_{11} & \cdots & A_{1,H-2} & A_{1,H-1}
\end{pmatrix}\varepsilon_{t-1}^{\bullet}\nonumber \\
 & \vdots\nonumber \\
 & \quad-\underbrace{\begin{pmatrix}I_{s} & \cdots & I_{s}\end{pmatrix}}_{(r+1)\text{ elements}}\left(\begin{array}{ccc|ccc}
0 &  & A_{00} & A_{01} & \cdots & A_{0,H-1-r}\\
 & \iddots & \vdots & \vdots &  & \vdots\\
A_{r,0} & \cdots & A_{r,r} & A_{r,r+1} & \cdots & A_{r,H-1}
\end{array}\right)\varepsilon_{t-r}^{\bullet}\nonumber \\
 & \vdots\nonumber \\
 & \quad-\underbrace{\begin{pmatrix}I_{s} & \cdots & I_{s}\end{pmatrix}}_{(H-1)\text{ elements}}\left(\begin{array}{ccc|c}
0 &  & A_{00} & A_{01}\\
 & \iddots & \vdots & \vdots\\
A_{H-2,0} & \cdots & A_{H-2,H-2} & A_{H-2,H-1}
\end{array}\right)\varepsilon_{t-(H-1)+1}^{\bullet}\nonumber \\
\hline  & \quad-\underbrace{\begin{pmatrix}I_{s} & \cdots & I_{s}\end{pmatrix}}_{H\text{ elements}}\left(\begin{array}{ccc}
0 &  & A_{00}\\
 & \iddots & \vdots\\
A_{H-1,0} & \cdots & A_{H-1,H-1}
\end{array}\right)\varepsilon_{t-(H-1)}^{\bullet}\nonumber \\
 & \vdots\nonumber \\
 & \quad-\underbrace{\begin{pmatrix}I_{s} & \cdots & I_{s}\end{pmatrix}}_{H\text{ elements}}\left(\begin{array}{ccc}
0 &  & A_{r0}\\
 & \iddots & \vdots\\
A_{H-1+r,0} & \cdots & A_{H-1+r,H-1}
\end{array}\right)\varepsilon_{t-(H-1)-r}^{\bullet}\nonumber \\
 & \vdots\nonumber \\
 & \quad-\underbrace{\begin{pmatrix}I_{s} & \cdots & I_{s}\end{pmatrix}}_{H\text{ elements}}\left(\begin{array}{ccc}
0 &  & A_{\Delta_{2}0}\\
 & \iddots & \vdots\\
A_{K,0} & \cdots & A_{K,H-1}
\end{array}\right)\varepsilon_{t-(H-1)+\Delta}^{\bullet}\nonumber \\
\hline  & \quad-\underbrace{\begin{pmatrix}I_{s} & \cdots & I_{s}\end{pmatrix}}_{(H-1)K\text{ elements}}\left(\begin{array}{c|ccc}
0 & 0 &  & A_{\Delta_{2}+1,0}\\
\vdots &  & \iddots & \vdots\\
0 & A_{K,0} & \cdots & A_{K,\left(H-1\right)-1}
\end{array}\right)\varepsilon_{t-K-1}^{\bullet}\nonumber \\
 & \vdots\nonumber \\
 & \quad-\underbrace{\begin{pmatrix}I_{s} & \cdots & I_{s}\end{pmatrix}}_{(H-r)\text{ elements}}\left(\begin{array}{ccc|ccc}
0 & \cdots & 0 & 0 &  & A_{\Delta_{2}+r,0}\\
\vdots &  & \vdots &  & \iddots & \vdots\\
0 & \cdots & 0 & A_{K,0} & \cdots & A_{K,\left(H-1\right)-r}
\end{array}\right)\varepsilon_{t-K-r}^{\bullet}\nonumber \\
 & \vdots\nonumber \\
 & \quad-\left(\begin{array}{ccc|c}
0 & \cdots & 0 & A_{K,0}\end{array}\right)\varepsilon_{t-K-(H-1)}^{\bullet}.\nonumber 
\end{align}
}In summary, we may write in the case $K\leq H-1,\ \Delta_{1}=H-1-K,$
that{\small{}
\[
\zeta_{t}=\underbrace{\begin{pmatrix}I_{s} & \cdots & I_{s}\end{pmatrix}}_{(H+K)\text{ elements}}\begin{pmatrix}\left(m_{0,\bullet}\right)\\
\left(m_{1,\bullet}\right)z\\
\vdots\\
\left(m_{K-1,\bullet}\right)z^{K-1}\\
\hline \left(m_{K,\bullet}\right)z^{K}\\
\vdots\\
\left(m_{K+\Delta_{1},\bullet}\right)z^{H-1}\\
\hline \left(m_{H,\bullet}\right)z^{H}\\
\vdots\\
\left(m_{H+K-1,\bullet}\right)z^{H+K-1}
\end{pmatrix}\varepsilon_{t}^{\bullet}=\begin{pmatrix}I_{s} & \cdots & I_{s}\end{pmatrix}\begin{pmatrix}\left(m_{0,\bullet}\right)\\
\left(m_{1,\bullet}\right)z\\
\vdots\\
\left(m_{K-1,\bullet}\right)z^{K-1}\\
\hline \left(\begin{array}{ccc|ccc}
A_{-K}^{*} & \cdots & A_{0}^{*} & A_{1}^{*} & \cdots & A_{\Delta_{1}}^{*}\end{array}\right)z^{K}\\
\left(\begin{array}{cccc}
0_{s} & A_{-K}^{*} & \cdots & A_{\Delta_{1}-1}^{*}\end{array}\right)z^{K+1}\\
\vdots\\
\left(\begin{array}{ccc|ccc}
0_{s} & \cdots & 0_{s} & A_{-K}^{*} & \cdots & A_{0}^{*}\end{array}\right)z^{H-1}\\
\hline \left(\begin{array}{c|c|ccc}
0_{s\times s\Delta_{1}} & 0_{s} & A_{-K}^{*} & \cdots & A_{-1}^{*}\end{array}\right)z^{H}\\
\vdots\\
\left(\begin{array}{cc}
0_{s\times s(H-1)} & A_{-K}^{*}\end{array}\right)z^{H-1+K}
\end{pmatrix}\varepsilon_{t}^{\bullet}
\]
}where $m_{i,\bullet}\in\mathbb{R}^{s\times sH}$ denotes the coefficient
matrix pertaining to the $i$-th lag of $\left(\varepsilon_{t}^{\bullet}\right)_{t\in\mathbb{Z}}$.
Note that $m_{i,\bullet}=0$ for\footnote{We repeat that $J_{0}=argmin_{i}\left\{ i\ |\ A_{i}^{*}\neq0\right\} $,
compare equation (\ref{eq:recursive}) on page \pageref{eq:recursive}. } $i\geq H-J_{0}$. In the case $K>H-1$ (where $\Delta_{2}=K-\left(H-1\right)$)
we obtain{\small{}
\[
\zeta_{t}=\underbrace{\begin{pmatrix}I_{s} & \cdots & I_{s}\end{pmatrix}}_{(H+K)\text{ elements}}\begin{pmatrix}\left(m_{0,\bullet}\right)\\
\vdots\\
\left(m_{H-2,\bullet}\right)z^{H-2}\\
\hline \left(m_{H-1,\bullet}\right)z^{H-1}\\
\vdots\\
\left(m_{H-1+\Delta_{2},\bullet}\right)z^{K-1}\\
\hline \left(m_{K,\bullet}\right)z^{K}\\
\vdots\\
\left(m_{K+H-1,\bullet}\right)z^{K+H-1}
\end{pmatrix}\varepsilon_{t}^{\bullet}=\begin{pmatrix}I_{s} & \cdots & I_{s}\end{pmatrix}\begin{pmatrix}\left(m_{0,\bullet}\right)\\
\vdots\\
\left(m_{H-2,\bullet}\right)z^{H-2}\\
\hline \left(m_{H-1,\bullet}\right)z^{H-1}\\
\vdots\\
\left(m_{H-1+\Delta_{2},\bullet}\right)z^{K-1}\\
\hline \left(\begin{array}{ccc}
A_{-K}^{*} & \cdots & A_{-\Delta_{2}}^{*}\end{array}\right)z^{K}\\
\left(\begin{array}{c|ccc}
0_{s} & A_{-K}^{*} & \cdots & A_{-\Delta_{2}}^{*}\end{array}\right)z^{K+1}\\
\vdots\\
\left(\begin{array}{cc}
0_{s\times s(H-1)} & A_{-K}^{*}\end{array}\right)z^{K+H-1}
\end{pmatrix}\varepsilon_{t}^{\bullet}
\]
}Similarly, note that if $-J_{0}\geq\Delta_{2}$, then $m_{i,\bullet}=0$
for $i\geq H-J_{0}$.

\paragraph{Step 3: Analyse equation (\ref{eq:main_proof_permutated_restrictions})
on page \pageref{eq:main_proof_permutated_restrictions}.}

We will now focus on the $k$-th $\left(H\times sH\right)$-dimensional
block in equation (\ref{eq:main_proof_permutated_restrictions}),
i.e. 
\[
\left(\mathbb{E}_{t}-\mathbb{E}_{t-1}\right)\left\{ \left[I_{H}\otimes\left(z^{-g_{k}}\mathfrak{P}_{k,\bullet}(z)\right)\right]\left[\begin{pmatrix}1\\
z^{-1}\\
\vdots\\
z^{-H+1}
\end{pmatrix}\otimes\left(\zeta_{t-J_{1}}-u_{t-J_{1}}\right)\right]\right\} =0.
\]
Without loss of generality we assume that $0\leq g_{1}\leq\cdots\leq g_{j}\leq J_{1}<g_{j+1}\leq\cdots\leq g_{s}$
and distinguish the cases $k\leq j$ and $k>j$. We will show that
the row rank of $C$ is bounded my $s(H-J_{1})-\sum_{k=1}^{j}g_{k}$.

\subparagraph{Step 3.a: Case $k\protect\leq j$.}

Defining $\delta_{k}=J_{1}-g_{k}\geq0$ and rewriting the equation
above as 
\[
\left(\mathbb{E}_{t}-\mathbb{E}_{t-1}\right)\left\{ \left[I_{H}\otimes\left(\mathfrak{P}_{k,\bullet}(z)\right)\right]\begin{pmatrix}\zeta_{t-\delta_{k}}-u_{t-\delta_{k}}\\
\vdots\\
\zeta_{t-1}-u_{t-1}\\
\hline \zeta_{t}-u_{t}\\
\vdots\\
\zeta_{t+H-\delta_{k}-1}-u_{t+H-\delta_{k}-1}
\end{pmatrix}\right\} =0,
\]
we obtain that 
\begin{equation}
\left(\begin{array}{c|c}
0_{\delta_{k}\times s\left(H-\delta_{k}\right)} & 0_{\delta_{k}\times s\delta_{k}}\\
\hline \begin{array}{ccc}
\mathfrak{P}_{k,\bullet|0}\\
\vdots & \ddots\\
\mathfrak{P}_{k,\bullet|H-1-\delta_{k}} & \cdots & \mathfrak{P}_{k,\bullet|0}
\end{array} & 0_{\left(H-\delta_{k}\right)\times s\delta_{k}}
\end{array}\right)\left[\begin{pmatrix}m_{0,\bullet}\\
\vdots\\
m_{H-1,\bullet}
\end{pmatrix}\varepsilon_{t}^{\bullet}-\left(\mathbb{E}_{t}-\mathbb{E}_{t-1}\right)\begin{pmatrix}u_{t}\\
\vdots\\
u_{t+(H-1)}
\end{pmatrix}\right]=0.\label{eq:main_proof_k_smaller_j}
\end{equation}
 This follows from considering, e.g., 
\begin{gather*}
\left(\mathfrak{P}_{k,\bullet}(z)\right)\zeta_{t-\delta_{k}+l}=\cdots\\
\cdots=\begin{pmatrix}\mathfrak{P}_{k,\bullet|0} & \cdots & \mathfrak{P}_{k,\bullet|N}\end{pmatrix}z^{\delta_{k}-l}\left[\begin{pmatrix}1 & z & z^{2} & \cdots & z^{H+K-1}\\
z & z^{2} & z^{3} & \cdots & z^{H+K}\\
\vdots &  &  &  & \vdots\\
z^{N} & z^{N+1} & z^{N+2} & \cdots & z^{N+H+K-1}
\end{pmatrix}\otimes I_{s}\right]\begin{pmatrix}m_{0,\bullet}\\
\vdots\\
m_{H+K-1,\bullet}
\end{pmatrix},
\end{gather*}
where $l\in\left\{ 0,\ldots,H-1\right\} $ corresponds to the respective
row of the matrix in equation (\ref{eq:main_proof_k_smaller_j}).
Obviously, only the elements pertaining to power zero of $z$ are
relevant when taking conditional expectations $\left[\mathbb{E}_{t}-\mathbb{E}_{t-1}\right]$.
In total, the rank deficiency of the part for which $g_{k}\leq J_{1}$
of the matrix obtained from equation (\ref{eq:main_proof_permutated_restrictions})
via (\ref{eq:main_proof_k_smaller_j}) is $\sum_{k=0}^{j}\delta_{k}$
(or equivalently the rank is $jH-\sum_{k=0}^{j}\delta_{k}$).

\subparagraph{Step 3.b: Case $k>j$.}

Similarly, defining $\gamma_{k}=g_{k}-J_{1}>0$ and rewriting the
equation above as 
\[
\left(\mathbb{E}_{t}-\mathbb{E}_{t-1}\right)\left[\left(I_{H}\otimes\left(\mathfrak{P}_{k,\bullet}(z)\right)\right)\begin{pmatrix}\zeta_{t-J_{1}+g_{k}}-u_{t-J_{1}+g_{k}}\\
\zeta_{t+\gamma_{k}+1}-u_{t+\gamma_{k}+1}\\
\vdots\\
\zeta_{t+\gamma_{k}+H-1}-u_{t+\gamma_{k}+H-1}
\end{pmatrix}\right]=0,
\]
we obtain that {\small{}
\[
\left(\begin{array}{ccc|ccc}
\mathfrak{P}_{k,\bullet|\gamma_{k}} & \cdots & \mathfrak{P}_{k,\bullet|1} & \mathfrak{P}_{k,\bullet|0}\\
\vdots &  & \vdots & \vdots & \ddots\\
\mathfrak{P}_{k,\bullet|\gamma_{k}+H-1} & \cdots & \mathfrak{P}_{k,\bullet|H} & \mathfrak{P}_{k,\bullet|H-1} & \cdots & \mathfrak{P}_{k,\bullet|0}
\end{array}\right)\left[\begin{pmatrix}m_{0,\bullet}\\
\vdots\\
m_{H-1+\gamma_{k},\bullet}
\end{pmatrix}\varepsilon_{t}^{\bullet}-\left(\mathbb{E}_{t}-\mathbb{E}_{t-1}\right)\begin{pmatrix}u_{t}\\
\vdots\\
u_{t+(H+\gamma_{k}-1)}
\end{pmatrix}\right]=0.
\]
}Although the row rank of the first matrix in the equation above is
always equal to $H$, we cannot conclude on the rank of the product
of the first matrix and $\begin{pmatrix}m_{0,\bullet}\\
\vdots\\
m_{H-1+\gamma_{k},\bullet}
\end{pmatrix}$ without making further assumptions.

\paragraph{Step 4: Obtain equation (\ref{eq:general_thm_constraints}).}

Stacking the matrices 
\[
\left(\begin{array}{c|c|c}
0_{\delta_{k}\times s\left(H-\delta_{k}\right)} & 0_{\delta_{k}\times s\delta_{k}} & 0_{\delta_{k}\times s\gamma_{k}}\\
\hline \begin{array}{ccc}
\mathfrak{P}_{k,\bullet|0}\\
\vdots & \ddots\\
\mathfrak{P}_{k,\bullet|H-1-\delta_{k}} & \cdots & \mathfrak{P}_{k,\bullet|0}
\end{array} & 0_{\left(H-\delta_{k}\right)\times s\delta_{k}} & 0_{\left(H-\delta_{k}\right)\times s\gamma_{k}}
\end{array}\right),\ k\in\left\{ 1,\ldots,j\right\} 
\]
and 
\[
\left(\begin{array}{ccc|ccc|c}
\mathfrak{P}_{k,\bullet|\gamma_{k}} & \cdots & \mathfrak{P}_{k,\bullet|1} & \mathfrak{P}_{k,\bullet|0} &  & \\
\vdots &  & \vdots & \vdots & \ddots &  & 0_{H\times\left(\gamma_{s}-\gamma_{k}\right)}\\
\mathfrak{P}_{k,\bullet|\gamma_{k}+H-1} & \cdots & \mathfrak{P}_{k,\bullet|H} & \mathfrak{P}_{k,\bullet|H-1} & \cdots & \mathfrak{P}_{k,\bullet|0}
\end{array}\right),\ k\in\left\{ j+1,\ldots,s\right\} 
\]
above each other, adding additional zero blocks to each such that
there are a total of $\gamma_{s}+H$ blocks, and denoting them by
$\mathfrak{P}^{k,\bullet}\in\mathbb{R}^{H\times s\left(H+\gamma_{k}\right)}$,
we obtain that 
\begin{equation}
\underbrace{\begin{pmatrix}\mathfrak{P}^{1,\bullet}\\
\vdots\\
\mathfrak{P}^{s,\bullet}
\end{pmatrix}\begin{pmatrix}m_{0,\bullet}\\
\vdots\\
m_{H-1+\gamma_{s},\bullet}
\end{pmatrix}}_{=C}\varepsilon_{t}^{\bullet}=\underbrace{\begin{pmatrix}\mathfrak{P}^{1,\bullet}\\
\vdots\\
\mathfrak{P}^{s,\bullet}
\end{pmatrix}}_{=D}\left(\mathbb{E}_{t}-\mathbb{E}_{t-1}\right)\begin{pmatrix}u_{t}\\
\vdots\\
u_{t+(H+\gamma_{s}-1)}
\end{pmatrix}.\label{eq:general_restrictions_in_detail}
\end{equation}

\paragraph{Step 5: Given that the rank of the matrix $C$ in equation (\ref{eq:general_thm_constraints})
is $w$, conclude that the $\left(Hs-w\right)q$ free parameters in
the SDE (\ref{eq:recursive_eq_in_dimension_thm}) generate distinct
causal solution (under the assumption that all non-zero zeros of $\pi(z)$
lie outside the unit circle).}

We start from equation 
\[
\pi(z)y_{t}=\pi(z)\left(\varepsilon_{t}^{0}+\varepsilon_{t-1}^{1}+\cdots+\varepsilon_{t-H+1}^{H-1}\right)+\zeta_{t-J_{1}}-u_{t-J_{1}}
\]
where the $H$ MDS of dimension $s$ satisfy the constraint\footnote{Note that this equation does not necessarily have a solution.}
(\ref{eq:general_restrictions_in_detail}), where $C\in\mathbb{R}^{sH\times sH}$
has rank $w$. 

Since the $\varepsilon_{t}^{j}$ are MDS with respect to $H_{\varepsilon}(t)$,
we may write them as $\varepsilon_{t}^{j}=h_{j}\varepsilon_{t},\ h_{j}\in\mathbb{R}^{s\times q}$.
Moreover, the Wold representation of $u_{t}$ is given as $u_{t}=\sum_{j=0}^{\infty}w_{j}\varepsilon_{t-j}=w(z)\varepsilon_{t}$
such that the SDE and the constraints are
\[
\pi(z)y_{t}=\pi(z)\left(h_{0}\varepsilon_{t}+h_{1}\varepsilon_{t-1}+\cdots+h_{H-1}\varepsilon_{t-H+1}\right)+\zeta_{t-J_{1}}-u_{t-J_{1}}
\]
and
\begin{equation}
C\begin{pmatrix}h_{0}\\
\vdots\\
h_{H-1}
\end{pmatrix}=D\begin{pmatrix}w_{0}\\
\vdots\\
w_{H-1}
\end{pmatrix}.\label{eq:constraints_for_identifiability_of_solutions-1}
\end{equation}

Pre-multiplying $\pi(z)^{-1}$ on the SDE, we obtain that the first
$H-1$ lags, i.e. the first $H$ coefficients when counting the contemporaneous
innovation as well, of the transfer function of $y_{t}$ are of the
form 
\[
h_{0}\varepsilon_{t}+h_{1}\varepsilon_{t-1}+\cdots+h_{H-1}\varepsilon_{t-\left(H-1\right)}
\]
because of the constraints (\ref{eq:constraints_for_identifiability_of_solutions-1}).
Since there are $\left(Hs-w\right)q$ free parameter in (\ref{eq:constraints_for_identifiability_of_solutions-1}),
it follows that each solution of this equation system corresponds
to a different transfer function.

\pagebreak{}

\section{Proof of Theorem \ref{thm:main_thm_A0_nonsingular}}

\paragraph{Step 1: Bound the rank of $C$ from below.}

Starting from equation (\ref{eq:general_restrictions_in_detail}),
here reproduced as 
\[
\begin{pmatrix}\mathfrak{P}^{1,\bullet}\\
\vdots\\
\mathfrak{P}^{s,\bullet}
\end{pmatrix}\begin{pmatrix}m_{0,\bullet}\\
\vdots\\
m_{H-1+\gamma_{s},\bullet}
\end{pmatrix}\varepsilon_{t}^{\bullet}=\begin{pmatrix}\mathfrak{P}^{1,\bullet}\\
\vdots\\
\mathfrak{P}^{s,\bullet}
\end{pmatrix}\left(\mathbb{E}_{t}-\mathbb{E}_{t-1}\right)\begin{pmatrix}u_{t}\\
\vdots\\
u_{t+(H+\gamma_{s}-1)}
\end{pmatrix},
\]
 and assuming w.l.o.g. that $0\leq g_{1}\leq\cdots\leq g_{j}\leq J_{1}<g_{j+1}\leq\cdots\leq g_{s}$,
we can conclude for $k\leq j$ that 
\[
\left(\begin{array}{c|c}
0_{\delta_{k}\times s\left(H-\delta_{k}\right)} & 0_{\delta_{k}\times s\delta_{k}}\\
\hline \begin{array}{ccc}
\mathfrak{P}_{k,\bullet|0}\\
\vdots & \ddots\\
\mathfrak{P}_{k,\bullet|H-1-\delta_{k}} & \cdots & \mathfrak{P}_{k,\bullet|0}
\end{array} & 0_{\left(H-\delta_{k}\right)\times s\delta_{k}}
\end{array}\right)\begin{pmatrix}m_{0,\bullet}\\
\vdots\\
m_{H-1,\bullet}
\end{pmatrix}
\]
has row rank $\left(H-\delta_{k}\right)$ because we assumed that
$\begin{pmatrix}m_{0,\bullet}\\
\vdots\\
m_{H-1,\bullet}
\end{pmatrix}\in\mathbb{R}^{sH\times sH}$ is non-singular. For $k>j$ and if $H>\gamma_{k}$, one can see in
\[
\left(\begin{array}{ccc|ccc|cc|c}
\mathfrak{P}_{k,\bullet|\gamma_{k}} & \cdots & \mathfrak{P}_{k,\bullet|1} & \mathfrak{P}_{k,\bullet|0} &  &  &  &  & 0_{1\times s\left(\gamma_{s}-\gamma_{k}\right)}\\
\vdots &  & \vdots & \vdots & \ddots &  &  &  & \vdots\\
 &  &  &  &  & \mathfrak{P}_{k,\bullet|0} &  & \\
\vdots &  & \vdots & \vdots &  & \vdots & \ddots &  & \vdots\\
\mathfrak{P}_{k,\bullet|\gamma_{k}+H-1} & \cdots & \mathfrak{P}_{k,\bullet|H} & \mathfrak{P}_{k,\bullet|H-1} & \cdots & \mathfrak{P}_{k,\bullet|\gamma_{k}} & \cdots & \mathfrak{P}_{k,\bullet|0} & 0_{1\times s\left(\gamma_{s}-\gamma_{k}\right)}
\end{array}\right)\begin{pmatrix}m_{0,\bullet}\\
\vdots\\
m_{\gamma_{k}-1,\bullet}\\
\hline m_{\gamma_{k},\bullet}\\
\vdots\\
m_{H-1,\bullet}\\
\hline m_{H,\bullet}\\
\vdots\\
m_{H+\gamma_{s}-1,\bullet}
\end{pmatrix}
\]
that the rank of this matrix product is at least $\left(H-\gamma_{k}\right)$
because the second block of the first matrix has at least row rank
$\left(H-\gamma_{k}\right)$ and $\begin{pmatrix}m_{\gamma_{k},\bullet}\\
\vdots\\
m_{H-1,\bullet}
\end{pmatrix}$ is of full row rank.

\paragraph{Step 2: For $g_{i}\protect\leq J_{1}$, the matrix $C$ has generically
has row rank $\left(H-J_{1}\right)s+\sum_{i=1}^{s}g_{i}$. }

From above we know that 
\[
\left(\begin{array}{c|c}
0_{\delta_{k}\times s\left(H-\delta_{k}\right)} & 0_{\delta_{k}\times s\delta_{k}}\\
\hline \begin{array}{ccc}
\mathfrak{P}_{k,\bullet|0}\\
\vdots & \ddots\\
\mathfrak{P}_{k,\bullet|H-1-\delta_{k}} & \cdots & \mathfrak{P}_{k,\bullet|0}
\end{array} & 0_{\left(H-\delta_{k}\right)\times s\delta_{k}}
\end{array}\right)\begin{pmatrix}m_{0,\bullet}\\
\vdots\\
m_{H-1,\bullet}
\end{pmatrix}
\]
has row rank $\left(H-J_{1}+g_{k}\right)$. Summing across all components
gives $\sum_{k=1}^{s}\left(H-J_{1}+g_{k}\right)=\left(H-J_{1}\right)s+\sum_{i=1}^{s}g_{i}$.

\paragraph{Step 3: For $g_{i}=0$, equation (\ref{eq:general_thm_constraints})
can be simplified.}

We proceed again from equation (\ref{eq:general_restrictions_in_detail}),
i.e. 
\[
\begin{pmatrix}\mathfrak{P}^{1,\bullet}\\
\vdots\\
\mathfrak{P}^{s,\bullet}
\end{pmatrix}\begin{pmatrix}m_{0,\bullet}\\
\vdots\\
m_{H-1+\gamma_{s},\bullet}
\end{pmatrix}\varepsilon_{t}^{\bullet}=\begin{pmatrix}\mathfrak{P}^{1,\bullet}\\
\vdots\\
\mathfrak{P}^{s,\bullet}
\end{pmatrix}\left(\mathbb{E}_{t}-\mathbb{E}_{t-1}\right)\begin{pmatrix}u_{t}\\
\vdots\\
u_{t+(H+\gamma_{s}-1)}
\end{pmatrix},
\]
where all $\mathfrak{P}^{k,\bullet}$ have the form 
\[
\left(\begin{array}{ccc}
\mathfrak{P}_{k,\bullet|0}\\
\vdots & \ddots\\
\mathfrak{P}_{k,\bullet|H-J_{1}-1} & \cdots & \mathfrak{P}_{k,\bullet|0}
\end{array}\right)\in\mathbb{R}^{\left(H-J_{1}\right)\times s\left(H-J_{1}\right)}.
\]
Thus, equation (\ref{eq:general_restrictions_in_detail}) is simplified
to 
\[
\begin{pmatrix}\mathfrak{P}^{1,\bullet}\\
\vdots\\
\mathfrak{P}^{s,\bullet}
\end{pmatrix}\begin{pmatrix}m_{0,\bullet}\\
\vdots\\
m_{H-J_{1}-1,\bullet}
\end{pmatrix}\varepsilon_{t}^{\bullet}=\begin{pmatrix}\mathfrak{P}^{1,\bullet}\\
\vdots\\
\mathfrak{P}^{s,\bullet}
\end{pmatrix}\left(\mathbb{E}_{t}-\mathbb{E}_{t-1}\right)\begin{pmatrix}u_{t}\\
\vdots\\
u_{t+(H-J_{1}-1)}
\end{pmatrix}.
\]
Furthermore, we pre-multiply the transpose of the permutation matrix
(\ref{eq:permutation_matrix}) on page \pageref{par:main_proof_step1_permute}
from step 1 in the proof of Theorem \ref{thm:general-solution-set}
, i.e. 
\[
U^{T}=\left(\begin{array}{cccc|c|cccc}
1 & 0 & \cdots & 0 & \\
 &  &  &  & \ddots\\
 &  &  &  &  & 1 & 0 & \cdots & 0\\
\hline  &  &  &  & \vdots\\
\hline 0 & \cdots & 0 & 1 & \\
 &  &  &  & \ddots\\
 &  &  &  &  & 0 & \cdots & 0 & 1
\end{array}\right)^{T},
\]
and obtain{\small{}
\[
\left(\begin{array}{ccc}
\mathfrak{P}_{\bullet,\bullet|0}\\
\vdots & \ddots\\
\mathfrak{P}_{\bullet,\bullet|H-J_{1}-1} & \cdots & \mathfrak{P}_{\bullet,\bullet|0}
\end{array}\right)\begin{pmatrix}m_{0,\bullet}\\
\vdots\\
m_{H-J_{1}-1,\bullet}
\end{pmatrix}\varepsilon_{t}^{\bullet}=\left(\begin{array}{ccc}
\mathfrak{P}_{\bullet,\bullet|0}\\
\vdots & \ddots\\
\mathfrak{P}_{\bullet,\bullet|H-J_{1}-1} & \cdots & \mathfrak{P}_{\bullet,\bullet|0}
\end{array}\right)\left(\mathbb{E}_{t}-\mathbb{E}_{t-1}\right)\begin{pmatrix}u_{t}\\
\vdots\\
u_{t+(H-J_{1}-1)}
\end{pmatrix}.
\]
}Since $P(z)$ and thus $P(z)^{-1}$ are unimodular it follows that
they are non-singular\footnote{Note that this does not follow from a genericity argument as stated
in \citep{BGS95} on page 255 after the first equation system.} for every $z\in\mathbb{C}$. Therefore, $\mathfrak{P}_{\bullet,\bullet|0}\in\mathbb{R}^{s\times s}$
is of full rank and we obtain the last result of Theorem \ref{thm:main_thm_A0_nonsingular}.

\pagebreak{}

\section{Proof of Theorem \ref{thm:predetermined_equivalence_of_solutions}}

\paragraph{Step 1: Take conditional expectations $\left(\mathbb{E}_{t-i}-\mathbb{E}_{t-(i+1)}\right),\ i\in\{0,\ldots,H-1\},$
of the left and right hand side of the SDE.}

In analogy to the proof of Theorem \ref{thm:multivariate-smith-constraints-1}
and Theorem \ref{thm:recursive_then_RE}, we obtain for the SDE with
the structured revision processes $\varepsilon^{j}$ that

{\footnotesize{}
\[
\begin{array}{l}
\underbrace{\begin{pmatrix}1\\
 & z\\
 &  & \ddots\\
 &  &  & z^{H-1}
\end{pmatrix}}_{=B(z)}\underbrace{\begin{pmatrix}\omega_{0} & 0 & \cdots & 0\\
\omega_{1} & \omega_{0} & \ddots & \vdots\\
\vdots &  & \ddots & 0\\
\omega_{H-1} & \cdots & \omega_{1} & \omega_{0}
\end{pmatrix}}_{=W}\left(\mathbb{E}_{t}-\mathbb{E}_{t-1}\right)\begin{pmatrix}y_{t}\\
y_{t+1}\\
\vdots\\
y_{t+H-1}
\end{pmatrix}=\\
\quad=\begin{pmatrix}1\\
 & z\\
 &  & \ddots\\
 &  &  & z^{H-1}
\end{pmatrix}\begin{pmatrix}\omega_{0} & 0 & \cdots & 0\\
\omega_{1} & \omega_{0} & \ddots & \vdots\\
\vdots &  & \ddots & 0\\
\omega_{H-1} & \cdots & \omega_{1} & \omega_{0}
\end{pmatrix}\begin{pmatrix}\varepsilon_{t}^{0}\\
\varepsilon_{t}^{1}\\
\vdots\\
\varepsilon_{t}^{H-1}
\end{pmatrix}+\cdots\\
\quad\quad\cdots+\begin{pmatrix}1\\
 & z\\
 &  & \ddots\\
 &  &  & z^{H-1}
\end{pmatrix}\left(\mathbb{E}_{t}-\mathbb{E}_{t-1}\right)\left(\alpha(z)^{-1}P(z)^{-1}\right)\begin{pmatrix}\zeta_{t-J_{1}}-u_{t-J_{1}}\\
\vdots\\
\zeta_{t-1}-u_{t-1}\\
\hline \zeta_{t}-u_{t}\\
\vdots\\
\zeta_{t+\left(H-J_{1}\right)-1}-u_{t+\left(H-J_{1}\right)-1}
\end{pmatrix}.
\end{array}
\]
}In short notation, this equation is 
\[
\begin{array}{l}
B(z)W\left(\mathbb{E}_{t}-\mathbb{E}_{t-1}\right)\begin{pmatrix}y_{t}\\
y_{t+1}\\
\vdots\\
y_{t+H-1}
\end{pmatrix}=\cdots\\
\quad\cdots=B(z)W\begin{pmatrix}\varepsilon_{t}^{0}\\
\varepsilon_{t}^{1}\\
\vdots\\
\varepsilon_{t}^{H-1}
\end{pmatrix}+B(z)\left(\mathbb{E}_{t}-\mathbb{E}_{t-1}\right)\left(\alpha(z)^{-1}P(z)^{-1}\right)\begin{pmatrix}\zeta_{t-J_{1}}-u_{t-J_{1}}\\
\vdots\\
\zeta_{t-1}-u_{t-1}\\
\hline \zeta_{t}-u_{t}\\
\vdots\\
\zeta_{t+\left(H-J_{1}\right)-1}-u_{t+\left(H-J_{1}\right)-1}
\end{pmatrix}.
\end{array}
\]

\paragraph{Step 2: Apply $B^{-1}(z)$ and use the facts that $\left(\mathbb{E}_{t}-\mathbb{E}_{t-1}\right)\left(y_{t+j}^{s_{i}}\right)=0$
and $\varepsilon_{t}^{j,s_{i}}=0$ for $i>j$.}

We define{\tiny{}
\[
R=\left(\begin{array}{cccc}
\left(\begin{array}{cc}
I_{s_{0}} & 0_{s_{0}\times s_{1}+\cdots+s_{H}}\end{array}\right)\\
 & \left(\begin{array}{cc}
I_{s_{0}+s_{1}} & 0_{s_{0}+s_{1}\times s_{2}+\cdots+s_{H}}\end{array}\right)\\
 &  & \ddots\\
 &  &  & \left(\begin{array}{cc}
I_{s_{0}+\cdots+s_{H-1}} & 0_{s_{0}+\cdots+s_{H-1}\times s_{H}}\end{array}\right)
\end{array}\right).
\]
}of dimension $\left(\underbrace{\left(\sum_{i=0}^{H-1}s_{i}\cdot\left(H-i\right)\right)}_{=P}\times Hs\right)$
and note that $R^{T}R\begin{pmatrix}\varepsilon_{t}^{0}\\
\varepsilon_{t}^{1}\\
\vdots\\
\varepsilon_{t}^{H-1}
\end{pmatrix}=\begin{pmatrix}\varepsilon_{t}^{0}\\
\varepsilon_{t}^{1}\\
\vdots\\
\varepsilon_{t}^{H-1}
\end{pmatrix}$ and $R^{T}R\left(\mathbb{E}_{t}-\mathbb{E}_{t-1}\right)\begin{pmatrix}y_{t}\\
y_{t+1}\\
\vdots\\
y_{t+H-1}
\end{pmatrix}=\left(\mathbb{E}_{t}-\mathbb{E}_{t-1}\right)\begin{pmatrix}y_{t}\\
y_{t+1}\\
\vdots\\
y_{t+H-1}
\end{pmatrix}$ hold. The selector matrix $R$ deletes all components in $\begin{pmatrix}\varepsilon_{t}^{0}\\
\varepsilon_{t}^{1}\\
\vdots\\
\varepsilon_{t}^{H-1}
\end{pmatrix}$ and $\left(\mathbb{E}_{t}-\mathbb{E}_{t-1}\right)\begin{pmatrix}y_{t}\\
y_{t+1}\\
\vdots\\
y_{t+H-1}
\end{pmatrix}$ which are zero due to the predeterminedness assumption, i.e. {\footnotesize{}
\[
\begin{array}{l}
\underbrace{\begin{pmatrix}1\\
 & z\\
 &  & \ddots\\
 &  &  & z^{H-1}
\end{pmatrix}}_{=B(z)}\underbrace{\begin{pmatrix}\omega_{0} & 0 & \cdots & 0\\
\omega_{1} & \omega_{0} & \ddots & \vdots\\
\vdots &  & \ddots & 0\\
\omega_{H-1} & \cdots & \omega_{1} & \omega_{0}
\end{pmatrix}}_{=W}\left(\mathbb{E}_{t}-\mathbb{E}_{t-1}\right)\begin{pmatrix}y_{t}\\
y_{t+1}\\
\vdots\\
y_{t+H-1}
\end{pmatrix}=\\
\quad=\begin{pmatrix}1\\
 & z\\
 &  & \ddots\\
 &  &  & z^{H-1}
\end{pmatrix}\begin{pmatrix}\omega_{0} & 0 & \cdots & 0\\
\omega_{1} & \omega_{0} & \ddots & \vdots\\
\vdots &  & \ddots & 0\\
\omega_{H-1} & \cdots & \omega_{1} & \omega_{0}
\end{pmatrix}\begin{pmatrix}\varepsilon_{t}^{0}\\
\varepsilon_{t}^{1}\\
\vdots\\
\varepsilon_{t}^{H-1}
\end{pmatrix}+\cdots\\
\quad\quad\cdots+\begin{pmatrix}1\\
 & z\\
 &  & \ddots\\
 &  &  & z^{H-1}
\end{pmatrix}\left(\mathbb{E}_{t}-\mathbb{E}_{t-1}\right)\left(\alpha(z)^{-1}P(z)^{-1}\right)\begin{pmatrix}\zeta_{t-J_{1}}-u_{t-J_{1}}\\
\vdots\\
\zeta_{t-1}-u_{t-1}\\
\hline \zeta_{t}-u_{t}\\
\vdots\\
\zeta_{t+\left(H-J_{1}\right)-1}-u_{t+\left(H-J_{1}\right)-1}
\end{pmatrix}.
\end{array}
\]
}
\[
R\begin{pmatrix}\varepsilon_{t}^{0}\\
\varepsilon_{t}^{1}\\
\vdots\\
\varepsilon_{t}^{H-1}
\end{pmatrix}=\left(\begin{array}{c}
\varepsilon^{0,s_{0}}\\
\hline \substack{\varepsilon^{1,s_{0}}\\
\varepsilon^{1,s_{1}}
}
\\
\hline \substack{\varepsilon^{2,s_{0}}\\
\varepsilon^{2,s_{1}}\\
\varepsilon^{2,s_{2}}
}
\\
\hline \substack{\varepsilon^{3,s_{0}}\\
\varepsilon^{3,s_{1}}\\
\varepsilon^{3,s_{2}}\\
\varepsilon^{3,s_{3}}
}
\\
\hline \vdots\\
\hline \substack{\varepsilon^{H-2,s_{0}}\\
\vdots\\
\varepsilon^{H-2,s_{H-2}}
}
\\
\hline \substack{\varepsilon^{H-1,s_{0}}\\
\vdots\\
\varepsilon^{H-1,s_{H-1}}
}
\end{array}\right)
\]
holds. We denote the $P$-dimensional vectors $R\begin{pmatrix}\varepsilon_{t}^{0}\\
\varepsilon_{t}^{1}\\
\vdots\\
\varepsilon_{t}^{H-1}
\end{pmatrix}$ and $R\left(\mathbb{E}_{t}-\mathbb{E}_{t-1}\right)\begin{pmatrix}y_{t}\\
y_{t+1}\\
\vdots\\
y_{t+H-1}
\end{pmatrix}$ by $\varepsilon$ and $y$ respectively. Thus, we obtain{\scriptsize{}
\[
WR^{T}\underbrace{\left[R\left(\mathbb{E}_{t}-\mathbb{E}_{t-1}\right)\begin{pmatrix}y_{t}\\
y_{t+1}\\
\vdots\\
y_{t+H-1}
\end{pmatrix}\right]}_{=y}=WR^{T}\underbrace{\left[R\begin{pmatrix}\varepsilon_{t}^{0}\\
\varepsilon_{t}^{1}\\
\vdots\\
\varepsilon_{t}^{H-1}
\end{pmatrix}\right]}_{=\varepsilon}+\left(\mathbb{E}_{t}-\mathbb{E}_{t-1}\right)\left(\alpha(z)^{-1}P(z)^{-1}\right)\begin{pmatrix}\zeta_{t-J_{1}}-u_{t-J_{1}}\\
\vdots\\
\zeta_{t-1}-u_{t-1}\\
\hline \zeta_{t}-u_{t}\\
\vdots\\
\zeta_{t+\left(H-J_{1}\right)-1}-u_{t+\left(H-J_{1}\right)-1}
\end{pmatrix}.
\]
}{\scriptsize\par}

\paragraph{Step 3: Pre-multiply $S$ defined in equation (\ref{eq:pseudo_inverse_selector})
and conclude about non-singularity of $\tilde{W}=SWR^{T}$. }

By pre-multiplying 
\[
S=\left(\begin{array}{cccccc}
\omega_{0,s_{0}}^{\dagger}\\
 & \omega_{0,s_{0}+s_{1}}^{\dagger}\\
 &  & \omega_{0,s_{0}+s_{1}+s_{2}}^{\dagger}\\
 &  &  & \ddots\\
 &  &  &  & \omega_{0,s_{0}+\cdots+s_{H-2}}^{\dagger}\\
 &  &  &  &  & \omega_{0,s_{0}+\cdots+s_{H-1}}^{\dagger}
\end{array}\right),
\]
we obtain from above that
\[
\begin{array}{l}
\underbrace{SWR^{T}}_{=\tilde{W}}\underbrace{R\left(\mathbb{E}_{t}-\mathbb{E}_{t-1}\right)\begin{pmatrix}y_{t}\\
y_{t+1}\\
\vdots\\
y_{t+H-1}
\end{pmatrix}}_{=y}=\\
\quad=\tilde{W}\underbrace{R\begin{pmatrix}\varepsilon_{t}^{0}\\
\varepsilon_{t}^{1}\\
\vdots\\
\varepsilon_{t}^{H-1}
\end{pmatrix}}_{=\varepsilon}+S\left(\mathbb{E}_{t}-\mathbb{E}_{t-1}\right)\left(\alpha(z)^{-1}P(z)^{-1}\right)\underbrace{\begin{pmatrix}\zeta_{t-J_{1}}-u_{t-J_{1}}\\
\vdots\\
\zeta_{t-1}-u_{t-1}\\
\hline \zeta_{t}-u_{t}\\
\vdots\\
\zeta_{t+\left(H-J_{1}\right)-1}-u_{t+\left(H-J_{1}\right)-1}
\end{pmatrix}}_{=\zeta}.
\end{array}
\]
\begin{equation}
\iff\tilde{W}y=\tilde{W}\varepsilon+S\left(\mathbb{E}_{t}-\mathbb{E}_{t-1}\right)\left(\alpha(z)^{-1}P(z)^{-1}\zeta\right).\label{eq:predetermined_coeff_comp}
\end{equation}
Since $W$ is a block lower-triangular matrix whose matrix in the
diagonal block is non-singular, it follows that any selection of columns
has full rank. Pre-multiplying the block diagonal matrix with the
corresponding Moore-Penrose pseudo-inverse $S$ of the selected columns
of the matrix in the diagonal block of $W$, we obtain a block lower-triangular
matrix whose diagonal blocks consist of identity matrices. Thus, $\tilde{W}$
is non-singular.

\paragraph{Step 4: Conclude that a solution of the RE model is also a solution
of the SDE and the restrictions hold.}

If $\left(y_{t}\right)_{t\in\mathbb{Z}}\in H_{\varepsilon}(t,\gamma)$
is a solution of the RE model, then it obviously solves the SDE whose
revision processes are of the special form stated in the theorem.
Moreover, if $\left(\mathbb{E}_{t}-\mathbb{E}_{t-1}\right)\left(y_{t+j}^{s_{i}}\right)=\varepsilon_{t}^{j,s_{i}}$
for $j\in\left\{ 0,\ldots,H-1\right\} $, $i\in\left\{ 0,\ldots,H\right\} $,
then it follows from equation (\ref{eq:predetermined_coeff_comp})
that 
\[
S\left(\mathbb{E}_{t}-\mathbb{E}_{t-1}\right)\left(\alpha(z)^{-1}P(z)^{-1}\right)\begin{pmatrix}\zeta_{t-J_{1}}-u_{t-J_{1}}\\
\vdots\\
\zeta_{t-1}-u_{t-1}\\
\hline \zeta_{t}-u_{t}\\
\vdots\\
\zeta_{t+\left(H-J_{1}\right)-1}-u_{t+\left(H-J_{1}\right)-1}
\end{pmatrix}=0.
\]

\paragraph{Step 5: Conclude that a solution of the SDE for which the MDS satisfy
the constraints is also a solution of the RE model.}

If $\left(y_{t}\right)_{t\in\mathbb{Z}}\in H_{\varepsilon}(t,\gamma)$
solves the SDE with the MDS of the special form outlined in the theorem
and if 
\[
S\left(\mathbb{E}_{t}-\mathbb{E}_{t-1}\right)\left(\alpha(z)^{-1}P(z)^{-1}\right)\begin{pmatrix}\zeta_{t-J_{1}}-u_{t-J_{1}}\\
\vdots\\
\zeta_{t-1}-u_{t-1}\\
\hline \zeta_{t}-u_{t}\\
\vdots\\
\zeta_{t+\left(H-J_{1}\right)-1}-u_{t+\left(H-J_{1}\right)-1}
\end{pmatrix}=0
\]
holds, it follows from equation (\ref{eq:predetermined_coeff_comp})
that

\[
\tilde{W}R\left(\mathbb{E}_{t}-\mathbb{E}_{t-1}\right)\begin{pmatrix}y_{t}\\
y_{t+1}\\
\vdots\\
y_{t+H-1}
\end{pmatrix}=\tilde{W}R\begin{pmatrix}\varepsilon_{t}^{0}\\
\varepsilon_{t}^{1}\\
\vdots\\
\varepsilon_{t}^{H-1}
\end{pmatrix}.
\]
 Thus, the MDS $\varepsilon_{t}^{j}$ correspond to the revision processes
of $\left(y_{t}\right)_{t\in\mathbb{Z}}$ and the RE model can be
reconstructed from the SDE.

\pagebreak{}

\section{Proof of Theorem \ref{thm:predetermined_dimension}}

\paragraph{Step 1: Start from equation (\ref{eq:predetermined_constraints_zeta})
and perform the same steps as in the case without predetermined variables
where we obtained equation (\ref{eq:general_restrictions_in_detail}). }

\subparagraph{Case $0\protect\leq g_{1}\protect\leq\cdots\protect\leq g_{j}\protect\leq J_{1}<g_{j+1}\protect\leq\cdots\protect\leq g_{s}$
and case $g_{i}\protect\leq J_{1}$ : }

From Section \ref{sec:Proof-of-Theorem General}, we know that (\ref{eq:predetermined_constraints_zeta})
is equivalent to
\[
SU^{T}U\left(\mathbb{E}_{t}-\mathbb{E}_{t-1}\right)\left\{ \begin{pmatrix}1\\
z^{-1}\\
\vdots\\
z^{-H+1}
\end{pmatrix}\otimes\left[\begin{pmatrix}z^{-g_{1}}\\
 & \ddots\\
 &  & z^{-g_{s}}
\end{pmatrix}\begin{pmatrix}\mathfrak{P}_{1,\bullet}(z)\\
\vdots\\
\mathfrak{P}_{s,\bullet}(z)
\end{pmatrix}\left(\zeta_{t-J_{1}}-u_{t-J_{1}}\right)\right]\right\} =0.
\]
where $U$ is given in equation (\ref{eq:permutation_matrix}). Performing
the same steps as in the proof of Theorem \ref{thm:general-solution-set},
we obtain that this equation is equivalent to
\[
SU^{T}\begin{pmatrix}\mathfrak{P}^{1,\bullet}\\
\vdots\\
\mathfrak{P}^{s,\bullet}
\end{pmatrix}\left[\begin{pmatrix}m_{0,\bullet}\\
\vdots\\
m_{H-1+\gamma_{s},\bullet}
\end{pmatrix}R^{T}\varepsilon_{t}^{p,\bullet}-\left(\mathbb{E}_{t}-\mathbb{E}_{t-1}\right)\begin{pmatrix}u_{t}\\
\vdots\\
u_{t+(H+\gamma_{s}-1)}
\end{pmatrix}\right]=0.
\]

\subparagraph{Case $g_{i}=\bar{g}\protect\leq J_{1}$ and case $g_{i}=0$:}

Since
\[
S=\begin{pmatrix}\begin{array}{lc}
S_{1,\bar{g}} & 0\\
0 & S_{2,\bar{g}}
\end{array}\end{pmatrix}=\left(\begin{array}{ccc|ccc}
\omega_{0,s_{0}}^{\dagger} &  & \\
 & \ddots & \\
 &  & \omega_{0,s_{0}+\cdots+s_{J_{1}-\bar{g}-1}}^{\dagger}\\
\hline  &  &  & \omega_{0,s_{0}+\cdots+s_{J_{1}-\bar{g}}}^{\dagger}\\
 &  &  &  & \ddots\\
 &  &  &  &  & \omega_{0,s_{0}+\cdots+s_{H-1}}^{\dagger}
\end{array}\right),
\]
 and since 
\[
U^{T}\begin{pmatrix}\mathfrak{P}^{1,\bullet}\\
\vdots\\
\mathfrak{P}^{s,\bullet}
\end{pmatrix}=\left(\begin{array}{ccc}
 & 0_{s\left(J_{1}-\bar{g}\right)\times s\left(H-J_{1}+\bar{g}\right)}\\
\hline \mathfrak{P}_{\bullet,\bullet|0}\\
\vdots & \ddots\\
\mathfrak{P}_{\bullet,\bullet|H-J_{1}+\bar{g}-1} & \cdots & \mathfrak{P}_{\bullet,\bullet|0}
\end{array}\right)
\]
we obtain that
\[
SU^{T}\begin{pmatrix}\mathfrak{P}^{1,\bullet}\\
\vdots\\
\mathfrak{P}^{s,\bullet}
\end{pmatrix}\left[\begin{pmatrix}m_{0,\bullet}\\
\vdots\\
m_{H-1+\gamma_{s},\bullet}
\end{pmatrix}R^{T}\varepsilon_{t}^{p,\bullet}-\left(\mathbb{E}_{t}-\mathbb{E}_{t-1}\right)\begin{pmatrix}u_{t}\\
\vdots\\
u_{t+(H+\gamma_{s}-1)}
\end{pmatrix}\right]=0
\]
simplifies to
\[
S_{2,\bar{g}}\left(\begin{array}{ccc}
\mathfrak{P}_{\bullet,\bullet|0}\\
\vdots & \ddots\\
\mathfrak{P}_{\bullet,\bullet|H-J_{1}+\bar{g}-1} & \cdots & \mathfrak{P}_{\bullet,\bullet|0}
\end{array}\right)\left[\begin{pmatrix}m_{0,\bullet}\\
\vdots\\
m_{H-J_{1}+\bar{g}-1,\bullet}
\end{pmatrix}R^{T}\varepsilon_{t}^{p,\bullet}-\left(\mathbb{E}_{t}-\mathbb{E}_{t-1}\right)\begin{pmatrix}u_{t}\\
\vdots\\
u_{t+(H-J_{1}+\bar{g}-1)}
\end{pmatrix}\right]=0.
\]
 $\left(P-\sum_{i=0}^{J_{1}-1}s_{i}\cdot(J_{1}-i)\right)\times P$,
so that the kernel of this equation system has dimension $\sum_{i=0}^{J_{1}-1}s_{i}\cdot\left(J_{1}-i\right).$ 

\paragraph{Step 2: Count the number of free parameters, using $\varepsilon_{t}^{j,s_{i}}=h_{j,s_{i}}\varepsilon_{t}$,
where $h_{j,s_{i}}\in\mathbb{R}^{s_{i}\times q}$.}

Since 
\[
R\begin{pmatrix}\varepsilon_{t}^{0}\\
\varepsilon_{t}^{1}\\
\vdots\\
\varepsilon_{t}^{H-1}
\end{pmatrix}=\left(\begin{array}{c}
\varepsilon_{t}^{0,s_{0}}\\
\hline \substack{\varepsilon_{t}^{1,s_{0}}\\
\varepsilon_{t}^{1,s_{1}}
}
\\
\hline \substack{\varepsilon_{t}^{2,s_{0}}\\
\varepsilon_{t}^{2,s_{1}}\\
\varepsilon_{t}^{2,s_{2}}
}
\\
\hline \substack{\varepsilon_{t}^{3,s_{0}}\\
\varepsilon_{t}^{3,s_{1}}\\
\varepsilon_{t}^{3,s_{2}}\\
\varepsilon_{t}^{3,s_{3}}
}
\\
\hline \vdots\\
\hline \substack{\varepsilon_{t}^{H-2,s_{0}}\\
\vdots\\
\varepsilon_{t}^{H-2,s_{H-2}}
}
\\
\hline \substack{\varepsilon_{t}^{H-1,s_{0}}\\
\vdots\\
\varepsilon_{t}^{H-1,s_{H-1}}
}
\end{array}\right)=\left(\begin{array}{c}
h_{0,s_{0}}\\
\hline \substack{h_{1,s_{0}}\\
h_{1,s_{0}}
}
\\
\hline \substack{h_{2,s_{0}}\\
h_{2,s_{1}}\\
h_{2,s_{2}}
}
\\
\hline \substack{h_{3,s_{0}}\\
h_{3,s_{1}}\\
h_{3,s_{2}}\\
h_{3,s_{3}}
}
\\
\hline \vdots\\
\hline \substack{h_{H-2,s_{0}}\\
\vdots\\
h_{H-2,s_{H-2}}
}
\\
\hline \substack{h_{H-1,s_{0}}\\
\vdots\\
h_{H-1,s_{H-1}}
}
\end{array}\right)\varepsilon_{t},
\]
it follows that there are $\left(\sum_{i=0}^{J_{1}-1}s_{i}\cdot\left(J_{1}-i\right)\right)q$
free parameters. 

\paragraph{Step 3: Conclude on dimension of solution set.}

Analogous to the case without predetermined variables.
\end{document}